\documentclass[%
 aps,
 prb,
 amsmath,amssymb,
 showpacs,
 reprint,%
 nofootinbib,
]{revtex4-2}

\usepackage{graphicx}
\usepackage{dcolumn}
\usepackage[T1]{fontenc} 
\usepackage{bm}
\usepackage{verbatim}

\begin{document}

\title{A mechanism for nonmonotonic $T_{c,{\rm max}}(n)$ in multilayer cuprates}

\author{Pavel Kornilovitch}
 \email{pavel.kornilovich@gmail.com}
 \affiliation{Department of Physics, Oregon State University, Corvallis, Oregon 97331, USA} 

\date{\today}  

\begin{abstract}

We propose an explanation of the observed dependence of the maximal critical temperature $T_{c,{\rm max}}$ on the number of conducting layers $n$ in layered copper-oxide superconductors within the preformed pair mechanism. Copper-oxygen planes fine-tune the lattice anisotropy and regulate the balance between the attractive and kinetic energies of carrier holes. To maximize the Bose-Einstein condensation temperature, real-space pairs must be compact and light at the same time. Generally, $T_{c,{\rm max}}$ increases between $n = 1$ and $n = 3$ because pairs become lighter. For $n > 3$, the rising kinetic energy weakens the pairs, leading to inflated pair volumes and reduced $T_{c,{\rm max}}$. By varying model parameters, the peak of $T_{c,{\rm max}}(n)$ can be tuned to $n = 2$, $n = 3$, or $n > 3$. We also discuss strategies for using this knowledge to boost $T_{c,{\rm max}}$ beyond the current record of 138 K.      

\end{abstract}



\maketitle

\section{\label{MLH:sec:one}
Introduction 
}

\begin{figure}[b]
\includegraphics[width=0.48\textwidth]{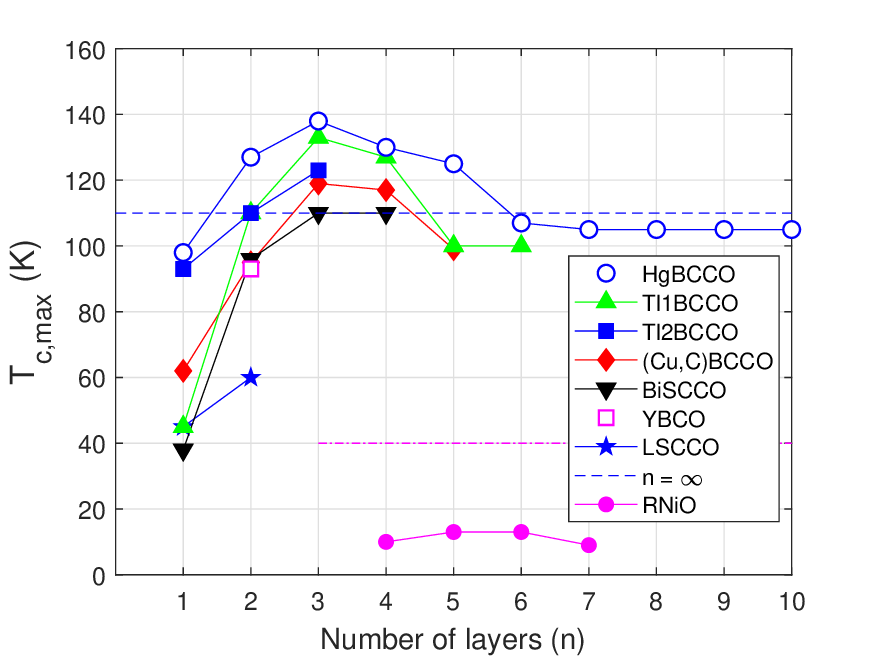}
\caption{The highest reported $T_c$ for the seven cuprate families and one nickelate family of superconductors. Acronyms and data sources are given in Table~\ref{MLH:tab:one}. The dashed line marks $T_c = 110$ K for the ``infinite-layer'' superconductor (Sr,Ca)CuO$_2$ \cite{Azuma1992}. The dot-dashed line marks $T_c = 40$ K of the ``infinite-layer'' superconductor ReNiO$_2$ \cite{Chow2025}.}  
\label{MLH:fig:zero}
\end{figure}

Dependence of {\em maximal} critical temperature $T_{c,{\rm max}}$ on the number of conducting layers $n$ in the unit cell is a long-standing puzzle of high-temperature superconductivity. $T_{c,{\rm max}}(n)$ is intimately linked with the search for an underlying microscopic mechanism of superconductivity and with continuing attempts to raise experimental $T_c$ toward room temperature. In most copper-oxide families, $T_{c,{\rm max}}$ peaks at $n = 3$ and $4$ and falls off on either side of the peak \cite{Chou1992,Cava1990,Liang1998,Tokunaga2000,Kotegawa2001,Ono2000,Eisaki2004,Tarascon1988,Saoudel2021,Manako1994,Nakajima1990,Iyo2001,Kotegawa2004,Hirai2007,Mukuda2016,Wagner1997,Shimakawa1989,Thompson1993,Putilin1993,Schilling1993,Scott1994,Tokiwa1996,Karpinski1999,Lokshin2001,Yamamoto2001,Antipov2002,Iyo2007,Azuma1992,Pan2022,Chow2025,Pan2026}, see Table~\ref{MLH:tab:one} and Fig.~\ref{MLH:fig:zero}. The highest reported $T_{c,{\rm max}}$ (at ambient pressure) is 138 K in the fluorinated $n = 3$ Hg family member HgBa$_2$Ca$_2$Cu$_3$O$_{8+\delta}$ \cite{Lokshin2001}. The largest reported number of layers is $n = 16$, also within the Hg family \cite{Iyo2007}. 

In discussing $T_{c,{\rm max}}(n)$, an important consideration is the existence of the ``infinite-layer'' ($n = \infty$) superconductor (Ca,Sr)CuO$_{2}$ with a remarkably high $T_c = 110$ K \cite{Azuma1992}.~\footnote{There has been hardly any confirmation of this result since the original report in 1992. Nonetheless, we chose to use it here because it fits well within the logic of cuprate superconductivity.} All other cuprates can be constructed \cite{Iyo2007} by replacing every $n$th Ca$^{2+}$ layer with a ``charge-reservoir layer'' comprised of Bi, Tl, Sr, Hg, Cu, C, O, rare-earth Re, or halogen ions.~\footnote{In the physical picture developed in this work, the charge-reservoir layers also play an important dynamic role by contributing to attraction between holes.} Therefore $n = \infty$ is a convenient starting point for understanding the entire $T_{c,{\rm max}}(n)$ dependence. The data of Table~\ref{MLH:tab:one} show that in most cases, insertion of charge reservoirs {\em decreases} $T_{c,{\rm max}}$. This is true for all $n = 1$ compounds and for all $n = 2$ compounds except Hg-1212. Only a handful of top-performing members of Cu, Tl, and Hg families with $n = 3$ and 4 overcome that drop and exceed the starting value of 110 K. Charge reservoirs both restrict and enhance superconductivity at the same time, and that requires an explanation. Simple logic also dictates that $T_{c,{\rm max}}$ must approach 110 K at large $n$. Indeed, $T_{c,{\rm max}}(n > 4)$ appears to stabilize within the Hg and 1-layer Tl families, albeit at values slightly below 110 K. For other cuprate families, the available data is insufficient to reach a conclusion.

\begin{table*}[t]
\begin{center}
\renewcommand{\tabcolsep}{0.2cm}
\renewcommand{\arraystretch}{1.5}
\begin{tabular}{|c||c|c|c|c|c|c|c||c|}
\hline\hline
    n    & LSCCO  &  YBCO  & (Cu,C)BCCO & BiSCCO & Tl$_1$BCCO & Tl$_2$BCCO & HgBCCO  & RNiO  \\ \hline
\hline 
    1    &   45   &        &      62    &    38  &      45    &      93    &    98   &       \\ \hline
    2    &   60   &   93   &      95    &    96  &     110    &     110    &   127   &       \\ \hline
    3    &        &        &     119    &   110  &     133    &     123    &   138   &       \\ \hline
    4    &        &        &     117    &   110  &     127    &            &   130   &   10  \\ \hline
    5    &        &        &      99    &        &     100    &            &   125   &   13  \\ \hline
    6    &        &        &            &        &     100    &            &   107   &   13  \\ \hline
    7    &        &        &            &        &            &            &   105   &    9  \\ \hline
   8-16  &        &        &            &        &            &            &   105   &       \\ \hline
$\infty$ &  110   &  110   &     110    &   110  &     110    &     110    &   110   &   40  \\ \hline
\hline 
\end{tabular}
\end{center}
\caption{
Critical temperatures of several multilayer families of oxide superconductors. For each $n$, the maximal $T_c$ reported in the literature is shown. References: La$_{2-x}$Sr$_{x}$Ca$_{n-1}$Cu$_{n}$O$_{y}$ \cite{Chou1992,Cava1990}, YBa$_2$Cu$_{3}$O$_{7-y}$ \cite{Liang1998}, (Cu,C)Ba$_2$Ca$_{n-1}$Cu$_{n}$O$_y$ \cite{Iyo2007,Tokunaga2000,Kotegawa2001,Hirai2007}, Bi$_2$Sr$_2$Ca$_{n-1}$Cu$_n$O$_y$ \cite{Ono2000,Eisaki2004,Tarascon1988,Saoudel2021}, TlBa$_2$Ca$_{n-1}$Cu$_n$O$_y$ \cite{Manako1994,Nakajima1990,Iyo2001,Kotegawa2004,Mukuda2016}, Tl$_2$Ba$_2$Ca$_{n-1}$Cu$_n$O$_y$ \cite{Wagner1997,Shimakawa1989,Thompson1993}, HgBa$_2$Ca$_{n-1}$Cu$_n$O$_y$ \cite{Putilin1993,Schilling1993,Scott1994,Tokiwa1996,Karpinski1999,Lokshin2001,Yamamoto2001,Antipov2002,Iyo2007}, R$_{n+1}$Ni$_{n}$O$_{y}$ \cite{Pan2022,Chow2025,Pan2026}. For cuprates, $n = \infty$ corresponds to the ``infinite-layer'' compound (Sr$_{1-x}$Ca$_x$)$_{1-y}$CuO$_2$ \cite{Azuma1992}. For nickelates, $n = \infty$ corresponds to the ``infinite-layer'' compound ReNiO$_2$ \cite{Chow2025}.   
} 
\label{MLH:tab:one}
\end{table*}

Since the discovery of superconductivity in nickel oxides in 2019 \cite{Li2019}, a similar picture has been emerging in nickelate compounds. Within the square-planer family Re$_{n+1}$Ni$_{n}$O$_{2n+2}$, the infinite-layer member ReNiO$_2$ has the largest $T_{c,{\rm max}} = 40$ K \cite{Chow2025}. All other members studied so far have $T_{c,{\rm max}}$ below that \cite{Pan2022,Pan2026}, see Table~\ref{MLH:tab:one}. In the nickelates, charge reservoirs appear to impede superconductivity more than enhance it. The nickelates are being actively studied and the experimental picture changes rapidly. For that reason, we will not be discussing square-planer nickelates in this paper. Also excluded is the octahedral nickelate family Re$_{n+1}$Ni$_{n}$O$_{3n+1}$ which in most cases becomes superconducting only under external pressure \cite{Sun2023,Zhu2024,Zhou2025}. Pressure-induced superconductivity is outside of this work's scope. The vast class of Fe-based superconductors with $T_c$ as high at 58 K are based on the same conducting layer FeA (A: As, P, Se, or Te), see for example \cite{Hosono2015} for a review. For the purposes of this work, all Fe-based materials correspond to $n = 1$. Their differences originate from different charge reservoirs rather than from the number of conducting layers within a unit cell. We therefore focus exclusively on the cuprates.       

Over the years, several explanations of the empirical $T_{c,{\rm max}}(n)$ dependence were proposed. Early on, Wheatley, Hsu, and Anderson \cite{Wheatley1988} derived a phenomenological formula based on the assumption than ``superconducting coupling'' between CuO$_2$ layers within an $n$-stack was larger than between neighboring stacks. With increasing $n$, superconductivity grows stronger ``on average''. Their formula predicted a monotonic increase of $T_{c,{\rm max}}(n)$, which was later invalidated by experiment. Leggett proposed that increasing $n$ systematically lowered the Coulomb energy within a stack of CuO$_2$ layers, and hence boosted superconductivity \cite{Leggett1999}. His formula also predicted a monotonically increasing $T_{c,{\rm max}}(n)$, which was criticized on the same experimental grounds \cite{Jansen2000}. A comprehensive review of site-selective nuclear magnetic resonance data by Mukuda {\em et al} \cite{Mukuda2012} attributed $T_{c,{\rm max}}(n)$ to the $n$-dependence of magnetic superexchange interaction within the $n$-stack of CuO$_2$ layers, but no reason for decreasing $T_{c,{\rm max}}$ at $n > 3$ was given. More recently, Ideta {\em et al} \cite{Ideta2025} argued that $T_{c,{\rm max}}$ peaks at $n = 3$ because the inner CuO$_2$ plane experiences a strong proximity effect from two optimally doped outer planes, which produces the most robust superconductivity. In contrast with all prior explanations, Rosenstein {\em et al} \cite{Rosenstein2025} attributed $T_{c,{\rm max}}(n)$ to the $n$-dependence of {\em phonon-induced} pairing. In their model, $T_{c,{\rm max}}$ rises between $n = 1$ and $n = 3$ but then saturates, the latter again being incompatible with the body of experimental data.         

In this paper, nonmonotonic $T_{c,{\rm max}}(n)$ is explained within the general framework of preformed pair superconductivity, also known as real-space pairing, Bose-Einstein condensation (BEC), or Blatt-Butler-Schafroth, superconductivity \cite{Ogg1946,Schafroth1954,Schafroth1957,Blatt1964,Eagles1969,Leggett1980,Nozieres1985,Alexandrov1981,Micnas1990,Alexandrov1993b,Alexandrov1994,Uemura1989,Uemura1991,Chen2005,Leggett2006,Chen2024}. We argue that $T_{c,{\rm max}}$ listed in Table~\ref{MLH:tab:one} is very simply embodied in the properties of real-space pairs through the anisotropic BEC formula \cite{Kornilovitch2015,Ivanov1994,Kornilovitch2024}: 
\begin{equation}
T_{c,{\rm max}} \propto \frac{1}{ ( m^{\ast}_x m^{\ast}_y m^{\ast}_z )^{1/3} \Omega^{2/3}_{p}} \: , 
\label{MLH:eq:zero}
\end{equation}
where $m^{\ast}_{i}$ is the pair effective mass and $\Omega_{p}$ is the pair volume. Preformed pair superconductivity benefits from small pairs: since the condensation temperature increases with density, a small $\Omega_{p}$ enables more pairs to be packed before they start to overlap and the BEC picture breaks down. Thus, a high $T_{c,{\rm max}}$ requires pairs to be both light and compact at the same time. This immediately introduces {\em anisotropy tradeoff} \cite{Kornilovitch2015}. At strong anisotropy, the out-of-plane mass $m^{\ast}_{z}$ is very large and, according to Eq.~(\ref{MLH:eq:zero}), superconductivity is suppressed due the lack of 3D phase coherence. In a fully isotropic system, kinetic energy is large and pairs are weakly bound, or unbound altogether, resulting in a large $\Omega_{p}$ and low close-packing density. According to Eq.~(\ref{MLH:eq:zero}), $T_{c,{\rm max}}$ is suppressed again. Therefore, there exists an {\em optimal level of anisotropy} at which $T_{c,{\rm max}}$ is maximized.       

\begin{figure}[t]
\includegraphics[width=0.48\textwidth]{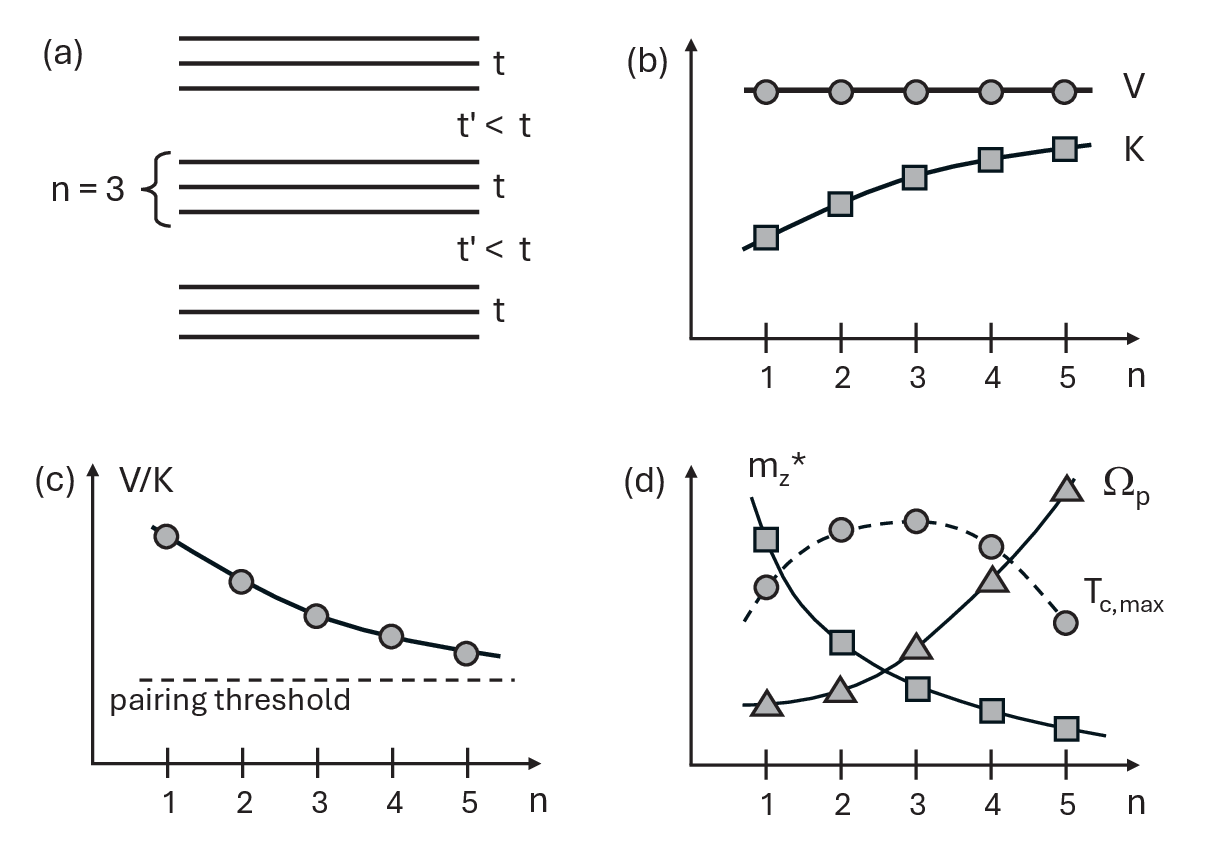}
\caption{Qualitative explanation of nonmonotonic $T_{c,{\rm max}}(n)$. (a) A multilayer lattice with $n = 3$. (b) Qualitative dependence of the attractive energy $V$ and kinetic energy $K$ on the number of layers $n$. (c) The ratio $V/K$ decreases with $n$, approaching the pairing threshold. (d) As the system approaches this threshold, the pair $z$ mass decreases but the pair volume increases. According to Eq.~(\ref{MLH:eq:two}), the close-packing temperature $T^{\ast}_{\rm cp} \propto T_{c,{\rm max}}$ exhibits a broad maximum. } 
\label{MLH:fig:one}
\end{figure}

A key observation made in this paper is that {\em copper-oxygen planes systematically change the out-of-plane kinetic energy and fine-tune the crystal to the optimal anisotro\-py}. In a multilayer cuprate, hopping between planes within a stack is ``faster'' than between stacks; see Fig.~\ref{MLH:fig:one}(a). Additional planes systematically increase the holes' kinetic energy $K$, which defines the energy penalty to be paid when the holes are localized in pairs, Fig.~\ref{MLH:fig:one}(b). At the same time, the attractive strength $V$ depends only weakly on $n$. The ratio $V/K$, which drives pair formation, monotonically decreases with $n$, Fig.~\ref{MLH:fig:one}(c). This is consistent with the doping dependence of the upper critical field \cite{Wang2003}. Implications for $T_{c,{\rm max}}$ are shown in Fig.~\ref{MLH:fig:one}(d). The out-of-plane mass $m^{\ast}_{z}$ decreases with $n$ because the system becomes more isotropic. At the same time, the pair volume $\Omega_{p}$ increases with $n$ because the attraction weakens relative to the kinetic energy, and the pairs become more extended. As a result, $T_{c,{\rm max}}(n)$ displays a broad maximum. The position of this maximum, $n_{\rm max}$, depends on details of the interaction, as discussed later in the paper. One can only say with reasonable confidence that $n_{\rm max} \geq 2$, because $T_{c,{\rm max}}( n = 2 ) > T_{c,{\rm max}}( n = 1 )$ is a universal property of preformed-pair superconductivity. This is discussed in Sec.~\ref{MLH:sec:twofour}.

We continue this paper with a more detailed exposition, in Sec.~\ref{MLH:sec:two}, of the qualitative picture outlined above. A multilayer pairing model is introduced and solved in Sec.~\ref{MLH:sec:three}. The pairs' effective masses, radii, and other properties are calculated exactly. From these results, $T_{c,{\rm max}}(n)$ is inferred and its consequences are discussed. The main result is shown in Fig.~\ref{MLH:fig:zeroeleven}(f), which displays a $T_{c,{\rm max}}(n)$ peaked at $n = 3$, consistent with Fig.~\ref{MLH:fig:zero}. All the technical details of the two-body solution are relegated to extensive Appendices. Section~\ref{MLH:sec:four} contains a discussion of how these findings could help in the search for higher $T_c$.

\section{\label{MLH:sec:two}
Physical picture 
}

\subsection{ \label{MLH:sec:twoone}
Anisotropic preformed pairs
}

The concept that electric superconductivity might arise from the Bose-Einstein condensation (BEC) of electron pairs originates in {\em pre}-BCS ideas of Ogg, Schafroth, Butler, and Blatt \cite{Ogg1946,Schafroth1954,Schafroth1957,Blatt1964}. After the development of BCS theory, the concept was further advanced by several researchers \cite{Eagles1969,Leggett1980,Nozieres1985}. At least one microscopic theory was proposed in which real-space pairs were identified as small phonon bipolarons \cite{Alexandrov1981}. Following the discovery of high-temperature superconductivity in 1986 \cite{Bednorz1986}, preformed pairs became a leading contender for the microscopic mechanism; see \cite{Micnas1990,Alexandrov1993b,Alexandrov1994} for early reviews. Key supporting evidence included the short coherence length \cite{Oh1988,Pan2000,Wang2003,Mangel2024}, Uemura plots \cite{Uemura1989,Uemura1991}, and the normal state pseudogap \cite{Warren1989,Homes1993,Norman2005}, the latter interpreted as the pair binding energy.~\footnote{One must add that the {\em absence} of a normal-state gap was the principal argument against the BEC mechanism in low-temperature superconductors \cite{Casimir1955}.} More recently, direct experimental evidence for real-space pairs in the normal state has come from tunneling experiments \cite{Zhou2019}.~\footnote{Real-space pairs were also reported in Fe-based superconductors \cite{Seo2019,Kang2020} and in BaBiO$_{3}$ \cite{Menushenkov2024}.} 

Nowadays, real-space superconductivity is usually discussed in the context of the ``BCS-BEC crossover''; see \cite{Chen2005,Leggett2006,Chen2024} for reviews. Despite its long history, this general framework has not been developed to the same depth as either of its weak-coupling and strong-coupling limits. Given the absence of a complete theory, in this paper we adopt the simplest approximation of an {\em ideal gas of real-space pairs}, whose $T_c$ is given by the ideal-gas BEC formula. It is important, however, to generalize the latter to anisotropic dispersion, yielding the result \cite{Prelovsek1987}
\begin{equation}
k_B T_c = \frac{3.3 \, \hbar^2 \, \rho_{2}^{2/3}}{ ( m^{\ast}_x m^{\ast}_y m^{\ast}_z )^{1/3} } \: , 
\label{MLH:eq:one}
\end{equation}
where $\rho_{2}$ is the volumetric density of pairs. Equation~(\ref{MLH:eq:one}) is valid in continuum space or in a lattice at low $\rho_{2}$ where the parabolic approximation of the pair dispersion is accurate. The treatment can be refined by replacing the parabolic dispersion with the exact pair dispersion $E_{\bf P}$ inside the Bose integral \cite{Kornilovitch2015}. This approach, however, lacks the physical transparency of Eq.~(\ref{MLH:eq:one}). For a qualitative explanation of $T_{c,{\rm max}}(n)$, which is the main goal of this paper, Eq.~(\ref{MLH:eq:one}) is adequate.            

\begin{figure}[t]
\includegraphics[width=0.48\textwidth]{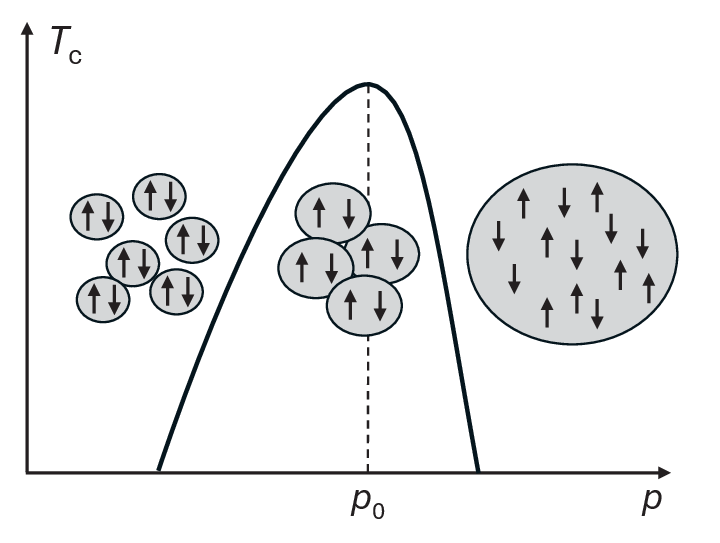}
\caption{Schematic evolution of $T_{c}(p)$ with doping in multilayer cuprates. The superconducting dome is asymmetric \cite{Honma2008}. Pairs are close-packed at optimal hole doping $p_{0}$. Note that the pairs are larger near $p_0$ than in the underdoped regime, $p < p_{0}$. In the overdoped regime, $p > p_{0}$, the pairs lose individuality and dissolve into a Fermi sea. }  
\label{MLH:fig:zeroone}
\end{figure}

For the preformed pair picture to be self-consistent, the pair size must be smaller that the distance between pairs. This condition is satisfied in {\em underdoped} cuprates: with a coherence length $\xi_{ab} \approx 1.6$ nm \cite{Oh1988,Wang2003}, a single pair occupies roughly 16 CuO$_{2}$ unit cells \cite{Li2023}. This implies that for {\em hole} densities $p < 0.12$ cell$^{-1}$, pairs overlap only weakly. Let us follow the evolution of $T_{c}$ with increasing $p$, as sketched in Fig.~\ref{MLH:fig:zeroone}. For $p < 0.12$, Eq.~(\ref{MLH:eq:one}) is approximately valid with $\rho_{2} = p/(2V_0)$, where $V_0$ is the unit cell volume, and $T_{c}$ increases with $p$. At $p > 0.12$, pairs begin to overlap and the simple BEC picture breaks down. With further doping, the constituent holes start to form a Fermi sea, the kinetic energy rises, and superconductivity crosses over to a BCS-like regime. Thus, $T_c$ reaches its maximum at {\em close-packing}, when the pair density becomes approximately equal to the inverse pair volume $\Omega_{p}$ \cite{Ivanov1994,Kornilovitch2015,Kornilovitch2024}:    
\begin{equation}
k_B T^{\ast}_{\rm cp} = 
\frac{3.3 \, \hbar^2 }{ ( m^{\ast}_x m^{\ast}_y m^{\ast}_z )^{1/3} \Omega^{2/3}_{p} } \: . 
\label{MLH:eq:two}
\end{equation}

In addition to the increasing density of pairs, individual pairs become progressively weaker with doping \cite{Kornilovitch2024,Wang2003}. The reason is the weakening of electron correlations. Near half-filling, $p < 0.1$, hole motion is restricted by short-range Coulomb repulsion and antiferromagnetic order, which reduces the hole kinetic energy by roughly a factor of 3 relative to the DFT band structure \cite{Harrison2023}. That effectively enhances the inter-hole attraction, to be discussed in Sec.~\ref{MLH:sec:twotwo}, and enables pair formation. With increasing doping, the correlations weaken, the delocalization kinetic energy gradually increases, and the relative attractive strength decreases, leading to a growth of the pair size. Around optimal doping, the Fermi surface transitions from $p$ to $(1-p)$ scaling, the delocalization energy rises sharply, and the pairs rapidly dissolve into the Fermi sea. Note that in this scenario the rise of $T_c(p)$ in the underdoped regime and the fall of $T_c(p)$ in the overdoped regime are governed by two physically distinct mechanisms. This may explain the asymmetry of the superconducting dome in multilayer cuprates \cite{Honma2008}.        

Returning to Eq.~(\ref{MLH:eq:two}), {\em we associate the close-packing BEC temperature $T^{\ast}_{\rm cp}$ with the maximum superconducting critical temperature $T_{c,{\rm max}}$ observed experimentally}. We note that an inverse correlation between $T_{c,{\rm max}}$ and the pair volume was recently reported in an empirical analysis of 14 different classes of unconventional superconductors \cite{Wang2025}. Although Eq.~(\ref{MLH:eq:two}) does not include pair-pair interaction and therefore cannot predict the absolute value of $T_{c}$, it is suitable for relative comparisons. In this paper, we apply Eq.~(\ref{MLH:eq:two}) to explain the dependence of $T_{c,{\rm max}}$ on $n$:
\begin{equation}
T^{\ast}_{\rm cp}(n) \propto T_{c,{\rm max}}(n) \: . 
\label{MLH:eq:twoone}
\end{equation}
A remarkable feature of $T^{\ast}_{\rm cp}$, as given by Eq.~(\ref{MLH:eq:two}), is that it depends solely on the properties of an individual real-space pair, which can be calculated exactly \cite{Kornilovitch2024}. Our analysis will be based on a multilayer pairing model introduced and solved in Sec.~\ref{MLH:sec:three}.

\subsection{ \label{MLH:sec:twotwo}
Sources of attraction
}

Real-space pairs can only form if there is a source of attraction. Cuprate superconductors are ionic crystals with low carrier density; therefore Fr\"ohlich-like interaction \cite{Froehlich1954} between holes and lattice polarization must be strong. A long-range electron-phonon interaction was directly measured in YBCO superlattices \cite{Driza2012}. Whether Fr\"ohlich interaction {\em alone} is sufficient to form pairs --- referred to in this case as {\em large bipolarons} --- remains an open question. An important aspect of large bipolarons is lattice retardation. Without retardation, two electric charges of the same sign always repel, whereas with retardation, bipolaron formation becomes possible \cite{Vinetskii1958,Vinetskii1961}. In this sense, large bipolarons are real-space analogues of Cooper pairs in BCS theory \cite{Cooper1956,Bardeen1957,Schrieffer1999}, which overcome Coulomb repulsion only if retardation is included \cite{Bogoliubov1959,Morel1962}. Verbist, Peeters and Devreese calculated \cite{Verbist1991} that large bipolarons in an {\em isotropic} 3D crystal require unrealistically large coupling constants $\alpha > 6.8$. However, their analysis excluded exchange effects, which would lower the bipolaron energy and reduce the critical $\alpha$. Morefover, cuprates are strongly anisotropic, which tends to promote pairing, as is known from studies of other model systems \cite{Kornilovitch2015,Alexandrov2002}. A rigorous analysis of bipolaron formation incorporating anisotropic hole bands \cite{Vinetskii1983}, anisotropic polarization, and anisotropic interaction would be valuable but has yet to be carried out. In any case, the long-range hole-polarization interaction cancels a large part of Coulomb repulsion between holes and thus promotes pairing.        

Another source of inter-hole attraction is short-range lattice effects that account for the detailed crystal structure of the CuO$_2$ planes and their surroundings. Emin suggested \cite{Emin1989a,Emin1989b} that large bipolarons can only form in the presence of a short-range lattice interaction. On the other hand, quantum chemical calculations indicate that short-range lattice distortion may stabilize a bipolaron even in the absence of retardation \cite{Zhang1991,Catlow1998,Edwards2023}. Jahn-Teller distortions, another short-range lattice effect, can also promote real-space pairing. The Jahn-Teller mechanism was the original motivation for investigating copper oxides \cite{Bednorz1988}. A more detailed Jahn-Teller pairing model was later developed by Mihailovic and Kabanov \cite{Mihailovic2001,Kabanov2002}. Structural distortions extending across four adjacent CuO$_2$ plaquettes have been observed in SrCuO$_2$ \cite{Zhong2018}, and rotational phonons have been detected in CaCuO$_2$ \cite{Wang2026}, indicating that short-range lattice effects must be part of the physics of the ``base'' $n = \infty$ cuprate superconductor. Local lattice distortions have also been observed in La$_{2-x}$Sr$_{x}$CuO$_{4}$, YBa$_{2}$Cu$_{4}$O$_{8}$ and Tl$_{2}$Ba$_{2}$CaCu$_{2}$O$_{8}$ \cite{Egami1995}. Nanoscale structural correlations were reported in HgBa$_2$CuO$_{4+\delta}$ \cite{Anderson2024}, which has the highest $T_c$ among the $n = 1$ cuprates. Apical oxygens may play an additional role. It has been proposed that $c$-axis displacements of apical oxygens can create anisotropic polarons of mixed character, in which in-plane polaron motion is Fr\"ohlich-like with a small mass \cite{Alexandrov1999}, while out-of-plane motion is Holstein-like with a large mass \cite{Kornilovitch1999}. This effect may explain why the spread of experimentally observed anisotropies among cuprate families is larger that what band-structure calculations predict. The Holstein-polaron character of $c$-axis transport in apical-containing cuprates is supported, for example, by the nonmonotonic temperature-dependence of $c$-axis resistivity \cite{Mino2024} and by the ``dynamic stabilization of superconductivity'' achieved through THz excitation of apical oxygens \cite{Hu2014,Kornilovitch2016,Kornilovitch2017}. Recently, a well-resolved Franck-Condon ladder was observed in X-ray scattering of HgBa$_2$Ca$_2$Cu$_3$O$_{8+\delta}$ \cite{Hong2025}, providing direct evidence of polaron distortion in the record-holding cuprate superconductor. 

Finally, one must mention the attraction mediated by spin fluctuations. This pairing mechanism received considerable attention after the discovery of cuprate superconductivity; see \cite{Scalapino2012} for review. Short-range antiferromagnetic order makes it energetically favorable for two holes to occupy nearest-neighbor sites, which is equivalent to a non-retarded attractive potential $\propto J = 4t^2/U$, where $U$ is the on-site Hubbard repulsion. In our physical picture, spin fluctuations act in concert with the lattice mechanisms reviewed above. More importantly, the strong Hubbard repulsion that produces antiferromagnetic order also restricts hole motion and suppresses the delocalization energy. This enables real-space pairs to form through the combined action of all pairing mechanisms. With increasing doping, correlations weaken, the delocalization energy rises, and the pairs disappear. Thus, destruction of antiferromagnetic order and disappearance of superconductivity proceed in parallel and share a common physical origin.        
   
The combined effect of all pairing mechanisms can be modeled by a short-range attractive {\em pseudopotential} $V$. The most popular choice is a $V$ restricted to nearest neighbors, supplemented by a strong on-site Hubbard repulsion $U$. This is the so-called $UV$ model \cite{Micnas1990,Kornilovitch2024}.~\footnote{Longer-range attractive potentials may appear more physical, especially when mediated by lattice polarization. Such pairing models are mathematically far more challenging, and only a small number of cases have been analyzed to date \cite{Kornilovitch1995,Bak1999,Kornilovitch2024,Adebanjo2024b}.} One advantage of $UV$ model is the ability to produce superconductivity with different orbital symmetries, including the $d$-symmetry observed experimentally in the cuprates \cite{Blaer1997,Bak1999,Adebanjo2024b,Kornilovitch2025,Kornilovitch2025b}. There are two versions of the $UV$ model. In the ``electron'' version, the $UV$ interaction acts between electrons, requiring the solution of a high-density many-body problem near half-filling \cite{Nayak2018,Jiang2021,Zhang2021,Qu2021,Singh2021,Huang2025}. In the ``hole'' version, an effective $UV$ interaction acts between holes near half-filling, reducing the problem to a low-density limit \cite{Micnas1990,Kornilovitch2024}. The hopping and interaction parameters differ between the two formulations. Recently, the $UV$ model received additional support from photoemission studies of one-dimensional cuprate chains that are structurally and chemically similar to superconducting cuprates \cite{Chen2021,Wang2021}. 

It is important to emphasize the role of lattice dimensionality and lattice structure in general. In elementary quantum mechanics, three-dimensional bound states require a finite attractive threshold, whereas two-dimensional bound states do not, albeit the latter have exponentially small binding energies. Generalizing this result to lattices, {\em given the same pairing mechanism, real-space pairing occurs more easily in anisotropic crystals than in cubic crystals}. Moreover, Eq.~(\ref{MLH:eq:two}) shows that there must be an optimal anisotropy \cite{Kornilovitch2015}. In the cubic case, where $m^{\ast}_{z} = m^{\ast}_{x}$, the holes posses large kinetic energy, pairs are weakly bound (or do not form at all), the pair volume $\Omega_{p}$ is large, and $T^{\ast}_{\rm cp}$ is low. Superconductivity is suppressed because the density of pairs is low. In the opposite case of very strong anisotropy, pairs are well-formed but $m^{\ast}_{z} \gg m^{\ast}_{x}$. Superconductivity is again suppressed --- this time due to the loss of phase coherence across the layers. Thus, $T^{\ast}_{\rm cp}$ is maximal at an intermediate anisotropy. This reasoning helps explain why, for example, Bi-based superconductors, which have cubic symmetry, reach only $T_c = 30$ K \cite{Cava1988} despite being structurally similar to the cuprates and despite the presence of real-space pairs \cite{Menushenkov2024}. 

Lattice structure may induce additional, and sometimes unexpected, pairing effects. For example, if the first-neighbor hopping and second-neighbor hopping have opposite signs, bound pairs can be {\em lighter} that an unbound hole \cite{Adebanjo2024b}. In a body-centered tetragonal lattice, a peak in the hole density of states can produce a zero binding threshold even in a three-dimensional crystal with well-developed kinetic energy \cite{Kornilovitch2025}. These effects lie outside the scope of the present work.    

Despite the appeal of $UV$ model, a rigorous analysis of pairing in a multi-layer crystal is highly nontrivial. While such a treatment is certainly possible, a full multilayer $UV$ pairing model is left for future research. In this work, we instead use the much simpler {\em multilayer attractive Hubbard model} introduced in Sec.~\ref{MLH:sec:three}. This model is sufficient to accomplish our main goal, which is to understand the dependence of $T_{c,{\rm max}}(n)$ in the framework of preformed-pair superconductivity.

\subsection{ \label{MLH:sec:twothree}
From $n = \infty$ to $n = 1$: $T_{c,{\rm max}}$ decreases
}

\begin{figure}[t]
\includegraphics[width=0.45\textwidth]{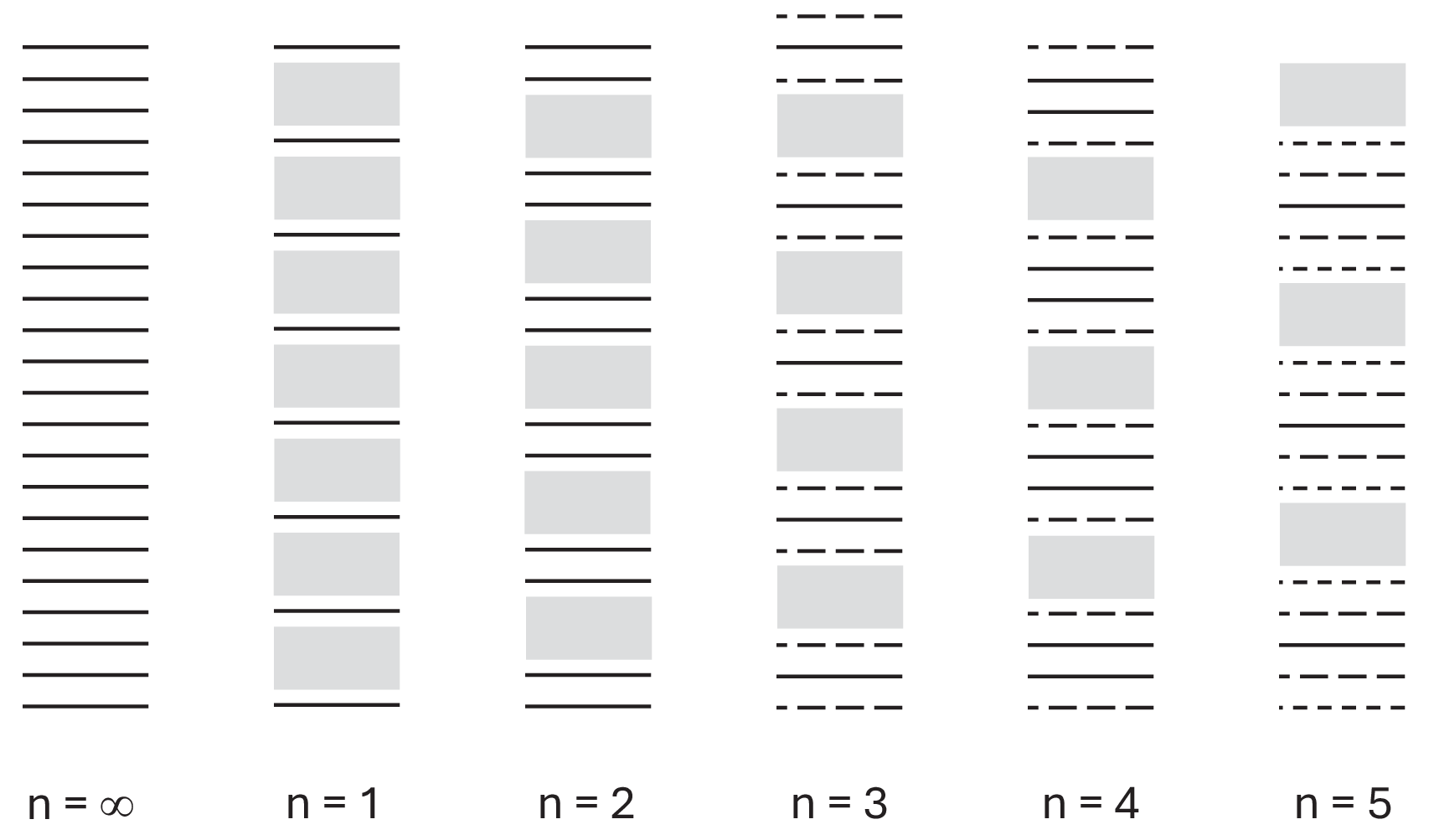}
\caption{Schematic structure of multilayer cuprates. Horizontal lines represent CuO$_2$ planes, and gray rectangles denote ``charge reservoir layers''. Note that the $n = 3, 4$ members contain two distinct types of planes, shown by solid and long-dashed lines, whereas the $n = 5$ member has three distinct types, shown by solid, long-dashed, and short-dashed lines.} 
\label{MLH:fig:zerotwo}
\end{figure}

Figure \ref{MLH:fig:zerotwo} schematically shows the structure of multilayer cuprates. The distance between CuO$_2$ planes within a stack is 3.17 \AA${}$, while the distance between stacks varies from 6.6 \AA${}$ in LSCO to 12.3 \AA${}$ in BSCCO. In the figure, the distance between stacks is 3 times the interplane distance, which roughly corresponds to HgSCCO (9.5 \AA ${}$). In comparing $n = \infty$ and $n = 1$ one notices that in both cases all the planes are equivalent. The $n = \infty$ member can also be regarded as an $n = 1$ system but of different anisotropy than the ``proper'' $n = 1$ member. The main difference is the presence of a ``charge reservoir layer'' (CRL) in $n = 1$.

The CRL alters the real-space pairing picture in two significant ways. First, the CRL increases the degree of lattice anisotropy, since the hopping integral across the CRL is smaller than the hopping between planes within a stack. This inevitably increases the out-of-plane pair mass $m^{\ast}_z$ and, according to Eq.~(\ref{MLH:eq:two}), decreases $T_{c,{\rm max}}$. This reduction is expected for {\em any} CRL. Therefore, the drop of $T_{c,{\rm max}}$ from $n = \infty$ to $n = 1$ is a {\em universal} feature of cuprate superconductors. It just occurs to different degrees among the various families because of differences in the CRL's chemical and electronic structure. Secondly, the CRL modifies the attractive strength $V$, particularly its lattice components. The long-range Fr\"ohlich interaction changes because the local polarizability of the CRL is not the same as that of an infinite stack of CuO$_2$ planes. The short-range lattice interaction also changes due to Jahn-Teller, small-polaron, and related effects discussed in Sec.~\ref{MLH:sec:twotwo}. Crucially, the induced change in $V$ is {\em not} universal. $V$ may increase --- promoting superconductivity through more compact pairs [smaller $\Omega_p$ in Eq.~(\ref{MLH:eq:two})] --- or it may decrease, thereby suppressing superconductivity. The data in Table~\ref{MLH:tab:one} suggest that $V$ increases only in the highest-$T_c$ families, namely the Hg, single-layer Tl, and (Cu,C) ones.         

One should also note that $n = 1$ superconductors are vulnerable to structural disorder introduced by the CRLs, which universally suppresses $T_{c,{\rm max}}$ \cite{Eisaki2004}. This effect lies outside the scope of the present work.   

\begin{figure*}[t]
\includegraphics[width=0.90\textwidth]{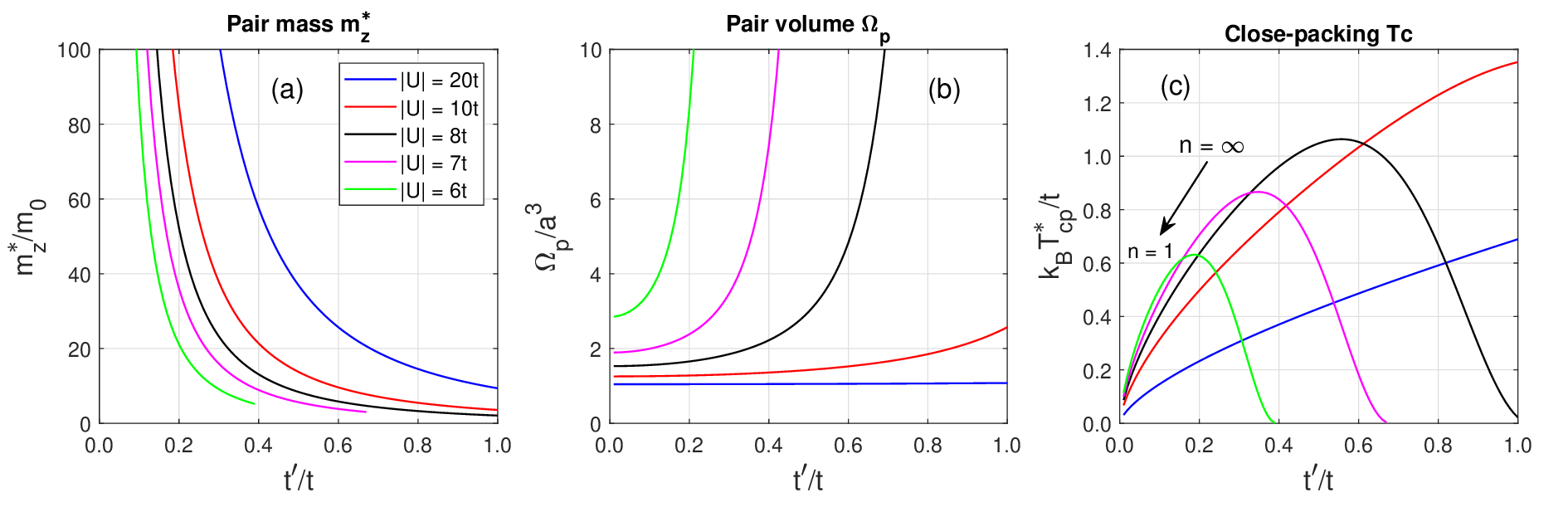}
\caption{(c) The close-packing BEC temperature, Eq.~(\ref{MLH:eq:two}), and its two constituents: (a) the $c$-axis pair mass $m^{\ast}_{z}$, and (b) the pair volume $\Omega_{p}$ in the $n = 1$ (``tetragonal'') attractive Hubbard model; see Appendix~\ref{MLH:sec:app:b} and \cite{Kornilovitch2024}. Note how the divergence of $m^{\ast}_{z}$ as $t^{\prime} \to 0$ and the divergence of $\Omega_{p}$ as $t^{\prime} \to t^{\prime}_{\rm pairing \: threshold}$ together produce an optimal $t^{\prime}$ at which $T^{\ast}_{\rm cp}$ is maximized \cite{Kornilovitch2015}. The transition from $n = \infty$ to $n = 1$ corresponds to moving away from this optimal $t^{\prime}$ toward smaller values of $t^{\prime}$. The legend in the left panel applies to all panels.} 
\label{MLH:fig:zerothree}
\end{figure*}

Figure~\ref{MLH:fig:zerothree} shows the pair mass $m^{\ast}_z$, pair volume $\Omega_{p}$, and $T^{\ast}_{\rm cp}$ computed for the $n = 1$ (``tetragonal'') attractive Hubbard model; see Appendix~\ref{MLH:sec:app:b} and \cite{Kornilovitch2024}. As the anisotropy increases, $t^{\prime} \to 0$, the pair mass diverges, which suppresses $T^{\ast}_{\rm cp}$. {\em This is the dominant effect.} In the isotropic limit, $t^{\prime} \to t$, the kinetic delocalization energy increases and the pairs become less strongly bound. For $|U| < 7.913552 \, t$ \cite{Kornilovitch2024}, there exists a finite value of $t^{\prime}/t$ above which no bound pairs form. As this threshold is approached from below, the pairs expand in size, as shown in the middle panel. This also limits $T^{\ast}_{\rm cp}$, producing a pronounced maximum at intermediate $t^{\prime}$. In our phenomenological picture, $n = \infty$ corresponds to the optimal values of $t^{\prime}$. The transition to $n = 1$ corresponds to a reduction of $t^{\prime}$ and, consequently, to a reduction of $T^{\ast}_{\rm cp}$. Thus, the drop in $T_{c,{\rm max}}$ is universal. Different cuprate families differ only in the degree of this drop, which depends on how much $t^{\prime}$ is reduced.

\subsection{ \label{MLH:sec:twofour}
From $n = 1$ to $n = 2$: $T_{c,{\rm max}}$ increases
}

Let us compare now $n = 1$ and $n = 2$, see Fig.~\ref{MLH:fig:zerotwo}. Since the hopping between the two CuO$_2$ planes within a bilayer is larger than the hopping between bilayers, the $n = 2$ system is {\em less} anisotropic than the $n = 1$ one. The $c$-axis transport is a mixture of fast intra-bilayer hopping and slow inter-bilayer hopping. As a result, the overall transport is ``easier'' in the $n = 2$ case: $m^{\ast}_{z}$ decreases and $T^{\ast}_{\rm cp}$ increases. Importantly, both CuO$_2$ planes remain physically equivalent because of the mirror symmetry. Even if attraction $V$ changes slightly between $n = 1$ and $n = 2$, it remains to be the same for all planes. The holes experience a uniform pairing potential throughout the entire lattice, with no tendency toward localization. This makes the reduction of $m^{\ast}_{z}$ the dominant effect. Hence, the increase of $T_{c,{}\rm max}$ from $n = 1$ to $n = 2$ is {\em universal} across all cuprate families.

\subsection{ \label{MLH:sec:twofive}
$T_{c,{\rm max}}(n \geq 3)$: it depends 
}

The behavior of $T_{c,{\rm max}}$ for $n \geq 3$ is governed by two physical effects. First, holes' kinetic energy $K$ continues to increase and the overall lattice anisotropy approaches its $n = \infty$ level. If this were the only effect, $m^{\ast}_{z}$ would continue to decrease, and $T_{c,{\rm max}}(n)$ would continue to rise until the growing $K/V$ weakened the pairs sufficiently to reduce $T_{c,{\rm max}}(n)$ through a lower close-packing density. However, this picture is complicated by a second effect: the inequivalence of the CuO$_2$ planes. There are two physically different plane types for $n = 3, 4$, three types for $n = 5, 6$, and so on. Importantly, not only the attractive strength $V$ but also the one-hole energy may vary from plane to plane. For example, outer planes may experience {\em polaron shifts} (energy lowering) due to displacements of apical oxygens \cite{Hong2025}. The polaron shift is arguably smaller for inner planes because they are farther from the apical oxygens. In the absence of first-principles calculations, the presence of multiple plane types proliferates the phenomenological parameters of the model, making a complete analysis challenging. We will discuss this issue further in the next section.     
   
The more complex energy landscape of $n \geq 3$ can either promote or impede pair transport. If resonant conditions are satisfied, $m^{\ast}_{z}$ will decrease further, since additional planes make the system less anisotropic on average. We conjecture that this is what occurs in cuprates with two plane types, i.e., for $n = 3, 4$. This is why $T_{c,{\rm max}}(3,4) > T_{c,{\rm max}}(2)$ in most families. Beginning with $n = 5$, however, resonant conditions must be simultaneously optimized across three plane types. While this is possible in principle, it requires more delicate fine-tuning of parameters. This condition does not appear to be satisfied in any cuprate family known at present; consequently, $T_{c,{\rm max}}(5,6) < T_{c,{\rm max}}(3,4)$. Nevertheless, if an $n = 5$ material were found in which this condition is met, one would expect $T_{c,{\rm max}}(5) > T_{c,{\rm max}}(3,4)$. This represents a plausible route toward increasing $T_{c,{\rm max}}$ beyond the current record.

\section{ \label{MLH:sec:three}
Multilayer pairing model 
}

\subsection{ \label{MLH:sec:threeone}
The Hamiltonian
}

In this section, we introduce and study a multilayer pairing model to support the qualitative ideas of Sec.~\ref{MLH:sec:two} with rigorous calculations. Our approach is to restrict consideration to two holes and solve the two-hole quantum-mechanical problem exactly for $n = 1, 2, 3, 4, 5$. Since the exact solution yields the pair masses and sizes, Eq.~(\ref{MLH:eq:two}) can be applied directly to compute $T^{\ast}_{\rm cp}$. Although all two-body lattice problem are in principle exactly solvable, the mathematical complexity increases sharply with both the interaction range \cite{Kornilovitch2024,Kornilovitch1995} and the number of energy bands. Because multiple bands are unavoidable in the present problem, we choose the simplest interaction that still captures the physics described in Sec.~\ref{MLH:sec:two}. It is essential to include {\em some} attractive interaction, and the simplest choice is a zero-range attractive potential, also known as the ``negative-$U$ Hubbard model''. Although less realistic than more general $UV$ potentials \cite{Micnas1990}, the attractive Hubbard interaction is fully adequate for the purposes of this paper.         

Our model Hamiltonian reads
\begin{eqnarray}
H_{n} & = & \sum_{i \alpha \sigma} \epsilon_{\alpha} \, p^{\dagger}_{i \alpha \sigma} p_{i \alpha \sigma} 
 - t \!\! \sum_{\langle i \alpha , j \beta \rangle \sigma} p^{\dagger}_{i \alpha \sigma} p_{j \beta  \sigma} 
\nonumber \\
  &   & - t^{\prime} \!\! \sum_{\langle k \gamma , l \delta \rangle \sigma} 
          p^{\dagger}_{k \gamma \sigma} p_{l \delta \sigma}
        - \vert U \vert \sum_{i \alpha} p^{\dagger}_{i \alpha \uparrow}   p_{i \alpha \uparrow} \, 
                                        p^{\dagger}_{i \alpha \downarrow} p_{i \alpha \downarrow} .     
\makebox[0.4cm]{}          
\label{MLH:eq:three}
\end{eqnarray}
Here, $p_{i \alpha \sigma}$ and $p^{\dagger}_{i \alpha \sigma}$ are hole operators for lattice unit cell $i$, CuO$_2$ layer $\alpha = 1 \ldots n$, and spin projection $\sigma = \uparrow, \downarrow$. Only two holes will be considered, one with spin up and the other with spin down. The problem will be solved directly in real space, and the second-quantized form of Eq.~(\ref{MLH:eq:three}) will not be used. Because the attraction is point-like, only one spin-singlet, $s$-symmetric bound state is expected.

\begin{figure}[t]
\includegraphics[width=0.40\textwidth]{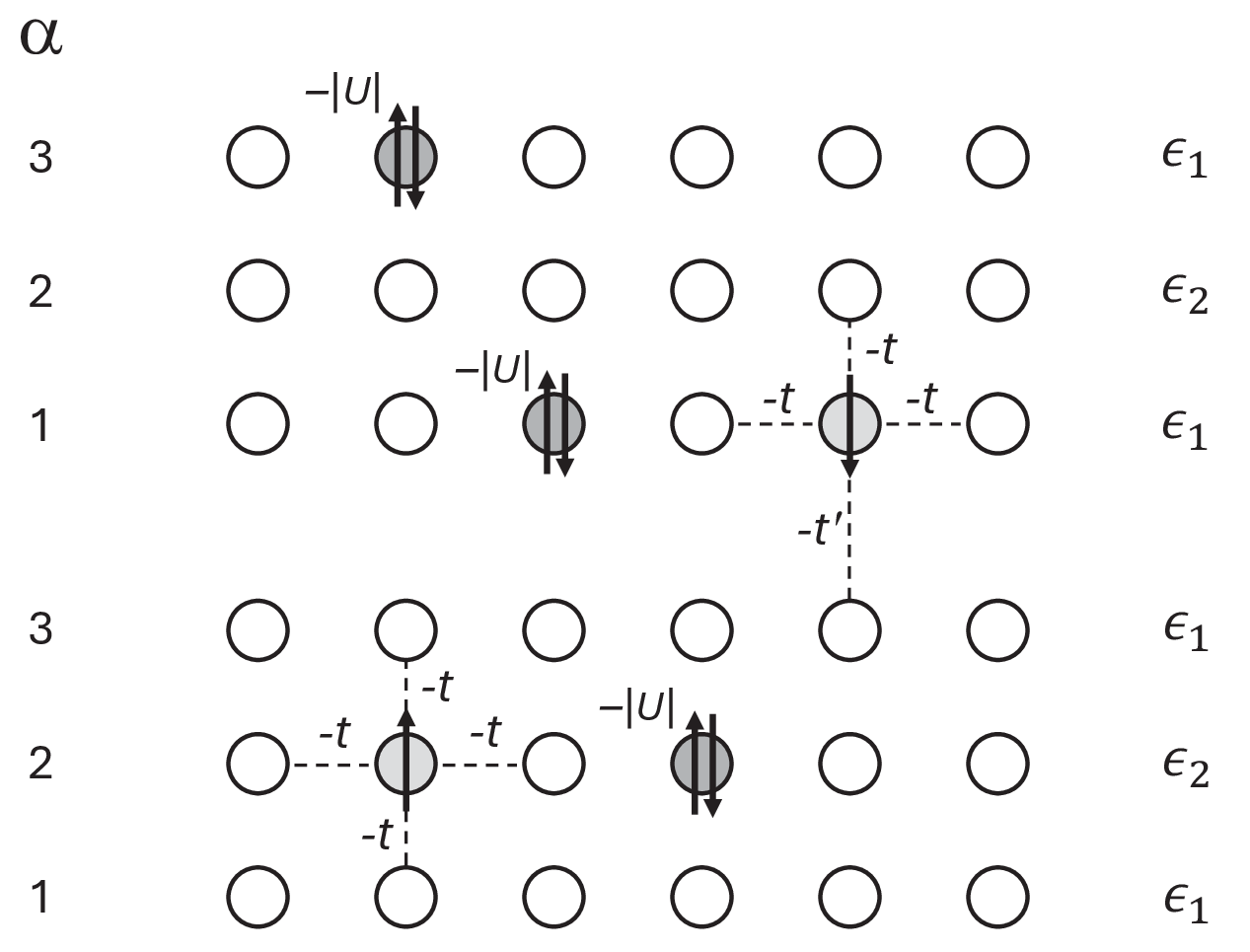}
\caption{The multilayer pairing model, Eq.~(\ref{MLH:eq:three}), for $n = 3$. The on-site hole energy $\epsilon$ is the same for the $\alpha = 1$ and $\alpha = 3$ planes because of mirror symmetry.} 
\label{MLH:fig:zerofive}
\end{figure}

Equation~(\ref{MLH:eq:three}) an effective low density cuprate model near half-filling. The first term describes plane-to-plane variations in the on-site energy of static holes. This variation will be essential for deriving a $T^{\ast}_{\rm cp}(n)$ curve peaked at $n = 3$. The second term describes hole motion within an $n$-layer stack of ${\rm CuO}_2$ planes. We assume only one orbital per each CuO$_2$; therefore the unit cell of the system contains $n$ orbitals. Consequently, $n$ energy bands are expected, both for the bare single-hole dispersion and for the bound-pair dispersion. The double sum $\langle i \alpha , j \beta \rangle$ runs over nearest-neighbor pairs within the same stack. We do not distinguish between in-plane and interplane hoppings within the stack and assign to both the same hopping amplitude $-t$. Although an exact solution can accommodate unequal hoppings, introducing such an extra parameter would clutter the results without adding new physical insights. The hopping amplitude is taken to be negative to ensure that the band minima occur at the $\Gamma$ point; thus, we explicitly write $-t$ with $t > 0$. The third term in Eq.~(\ref{MLH:eq:three}) describes hole transport between $({\rm CuO}_2)_n$ stacks, characterized by a negative amplitude $-t^{\prime}$, with $t^{\prime} > 0$. We consider only the physically relevant regime where $t^{\prime} \leq t$. The double sum $\langle k \gamma , l \delta \rangle$ runs over nearest-neighbor pairs belonging to adjacent stacks. More general inter-stack hopping patterns can be included, but they would complicate the analysis without adding new physics. Finally, the fourth term describes short-range (Hubbard) attraction between holes. We take $U$ to be the same for all planes. Although a more general interaction could be implemented, the constant-$U$ model is sufficient for the purposes of this work. Figure~\ref{MLH:fig:zerofive} illustrates the model for $n = 3$.

Model parameters are now discussed. For the in-plane hopping integral, we assume a correlation-reduced value $t = 0.1$ eV, consistent with thermodynamic measurements \cite{Harrison2023}. For the inter-stack hopping, we take $t^{\prime} \leq 0.1 t$. Recent photoemission and X-ray measurements \cite{Chen2021,Wang2021,Padma2025} indicated the presence of a {\em nearest-neighbor} attraction of approximately $1.0 t$. It is known from model calculations \cite{Kornilovitch2024} that the equivalent on-site attraction is at least $z$ times larger, where $z$ is the number of nearest neighbors. Thus, in our model, $|U| \geq 6 t$. In addition, the long-range nature of the hole-polarization interaction suggests that some attraction persists beyond the first nearest neighbors \cite{Alexandrov2002}, which further increases the effective $|U|$. We therefore assume that physically relevant values of the attraction are $|U| = (6 - 10) \, t$.  

To the best of our knowledge, the model given by Eq.~(\ref{MLH:eq:three}) has not yet been considered in the literature yet. A {\em single} $n$-stack attractive Hubbard model (that is, a system of finite extent along the $c$ axis) was recently introduced and solved by Bak \cite{Bak2024}. It was found that the crossover from the zero binding threshold in a finite-$n$ stack of 2D planes to a finite threshold in the fully 3D model is rather sharp.

\subsection{ \label{MLH:sec:threetwo}
The method
}

{\em Any} quantum-mechanical two-body lattice problem is exactly solvable, at least in principle, but the mathematical complexity of the solution increases rapidly with the spacial extent of the interaction and with the number of orbitals in the unit cell (i.e., the number of bands in the single-particle dispersion). There is now a sizeable literature on {\em one-band} two-fermion problems, recently reviewed in \cite{Kornilovitch2024}. By contrast, only a few multiband two-body problems have been studied \cite{Ivanov1994,Alexandrov1992,Alexandrov1993,Kornilovitch1997,Iskin2024}, so the present work represents a significant advance in this area. The overall solution procedure follows the same steps as in one-band models: (i) the two-body Schr\"odinger equation is transformed into momentum space; (ii) the interaction term is expressed as a linear combination of auxiliary integrals that depend only on the total pair momentum ${\bf P}$ and not on the individual momenta; (iii) momentum conservation reduces the Schr\"odinger equation to a system of linear equation for these auxiliary integrals; (iv) the eigenvalues of this system give the bound-pair energy, while the eigenvectors provide information about the pair wave function, from which the pair size can be extracted.

\begin{figure*}[t]
\includegraphics[width=0.90\textwidth]{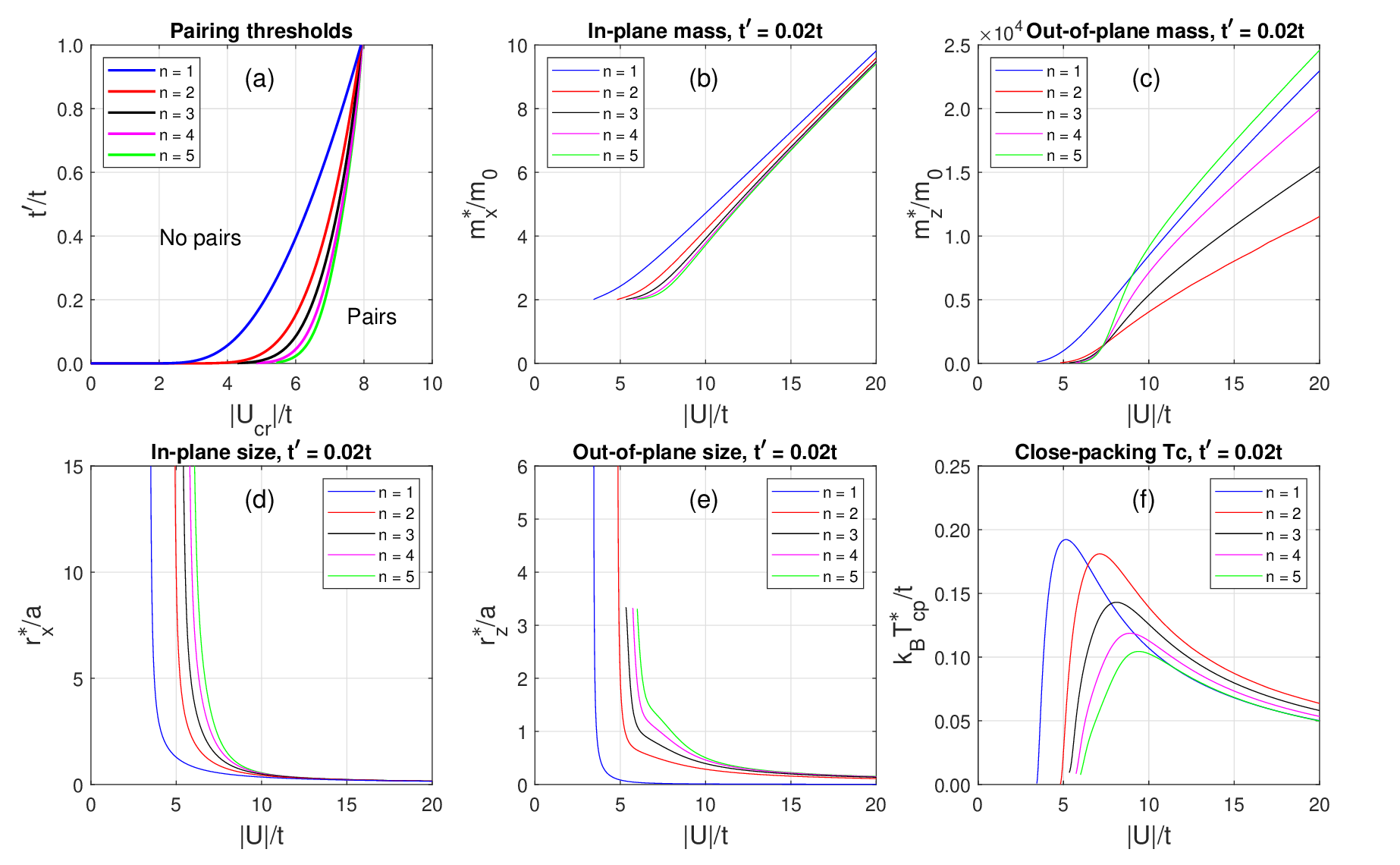}
\caption{Physical properties of the ``uniform'' attractive Hubbard model, $\epsilon_{\alpha} = 0$. Panels (b)-(f) show data for only one value of the anisotropy parameter. Note the large $y$-axis scale in panel (c). Data for other values of $t^{\prime}$ are provided in the Appendices.} 
\label{MLH:fig:zeroseven}
\end{figure*}

The technical details of the exact solutions are too cumbersome to present here and are provided in the Appendices. We comment only on several key points. The pairing threshold $|U_{\rm cr}|$ is determined by equating the pair energy $E_{2}({\bf P})$, obtained from the exact solution, to the minimum energy of two free fermions with the same total momentum: 
\begin{equation}
E_{2}({\bf P}) = 2 \varepsilon_{1}({\bf P}/2) \: , 
\label{MLH:eq:four}
\end{equation}
where $\varepsilon_{1}({\bf k})$ is the lowest band of the one-particle spectrum. In general, $|U_{\rm cr}|$ is a decreasing function of the pair momentum ${\bf P}$, indicating that moving pairs are formed more easily than stationary ones \cite{Kornilovitch2024,Kornilovitch2004}. In this paper, we ignore this effect and consider only ground-state pairs, i.e., those with ${\bf P} = 0$. 

Pair masses $m^{\ast}_{i}$ are computed numerically as second derivatives of the pair energy and are expressed in units of the bare in-plane hole mass $m_0 \equiv \hbar^2/(2ta^2)$, where $a$ is the $xy$ lattice constant. If $\vartriangle \!\! P$ is a small pair momentum with $a \! \vartriangle \!\! P \ll 1$, then 
\begin{eqnarray}
\frac{m^{\ast}_{x}}{m_0} = \frac{m^{\ast}_{y}}{m_0} & = &  
\frac{ ( a \! \vartriangle \!\! P )^2 t }{ E( a \! \vartriangle \!\! P , 0 , 0 ) - E_{0}} \: , 
\label{MLH:eq:zerosix} \\
\frac{m^{\ast}_{z}}{m_0}  & = &  
\frac{( a \! \vartriangle \!\! P )^2 t}{ E( 0 , 0 , a \! \vartriangle \!\! P ) - E_{0}} \: . 
\label{MLH:eq:zeroseven} 
\end{eqnarray}

The effective pair radii $r^{\ast}_{i}$ are computed from the full coordinate-space wave function as the mean-squared distance between the constituent holes. Using Fourier transformations, integrals over the system volume are converted into derivatives of momentum-dependent wave functions. Explicit formulae are provided in the Appendices. Here we only note that all $n^2$ components of the wave function are required to compute $r^{\ast}_{i}$. For computational simplicity, the distance between two neighboring multilayer stacks is taken to be $a$. Although the physical inter-stack separations are $(2-4)a$, we assume that the anisotropy is fully captured by the value of the hopping integral $t^{\prime}$, so using the inter-stack distances is unnecessary.  

Finally, we discuss the calculation of the pair volume $\Omega_{p}$. The most natural definition would be $\Omega_{p} = r^{\ast}_{x} r^{\ast}_{y} r^{\ast}_{z}$. However, the oversimplified attractive Hubbard model is pathological in the strong-coupling limit $|U| \gg t$: the pair collapses to a point, yielding unphysically large close-packing densities and correspondingly unphysically values of $T^{\ast}_{\rm cp}$. In reality, the pair volume cannot be smaller than $a^3$ because the pair consists of two fermions. To enforce this lower bound, we use the regularized expression
\begin{equation}
\Omega_{p} = \sqrt{ ( a^2 + r^{\ast 2}_{x} ) ( a^2 + r^{\ast 2}_{y} ) ( a^2 + r^{\ast 2}_{z} ) } \: . 
\label{MLH:eq:zeroeight}
\end{equation}
Using the pair masses from Eqs.~(\ref{MLH:eq:zerosix}) and (\ref{MLH:eq:zeroseven}) together with the regularized pair volume of Eq.~(\ref{MLH:eq:zeroeight}), the close-packing critical temperature is then computed from Eq.~(\ref{MLH:eq:two}).

\section{ \label{MLH:sec:four}
Results
}

\begin{figure*}[t]
\includegraphics[width=0.90\textwidth]{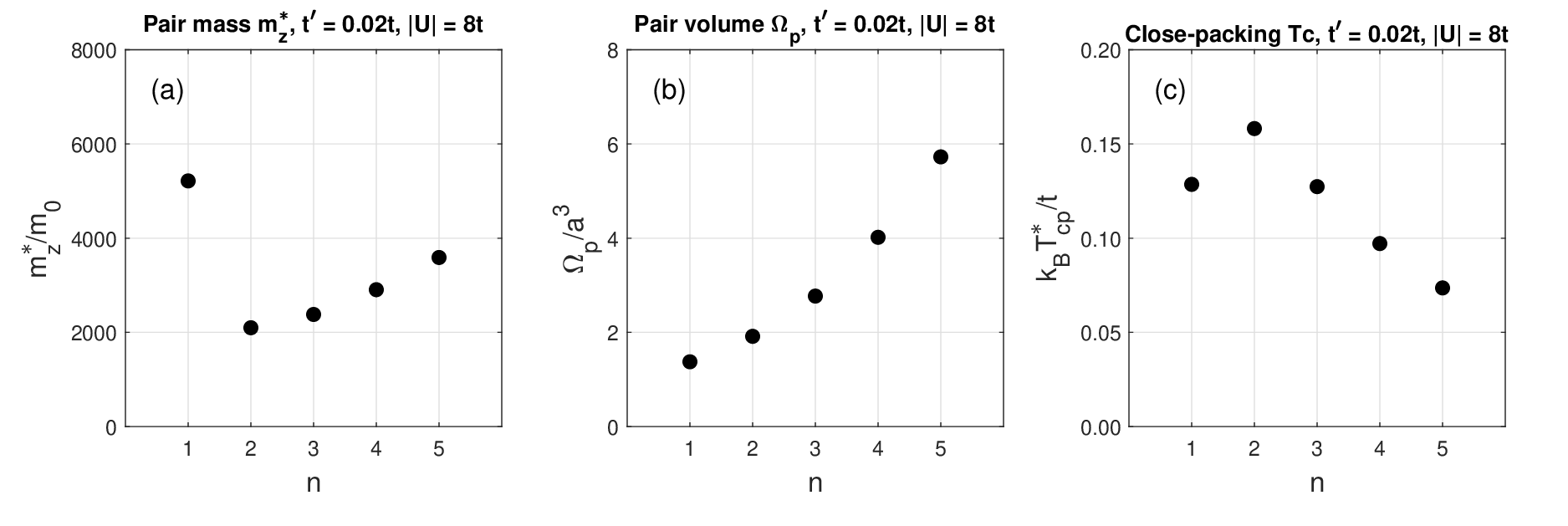}
\caption{$n$-dependence of the key physical properties of the uniform model for $t^{\prime} = 0.02 t$ and $|U| = 8.0 t$. Panel (c) shows the maximal critical temperature, which peaks at $n = 2$.} 
\label{MLH:fig:zeroeight}
\end{figure*}

\subsection{ \label{MLH:sec:fourone}
Uniform model, $\epsilon_{\alpha} = 0$ 
}

First, we present results for a model in which the on-site energies are equal for all layers. Hereafter, we refer to this this model as ``uniform.'' The on-site energies can be set to zero without loss of generality, $\epsilon_{\alpha} = 0$. The uniform model still captures the essential physics while involving only two parameters: attraction strength $|U|$ and the anisotropy $t^{\prime}$. It is therefore a minimal model and the natural first candidate to explain a nonmonotonic $T_{c,{\rm max}}(n)$.  

A selection of physical properties of the uniform model is shown in Fig.~\ref{MLH:fig:zeroseven}. Additional data and plots are provided in the Appendices. 

Figure~\ref{MLH:fig:zeroseven}(a) shows the pairing thresholds. For a fixed $t^{\prime}$, $|U_{\rm cr}|$ systematically increases with $n$, reflecting the rising kinetic energy that the attraction must overcome. Notice that the largest jump occurs between $n = 1$ and $n = 2$, where the {\em relative} change in kinetic energy is greatest. As $n$ increases further, the growth of the threshold slows. At $t^{\prime} = t$, all models reduce to the isotropic 3D attractive Hubbard model, so all curves converge to its threshold value, $\vert U_{\rm cr}(t^{\prime} = t) \vert = ( 7.913552 \ldots ) \, t$, \cite{Kornilovitch2024}. Another common feature of the threshold curves is the logarithmic singularity at extreme anisotropy, $t^{\prime} < 0.002 \, t$. In this regime, pairing is dominated by long-range fluctuations and the pair may be regarded as approximately two-dimensional. At $t^{\prime} > 0.002 \, t$, the pairs are fully three-dimensional, albeit strongly anisotropic \cite{Kornilovitch2024}. 

Figures~\ref{MLH:fig:zeroseven}(b)-(e) show the pair properties that determine the close-packing temperature via Eqs.~(\ref{MLH:eq:two}) and (\ref{MLH:eq:zeroeight}). The in-plane mass $m^{\ast}_{x}$ depends on $n$ monotonically and only weakly. For the physically most relevant regime, $|U| = (7-9)t$, the in-plane mass lies between 2.5 $m_0$ and 4.2 $m_0$, an interval consistent with cuprate estimates \cite{Bangura2008} if $m_0$ is interpreted as the free-electron mass. Thus, the degree of anisotropy affects in-plane dynamics only weakly, which is unsurprising. The out-of-plane mass $m^{\ast}_{z}$ exhibits a more intricate dependence on $n$, as shown in panel (c). Closer to threshold, $m^{\ast}_{z}$ decreases monotonically with $n$, similar to the in-plane mass. However, for $|U| > 7t$ the behavior becomes nonmonotonic: while $m^{\ast}_{z}$ still drops between $n = 1$ and $n = 2$, for $n \geq 3$ the trend reverses and $m^{\ast}_{z}$ begins to increase. This growth is attributed to increasing localization of the pair wave function on the inner planes of the stack. As discussed below, {\em the sharp increase of out-of-plane mass for $n \geq 3$ is the single most important factor causing $T^{\ast}_{\rm cp}$ to peak at $n = 2$ within the uniform model.} The in-plane and out-of-plane pair sizes are shown in panels (d) and (e), respectively. Both exhibit similar shapes: they diverge near threshold and approach zero in the strong-coupling limit where the pair collapses to a single site. Note that generally $r^{\ast}_{z} < r^{\ast}_{x}$, consistent with measurements of the coherence length in cuprates \cite{Mangel2024}. If $|U|$ is close to threshold, changing $n$ can produce a large, discontinuous increase in the pair volume $\Omega_{p}$ and a corresponding reduction in the close-packing density, which would suppress the maximum critical temperature. Finally, the close-packing temperature $T^{\ast}_{\rm cp}$ appears in Fig.~\ref{MLH:fig:zeroseven}(f). Its most striking feature is the well-defined maximum at an intermediate interaction strength. This peak reflects the simple physics outlined in Sec.~\ref{MLH:sec:two}. If the coupling is too strong, pairs are heavy and $T^{\ast}_{\rm cp}$ is low. Conversely, if the coupling is weak and near threshold, the pairs are diffuse and the packing density is low, again reducing $T^{\ast}_{\rm cp}$.

\begin{figure}[b]
\includegraphics[width=0.40\textwidth]{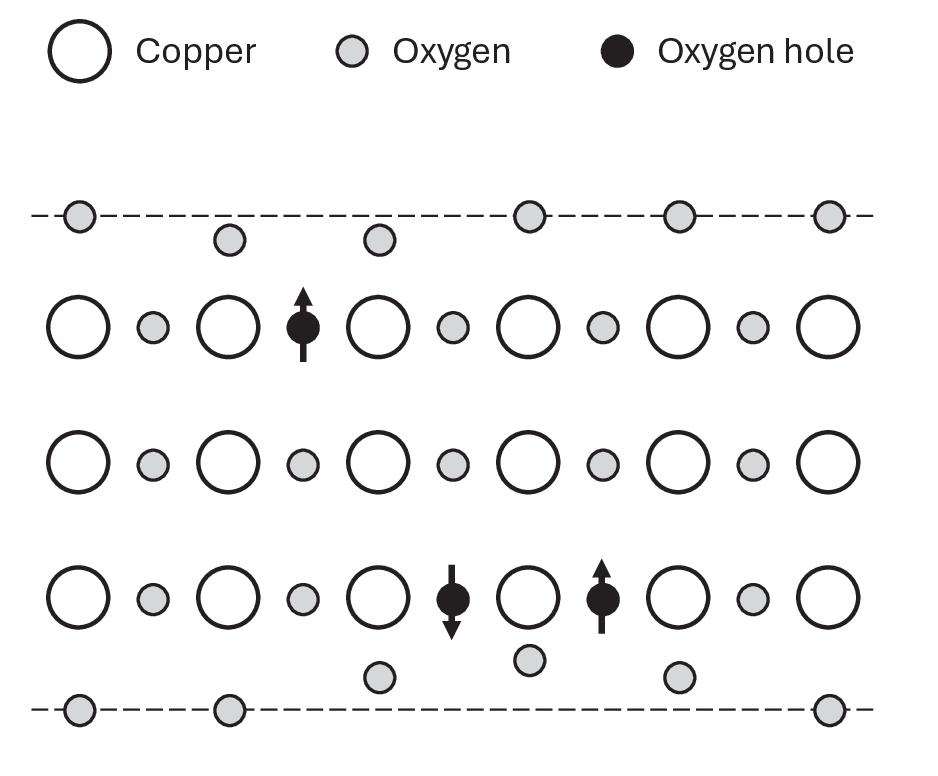}
\caption{Physical origin of the $n = 3$ low-outer-planes model. Dashed lines indicate the equilibrium positions of the apical oxygens.} 
\label{MLH:fig:zeronine}
\end{figure}
\begin{figure*}[t]
\includegraphics[width=0.95\textwidth]{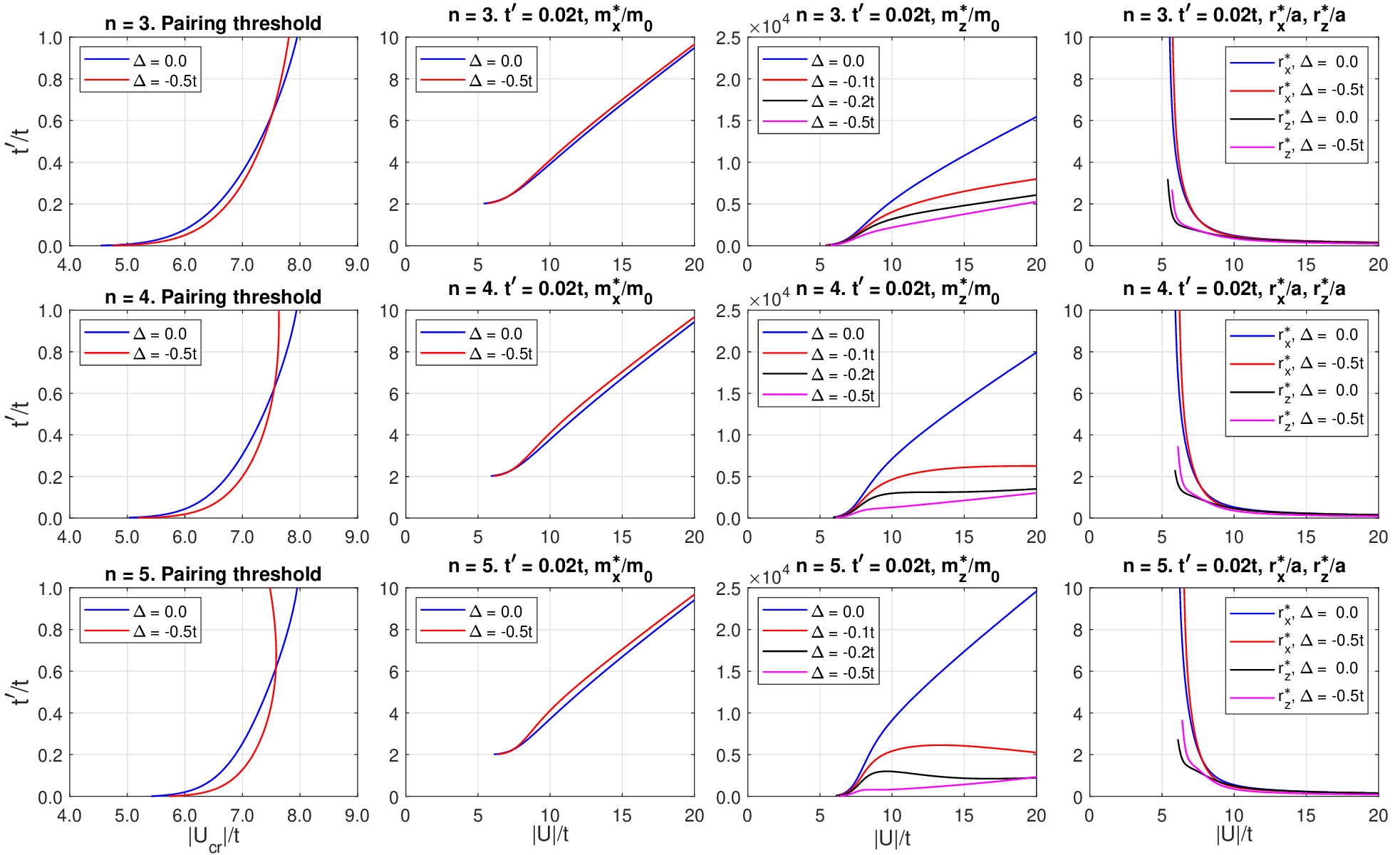}
\caption{Physical properties of the ``low-outer-planes'' attractive Hubbard model. The top, middle, and bottom rows show data for $n = 3$, $4$, and $5$, respectively. The first column displays the pairing threshold; the second, the in-plane mass $m^{\ast}_{x}/m_{0}$; the third, the out-of-plane mass $m^{\ast}_{z}/m_{0}$; and the forth, both the in-plane and out-of-plane pair sizes, $r^{\ast}_{x}$ and $r^{\ast}_{z}$, in units of $a$.} 
\label{MLH:fig:zeroten}
\end{figure*}

We now convert the above results into $n$-dependencies, shown in Fig.~\ref{MLH:fig:zeroeight}, which can be directly compared with empirical data. Panels (a) and (b) display the two key ingredients of Eq.~(\ref{MLH:eq:two}): the out-of-plane mass and the pair volume. Between $n = 1$ and $n = 2$, $m^{\ast}_{z}$ exhibits a sharp drop, consistent with the physical picture discussed in Sec.~\ref{MLH:sec:twofour}. This reduction leads to the increase in $T^{\ast}_{\rm cp}$ seen in panel (c). For $n > 2$, $m^{\ast}_{z}$ begins to increase, while $\Omega_{p}$ increases monotonically for all $n$. Those two effects combine to suppress $T^{\ast}_{\rm cp}$ for $n \geq 3$. {\em Thus, while the uniform model successfully achieves the key goal of explaining a nonmonotonic $T_{c,{\rm max}}(n)$, it predicts that the highest critical temperature always occurs at $n = 2$.} In the next section, we consider a more general model that yields a $T_{c,{\rm max}}(n)$ peaked at $n \geq 3$, consistent with the physics of cuprates.

\subsection{\label{MLH:sec:fourtwo}
``Low-outer-planes'' model, $\epsilon_{1} = \epsilon_{n} = \Delta < 0$ 
}

In this section, we introduce a pairing model in which the one-hole energies in the outer planes are {\em lower} than those in the inner planes. We refer to this as the ``low outer planes'' model. The physical justification for such an energy landscape lies in the interaction between holes and apical oxygen ions. It has long been proposed that displaced apical oxygens form {\em mobile} polarons within the copper-oxygen planes \cite{Alexandrov1999,Kornilovitch1999}, thereby promoting the formation of real-space pairs (i.e., {\em bipolarons}). Strong coupling between apical oxygens and the holes in the {\em outer} CuO$_{2}$ planes was recently confirmed in \cite{Hong2025}. 

The physics is illustrated in Fig.~\ref{MLH:fig:zeronine}. A positive hole residing in an outer CuO$_2$ plane attracts the negatively charged apical oxygens. The displacement of the latter lowers the hole's energy by an amount known as {\em polaron shift} \cite{Alexandrov1994,Lang1963}. Since a hole located in an inner plane is farther from the apical oxygens, its polaron shift is smaller. This difference can be modeled as a difference in single-hole energies, $\epsilon_{\rm outer} - \epsilon_{\rm inner} = \Delta$. Figure~\ref{MLH:fig:zeronine} also illustrates a possible mechanism of attraction based on the {\em bipolaron effect}. Two co-located holes deform the apical oxygens more effectively than two separated holes \cite{Alexandrov2002}, which is equivalent to a short-range attraction $-|U|$. In real cuprates, this bipolaron attraction receives contributions from many distant ions, and its proper calculation is complex \cite{Alexandrov2002}. Here, we ignore this complications, as well as potential variations of $|U|$ with hole location within the stack. The crucial element of the model is a negative $\Delta$, which increases the probability weight of holes on the outer planes and thereby reduces the out-of-plane pair mass. The same effect leads to an apparent overdoping of the outer planes relative to the inner ones. This property is consistent with cuprate experiments \cite{Horio2025,Chen2025}.

\begin{figure*}[t]
\includegraphics[width=0.90\textwidth]{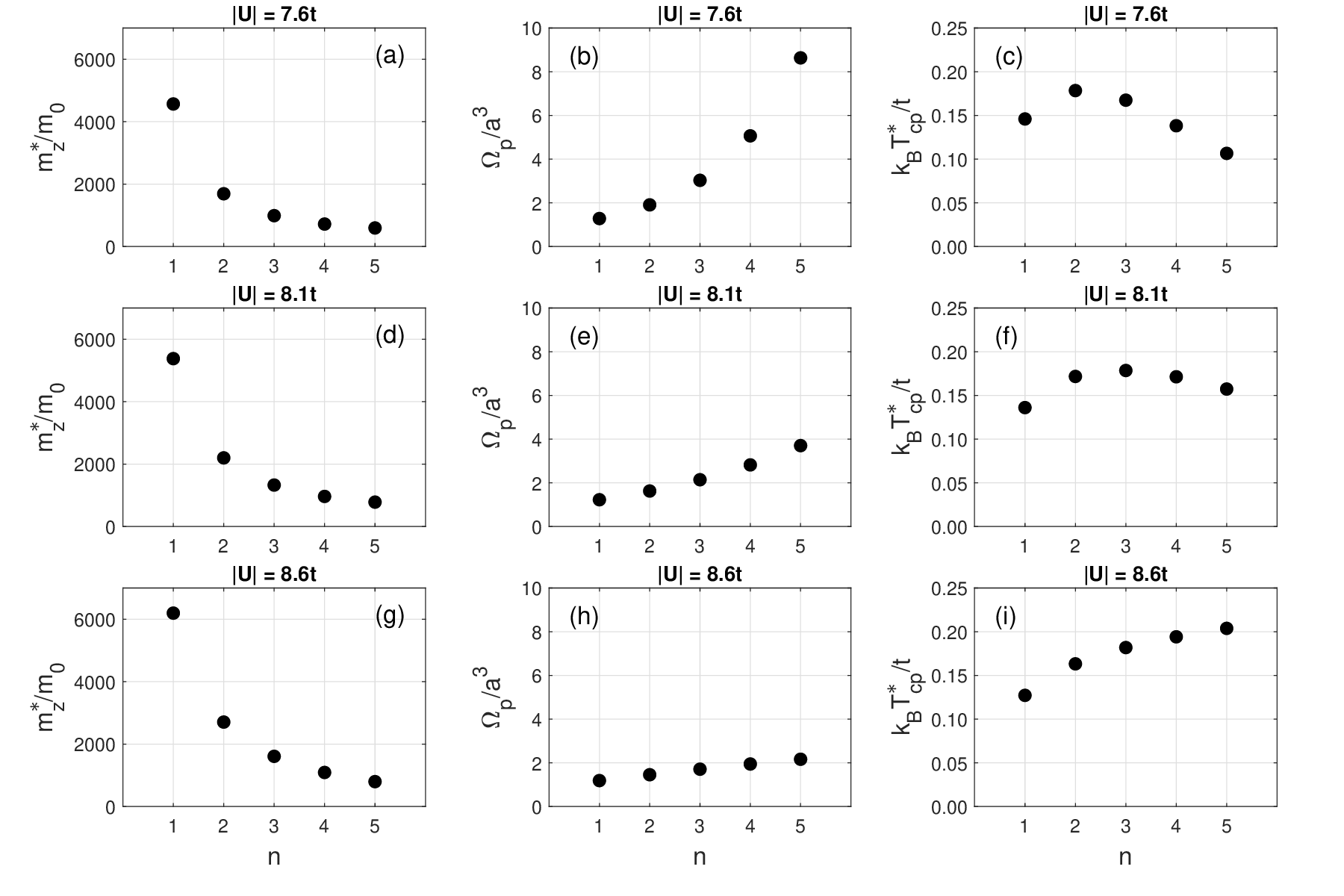}
\caption{$n$-dependence of the key physical properties of the low-outer-planes model for $t^{\prime} = 0.02t$, $\Delta = -0.5 t$, and three different values of $|U|$. The right column shows the close-packing critical temperature. Compare panel (f) with Fig.~\ref{MLH:fig:zero}. Compare the middle row with Fig.~\ref{MLH:fig:one}(d). } 
\label{MLH:fig:zeroeleven}
\end{figure*}

$n$-dependence of the key physical properties of the uniform model for $t^{\prime} = 0.02 t$ and $|U| = 8.0 t$. Panel (c) shows the maximal critical temperature, which peaks at $n = 2$.

Thus, the low-outer-planes pairing model is described by Eq.~(\ref{MLH:eq:three}) with $\epsilon_{1} = \epsilon_{n} = \Delta$ and all other $\epsilon_{\alpha}$ set to zero. We consider only $\Delta < 0$. The introduction of an additional parameter makes complete characterization more challenging, so here we present results for a single representative parameter set that supports the main message of the paper. Additional data are provided in the Appendices. 

The first column of Fig.~\ref{MLH:fig:zeroten} compares the pairing thresholds for $\Delta = 0$ and $\Delta = - 0.5t$ in the $n = 3$, $4$, and $5$ models. One sees that the shift in the threshold is less than $0.5 t$ for any value of $t^{\prime}$, and less than $0.3 t$ in the physically most relevant region, $t^{\prime} < 0.1 t$. Overall, threshold variation with $\Delta$ is not a significant factor in our model. The other three columns of Fig.~\ref{MLH:fig:zeroten} show the effective pair masses and sizes for a single anisotropy value, $t^{\prime} = 0.02 t$. Among the four properties, only $m^{\ast}_{z}$ exhibits a significant dependence on $\Delta$, as seen in the third column. The out-of-plane pair mass decreases systematically with increasingly negative $\Delta$, due to the growing accumulation of hole weight on the outer planes, which enhances the probability of tunneling across the charge-reservoir layers. {\em It is this reduction of out-of-plane mass at negative $\Delta$ that raises the critical temperature in the $n \geq 3$ models above their $n = 2$ values.}        

Close-packing temperature $T^{\ast}_{\rm cp}$ versus $n$ for $t^{\prime} = 0.02 \, t$, $\Delta = -0.5 \, t$, and three different values of the attraction is shown in Fig.~\ref{MLH:fig:zeroeleven}. The $U$ values were chosen to illustrate the flexibility of the model. At $|U| = 7.6t$, $T^{\ast}_{\rm cp}(n)$ still peaks at $n = 2$, as in the uniform model. However, the rate of decline for $n \geq 3$ is smaller than in the uniform model, reflecting an important difference in the underlying physics. As seen in Fig.~\ref{MLH:fig:Hsix}, for $n \geq 3$, the value of $|U| = 7.6t$ lies on the {\em left} side of the $T^{\ast}_{\rm cp}(|U|)$ curves, meaning it is very close to the threshold. As $n$ increases, the threshold moves even closer to $|U| = 7.6t$, weakening the binding. In other words, {\em the decline of $T^{\ast}_{\rm cp}(n \geq 3)$ is caused by ballooning pair volumes rather than by an increase in $m^{\ast}_{z}$}. At $|U| = 8.1 t$ (middle row of Fig.~\ref{MLH:fig:zeroeleven}), decreasing $m^{\ast}_{z}(n)$ and increasing $\Omega_{p}(n)$ combine to shift the peak to $n = 3$. At $n < 3$, the system is far from threshold, so the increase of $T^{\ast}_{\rm cp}$ is driven by decreasing out-of-plane mass $m^{\ast}_{z}$. For $n \geq 3$, the system lies close to threshold, and the decrease of $T^{\ast}_{\rm cp}$ is driven by primarily by increasing pair volume $\Omega_{p}$. {\em This physical mechanism is the central result of the paper.} At $|U| = 8.6 t$ (bottom row of Fig.~\ref{MLH:fig:zeroeleven}), the system remains far from threshold for all $n$. As a result, the pair volume is relatively stable, and $T^{\ast}_{\rm cp}$ is determined mainly by $m^{\ast}_{z}$. Because the latter decreases for all $n$ in this regime, $T^{\ast}_{\rm cp}(n)$ continues to rise beyond $n = 3$.       

Thus, the low-outer-planes pairing model is capable of reproducing a $T^{\ast}_{\rm cp}(n)$ curve peaked at $n = 3$, in agreement with empirical observations for the cuprates. Compare Fig.~\ref{MLH:fig:zeroeleven}(f) with Fig.~\ref{MLH:fig:zero}. In addition, the model predicts scenarios in which $T^{\ast}_{\rm cp}$ peaks at $n > 3$. Such cuprates families have not yet been discovered.

\section{ \label{MLH:sec:five}
Path to higher $T_{c}$ 
}

Increasing the critical temperature of superconductors remains a central objective of the condensed matter community \cite{Prasankumar2026}. The current ambient-pressure record of 138 K was set in 2001 \cite{Lokshin2001} and has not been surpassed since.~\footnote{The above-200 K superconductivity reported in metal hydrides under high pressure \cite{Drozdov2015,Drozdov2019} is unstable under ambient conditions and therefore is not discussed here.} Comprehensive theoretical analyses of {\em conventional} superconductivity in the McMillan \cite{McMillan1968} and Migdal-Eliashberg formulations \cite{Eliashberg1960,Eliashberg1961,Nambu1960} still limit $T_{c, {\rm max}}$ to below 100 K, even when reasonable extensions such as lattice anisotropy, phonon anharmonicity, and multiple gaps are included \cite{Gao2025}.~\footnote{We note that analysis of Migdal-Eliashberg superconductivity at strong electron-phonon coupling must be done with great care because of spontaneous polaron formation \cite{Alexandrov1992b,Alexandrov2001,Alexandrov2007}.}

It must be appreciated then that preformed pair superconductivity faces the {\em opposite} challenge: its critical temperature tends to be too high! Substituting the {\em pair} density~\footnote{We assume lattice parameters $a = b = 3.853$ \AA, $c = 15.78$ \AA, $n = 3$, and a hole density of $0.3$ per CuO$_{2}$ unit cell.} $\rho_2 = 1.9 \times 10^{21}$ cm$^{-3}$ and effective masses $m^{\ast}_{i} \sim 10 \, m_{e}$ into Eq.~(\ref{MLH:eq:one}) yields $T_c \approx 450$ K. Interaction between pairs is expected to reduce $T_{c}$, but not dramatically. Even in liquid $^{4}$He --- arguably a very strongly interacting quantum liquid --- the observed $T_{c}$ is suppressed relative to the ideal-BEC value by only $\sim 30$\%. Assuming a comparable reduction for a liquid of preformed pairs still leaves $T_{c} > 300$ K.~\footnote{The original BEC ideas of Schafroth {\em el al.} \cite{Schafroth1954,Schafroth1957} encountered a similar difficulty: the predicted $T_c$ was far too high compared with the $T_{c} \sim 10$ K of simple metals. Another early criticism \cite{Casimir1955} concerned the prediction of a gap in the {\em normal} state (due to pair binding), which was not observed in simple metals above $T_{c}$. In underdoped cuprates, such a gap {\em is} seen and is typically referred to as the {\em pseudogap}.}

The reasons why $T_{c, {\rm max}}$ of cuprates is ``mere'' $\sim 130$ K can be deduced from Eq.~(\ref{MLH:eq:two}). First, the out-of-plane mass $m^{\ast}_{z}$ is much larger than $10 \, m_{e}$, as discussed in Sec.~\ref{MLH:sec:four}. The large mass arises both from the intrinsic anisotropy of the cuprates ($t^{\prime} \ll t$) and from polaron mass enhancement associated with the displacement of apical oxygens. These observations suggest several possible strategies for boosting $T_{c}$ through a reduction of $m^{\ast}_{z}$. The most obvious method would be to increase $t^{\prime}$ by narrowing the charge reservoir layers (CRL). However, strong anisotropy is essential for pair formation, and the ionic CRLs also contribute to the effective attraction between holes. Thus, CRLs must be carefully optimized with respect to several competing tradeoffs. Such optimization has been pursued empirically since the discovery of the cuprates and has led to the highest-performing families known today.       

Another approach would be to reduce $m^{\ast}_{z}$ without changing CRL composition. For example, external pressure applied along the $c$-axis constrains the motions of apical oxygens and increases their vibrational frequency, as confirmed by Raman spectroscopy \cite{Mark2022}. According to small-polaron theory \cite{Lang1963,Alexandrov1994}, higher phonon frequencies reduce the bipolaron effective mass and, by virtue of Eq.~(\ref{MLH:eq:two}), enhance $T_{c}$. This provides a natural explanation for the observed pressure dependence of $T_{c}$ in cuprates \cite{Mark2022}. In general, {\em any action that increases the apical-oxygen vibration frequency without weakening the attraction between holes would raise $T_{c}$.} This may be what occurs in the recently reported ``pressure-quench'' experiments \cite{Deng2026}. Another indirect support for the polaron mechanism comes from ``dynamically stabilized superconductivity'' \cite{Hu2014}. In these experiments, apical-oxygen vibrations are resonantly enhanced by THz laser pulses, leading to superconducting-like correlations up to $T \sim 250$ K. This enhancement has been interpreted in terms of ``polaron undressing'' \cite{Kornilovitch2016,Kornilovitch2017}: higher-amplitude oscillations move ions closer to their equilibrium positions, enabling a hole pair to tunnel between conductive stacks without carrying the accompanying lattice distortion. In this regime, the pair's $m^{\ast}_{z}$ is reduced toward its bare band-structure value. Because this mass reduction is exponential, the {\em apparent} critical temperature rises dramatically \cite{Hu2014}.      
      
A second reason why cuprate $T_{c,{\rm max}}$ may appear ``too low'' is the pair volume $\Omega_{p}$. The in-plane correlation length is $1.5 - 2.3$ nm \cite{Wang2003,Oh1988,Pan2000,Mangel2024}. The ``1/8'' $T_c$ anomaly suggests that the in-plane area occupied by a hole pair is about 16 CuO$_2$ unit cells \cite{Emin1995,Li2023}. As shown by the the size calculations in Sec.~\ref{MLH:sec:four}, near the pairing threshold pair volume is extremely sensitive to the attractive interaction strength. Even a modest increase in $|U|$ can cause $\Omega_{p}$ to collapse, producing a substantial increase in $T^{\ast}_{\rm cp}$. In other words, more compact pairs enable denser close packing and therefore a higher $T_{c}$. This mechanism is limited, at least conceptually, by the possibility of phase separation: an excessively strong attraction would favor the formation of hole trions \cite{Kornilovitch2013,Kornilovitch2014,Kornilovitch2020}, hole quads \cite{Kornilovitch2023}, and larger clusters, thereby undermining the picture of a stable liquid of hole pairs. However, moderate reductions of the pair volume on the order of 50\% would not induce phase separation while proportionally increasing $T_{c,{\rm max}}$ at the same time. A similar conclusion was reached in a large-scale empirical study of unconventional superconductivity \cite{Wang2025}.        

Based on the above reasoning, one can formulate the following recommendations for increasing $T_{c,{\rm max}}$. (i) Ionic crystals are preferred, as they induce attraction between holes through long-range electron-ion or hole-ion interactions. (ii) Lattice anisotropy helps reduce kinetic energy and promotes pairing. (iii) Strong correlations further suppress kinetic energy and enhance pairing. (iv) Both cuprates and nickelates have all of these features and thus serve as strong starting points. Quasi-one-dimensional crystals exhibit similar physics and represent another promising materials class. (v) One must seek ways to decrease out-of-plane mass, for example by restricting the motion of apical oxygens, but without diminishing the attractive potential. (vi) One must seek ways to increase the attracting interaction to produce pairs more compact, but without causing phase separation \cite{Kornilovitch2023}. (vii) The maximum $T_{c}$ is given by Eq.~(\ref{MLH:eq:two}) and its finite-interaction corrections.  

Like in conventional superconductivity, modern theory can provide useful guidance. State-of-the-art DFT calculations may help clarify phonon frequencies, polaron effects, and pressure responses \cite{Park2020,Houtput2025}. Lattice-energy minimization techniques \cite{Zhang1991,Catlow1998,Edwards2023} may aid in determining $|U|$ and, more generally, the full attractive potential profile. Finally, there is a need for a general theory of real-space superconductivity in which particle-particle pairing is treated rigorously using a Bethe-Salpeter approach, analogous to the treatment of excitons in optical response \cite{Strinati1988,Rohlfing2000}.

\begin{acknowledgments}

The author wishes to thank Maciej Bak, David Blatt, and James Hague for helpful and insightful discussions.     

\end{acknowledgments}
%
%


\begin{appendix}

\begin{widetext}

\section{\label{MLH:sec:app:a}
Preliminaries
}

In these Appendices, two-particle solutions of multi-layer attractive Hubbard models with increasing number of layers $n$ are developed. Multi-band pairing models are significantly more complex than single-band pairing models recently reviewed in [\onlinecite{Kornilovitch2024}]. Beyond our own old works [\onlinecite{Alexandrov1992,Alexandrov1993,Ivanov1994,Kornilovitch1997}] and a recent paper by Iskin \cite{Iskin2024}, we are not aware of any other multi-band models studied so far. These Appendices are the first systematic exposition of the subject, albeit for the simplest form of attractive interaction. 

Throughout the Appendices, nearest-neighbor distances are set to unity along all lattice axes, $a = 1$. Layer-to-layer distances within an $n$-layer stack and between stacks are the same and equal to $1$. All three components of lattice momentum $k_i$ or $q_i$, $i = x, y, z$ are measured in units of $1/a$ and are dimensionless. The total pair lattice momentum ${\bf P}$ is a constant of motion. Vectors ${\bf m}$ mark the common reference point in all unit cells. The nearest neighbor vectors along the three axes are denoted ${\bf x}$, ${\bf y}$, and ${\bf z}$, respectively. Pair size is measured in units of $a$. Pair masses are measured in units of bare $(xy)$ mass, $m_0 \equiv \hbar^2/(2 t a^2)$. Only one spin-singlet $s$-symmetrical pair state is formed by the Hubbard attraction. Therefore, (anti)symmetrization of the wave function (Ref. [\onlinecite{Kornilovitch2024}], sections 2.3 and 2.4) is not needed, and the unsymmetrized formalism (Ref. [\onlinecite{Kornilovitch2024}], section 2.2) is utilized.      

In the limit $t^{\prime} = t$, the multi-layer attractive Hubbard model reduces to the isotropic {\em cubic} attractive Hubbard model for {\em all} $n$. Thus, the single-band cubic model serves as a useful reference.

\section{\label{MLH:sec:app:b}
One layer, $n = 1$
}

The $n = 1$ attractive Hubbard model was solved earlier in [\onlinecite{Kornilovitch2024}], section 3.6, where it was called {\em tetragonal} attractive Hubbard model. Here, theory of the $n = 1$ case is presented for completeness. New numerical results are also given. Two-body wave function has one component $\Psi_{11}({\bf m}_1 , {\bf m}_2)$. Two-body Schr\"odinger equation reads   
\begin{eqnarray}
&& - t \left[ 
\Psi_{11}( {\bf m}_1 - {\bf x} , {\bf m}_2 ) + \Psi_{11}( {\bf m}_1 + {\bf x} , {\bf m}_2 ) + 
\Psi_{11}( {\bf m}_1 - {\bf y} , {\bf m}_2 ) + \Psi_{11}( {\bf m}_1 + {\bf y} , {\bf m}_2 ) \right]
\nonumber \\              
&& - t \left[ 
\Psi_{11}( {\bf m}_1 , {\bf m}_2 - {\bf x} ) + \Psi_{11}( {\bf m}_1 , {\bf m}_2 + {\bf x} ) + 
\Psi_{11}( {\bf m}_1 , {\bf m}_2 - {\bf y} ) + \Psi_{11}( {\bf m}_1 , {\bf m}_2 + {\bf y} ) \right]              
\nonumber \\
&& - t^{\prime} \left[ 
\Psi_{11}( {\bf m}_1 - {\bf z} , {\bf m}_2 ) + \Psi_{11}( {\bf m}_1 + {\bf z} , {\bf m}_2 ) + 
\Psi_{11}( {\bf m}_1 , {\bf m}_2 - {\bf z} ) + \Psi_{11}( {\bf m}_1 , {\bf m}_2 + {\bf z} ) \right]              
\nonumber \\
&& - \vert U \vert \: \delta_{ {\bf m}_1 , {\bf m}_2 } \Psi_{11}({\bf m}_1 , {\bf m}_2) 
= E \Psi_{11}({\bf m}_1 , {\bf m}_2) \: .   
\label{MLH:eq:bone}
\end{eqnarray}
Here $E$ is total energy of the system, i.e., pair energy. The equation is transformed in momentum space by applying Fourier transformation  
\begin{eqnarray}
\Psi_{11}({\bf m}_1 , {\bf m}_2) & = & \frac{1}{N_1} \sum_{{\bf k}_1 {\bf k}_2} 
\psi_{11}({\bf k}_1 , {\bf k}_2) \, e^{ i {\bf k}_1 {\bf m}_1 + i {\bf k}_2 {\bf m}_2 }   \, ,  
\label{twopart:eq:five} \\
\psi_{11}({\bf k}_1 , {\bf k}_2) & = & \frac{1}{N_1} \! \sum_{{\bf m}_1 {\bf m}_2} \!\! 
\Psi_{11}({\bf m}_1 , {\bf m}_2) \, e^{ - i {\bf k}_1 {\bf m}_1 - i {\bf k}_2 {\bf m}_2 } \, ,
\label{twopart:eq:fiveone}
\end{eqnarray}
where $N_1$ is the number of units cells of unit cell volume $\Omega_0 = 1a^3$ in the system. A transformed equation reads
\begin{equation}
( E - \varepsilon_{{\bf k}_1} - \varepsilon_{{\bf k}_2} ) \, \psi_{11}( {\bf k}_1 , {\bf k}_2 ) =  
- \vert U \vert \, \frac{1}{N_1} \sum_{\bf q} \psi_{11}( {\bf q}, {\bf k}_1 + {\bf k}_2 - {\bf q} ) \: ,  
\label{twopart:eq:six}
\end{equation}
where
\begin{equation}
\varepsilon_{\bf k} = - 2 t \left( \cos{k_x} + \cos{k_y} \right) - 2 t^{\prime} \cos{k_z}   
\label{MLH:eq:bfive}
\end{equation}
is the bare one-particle dispersion. The Schr\"odinger equation, Eq.~(\ref{twopart:eq:six}), is solved by introducing auxiliary function
\begin{equation}
\Phi_{11}( {\bf k}_1 + {\bf k}_2 ) = \Phi_{11}( {\bf P} )
\equiv \frac{1}{N_1} \sum_{\bf q} \psi_{11}( {\bf q}, {\bf k}_1 + {\bf k}_2 - {\bf q} ) 
=      \frac{1}{N_1} \sum_{\bf q} \psi_{11}( {\bf q}, {\bf P} - {\bf q} ) \: ,  
\label{MLH:eq:bsix}
\end{equation}
where
\begin{equation}
{\bf P} = {\bf k}_1 + {\bf k}_2   
\label{MLH:eq:bseven}
\end{equation}
is the pair lattice momentum. The pair wave function follows from Eq.~(\ref{twopart:eq:six}):
\begin{equation}
\psi_{11}( {\bf k}_1 , {\bf k}_2 ) = 
\frac{- \vert U \vert}{ E - \varepsilon_{{\bf k}_1} - \varepsilon_{{\bf k}_2} } \: 
\Phi_{11}( {\bf P} ) \: .  
\label{MLH:eq:beight}
\end{equation}
Note that $\psi_{11}( {\bf k}_2 , {\bf k}_1 ) = \psi_{11}( {\bf k}_1 , {\bf k}_2 )$. Equation~(\ref{MLH:eq:bsix}) describes a symmetric pair state, i.e., spin-singlet. Substitution of $\psi_{11}( {\bf k}_1 , {\bf k}_2 )$ back into the definition of $\Phi_{11}$, Eq.~(\ref{MLH:eq:bsix}), yields 
\begin{equation}
\Phi_{11}( {\bf P} ) = \left[ - \vert U \vert \: \frac{1}{N_1} \sum_{\bf q} 
\frac{1}{ E - \varepsilon_{{\bf q}} - \varepsilon_{{\bf P} - {\bf q}} } \right]  
\Phi_{11}( {\bf P} ) \: .  
\label{MLH:eq:bnine}
\end{equation}
Crucially, $\Phi$ has been moved outside of the ${\bf q}$ integral because $\Phi({\bf P})$ depends on the sum of two momenta rather than on ${\bf k}_1$ and ${\bf k}_2$ separately. This reduces the integral equation to an algebraic one. In other words, momentum conservation ensures the integrability of the problem. For the following, it is convenient to shift the internal variable by half the total momentum: ${\bf q} \to {\bf q} + ({\bf P}/2)$. This does not change the integral but makes the integrand explicitly even in both ${\bf q}$ and ${\bf P}$. Equation~(\ref{MLH:eq:bnine}) takes the form
\begin{equation}
\vert U \vert M_{11} \cdot  \Phi_{11}( {\bf P} ) = \Phi_{11}( {\bf P} ) \: ,  
\label{MLH:eq:bten}
\end{equation}
\begin{equation}
M_{11} = M_{11}(E_{2},{\bf P}) \equiv - \frac{1}{N_1} \sum_{\bf q} 
\frac{1}{E - \varepsilon_{ \frac{\bf P}{2} + {\bf q} } - \varepsilon_{ \frac{\bf P}{2} - {\bf q} } } \: .  
\label{MLH:eq:beleven}
\end{equation}
Equation~(\ref{MLH:eq:bten}) is a $( 1 \times 1 )$ linear ``matrix equation'' for $\Phi_{11}({\bf P})$. (Recall that pair momentum ${\bf P}$ is treated as a parameter.) Its ``consistency condition'' 
\begin{equation}
\vert U \vert M_{11}( E , {\bf P} ) = 1    
\label{MLH:eq:btwelve}
\end{equation}
implicitly determines pair dispersion $E({\bf P})$. The quantity $M_{11}$ is the {\em two-body} Green's function of the tetragonal lattice, which is equivalent, for arbitrary ${\bf P}$, to the {\em one-body} Green's function of the {\em orthorhombic} lattice. Mathematically, $M_{11}$ is a Watson integral [\onlinecite{Zucker2011}]. Note that we will be dealing exclusively with negative energies, $E < 0$, so that $M_{11} > 0$. The pair wave function follows from Eq.~(\ref{MLH:eq:beight}) subject to the constraint ${\bf k}_1 + {\bf k}_2 = {\bf P} = {\rm const}$.

\begin{figure}[t]
\includegraphics[width=0.48\textwidth]{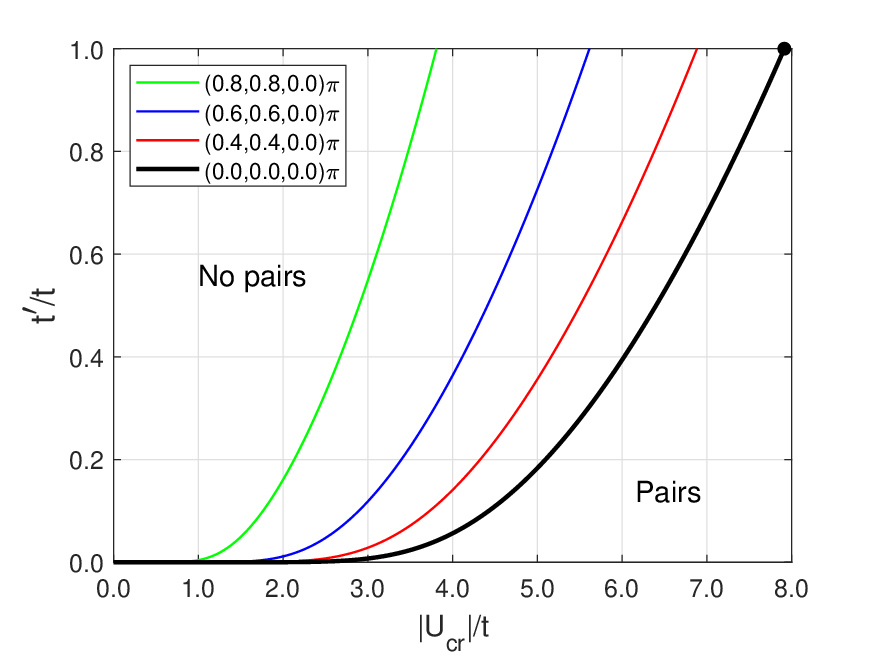}
\caption{Binding thresholds in $n = 1$ attractive Hubbard model for several pair momenta {\bf P} whose values are listed in the legend. Bound pairs are stable to the right of respective lines. Since $P_x = P_y$, the lines are described by an analytical formula, the inverse of Eq.~(\ref{twopart:eq:sixtythree}). Notice that moving pairs are formed at smaller $\vert U \vert$. The ground state threshold (${\bf P} = 0)$ is shown by the thick black line. The isotropic threshold $\vert U_{\rm cr}({\bf P} = 0, t^{\prime} = t) \vert = ( 7.913552 \ldots ) \, t$ is a useful reference value [\onlinecite{Kornilovitch2024}], section 3.5. It is marked by a circle.} 
\label{MLH:fig:Bone}
\end{figure}

We begin analysis with deriving a binding threshold $|U_{\rm cr}|$. The lowest energy of two free particles with total mo\-mentum ${\bf P}$ is $E_{11} = 2 \varepsilon_{\frac{\bf P}{2}}$. Substituting $E = E_{11}$ in the dispersion equation, Eq.~(\ref{MLH:eq:btwelve}), yields $\vert U_{\rm cr} \vert$ for given pair momentum ${\bf P}$. Utilizing $\varepsilon_{\bf k}$ from Eq.~(\ref{MLH:eq:bfive}) one obtains
\begin{equation}
M_{11}(E_{2} = E_{11},{\bf P}) = 
\int\limits^{\pi}_{-\pi} \!\! \int\limits^{\pi}_{-\pi} \!\! \int\limits^{\pi}_{-\pi}     
\frac{ {\rm d} q_x {\rm d} q_y {\rm d} q_z}{ ( 2 \pi )^3 }
\frac{1}{  4t \cos{\frac{P_x}{2}} ( 1 - \cos{q_x} ) + 4t \cos{\frac{P_y}{2}} ( 1 - \cos{q_y} ) 
+ 4t^{\prime} \cos{\frac{P_z}{2}} ( 1 - \cos{q_z} ) } \: .  
\label{MLH:eq:bthirteen}
\end{equation}
If $P_x = P_y$, this integral is known analytically [\onlinecite{Montroll1956}]: 
\begin{equation}
M_{11}( E = E_{11} , P_x = P_y , P_z) = 
\frac{1}{\pi^2 t \cos{\frac{P_x}{2}} } \sqrt{ \frac{ 2 \kappa_1 \kappa_2 }{ \xi } } \, 
{\bf K}(\kappa_1) {\bf K}(\kappa_2) \: ,   
\label{twopart:eq:sixtythree}
\end{equation}
where 
\begin{equation}
\kappa_{1,2} = \frac{1}{2\xi} \left( \sqrt{ 4 + 2\xi } \pm 2 \right) \! 
\left( 2 \sqrt{ 1 + \xi } - \sqrt{ 4 + 2\xi } \right) ,
\hspace{1.0cm}
\xi \equiv \frac{ t^{\prime} \cos{\frac{P_z}{2}} }{ t \cos{\frac{P_x}{2} } } \: ,   
\hspace{1.0cm}
{\bf K}(\kappa) = \int\limits^{\pi/2}_{0}  
\frac{{\rm d} \phi}{\sqrt{ 1 - \kappa^2 \sin^2{\! \phi} }} \: .
\label{twopart:eq:sixtyfour}
\end{equation}
${\bf K}(\kappa)$ is the complete elliptic integral of the first kind. For arbitrary ${\bf P}$, only two integrations in Eq.~(\ref{MLH:eq:bthirteen}) can be evaluated analytically while the remaining one has to be computed numerically, see for example Appendix D3 in [\onlinecite{Kornilovitch2024}]. From Eq.~(\ref{MLH:eq:btwelve}), $\vert U_{\rm cr} \vert = M^{-1}_{11}$. $\vert U_{\rm cr} \vert$-vs.-$t^{\prime}$ curves for different pair momenta including ${\bf P} = 0$ are shown in Fig.~\ref{MLH:fig:Bone}. Two properties of these curves should be noticed. The first is the overall shape. For anisotropy levels of order one, $t^{\prime} \sim t$, the triple integral in Eq.~(\ref{MLH:eq:bthirteen}) converges, which leads to finite $\vert U_{\rm cr} \vert \sim {\cal{O}}(t)$. In this regime, the pair is still three-dimensional but anisotropic. But when $t^{\prime} \to 0$, $M_{11}$ starts to diverge logarithmically, and the threshold tends to zero as $\vert U_{\rm cr} \vert \propto 1/\log(t/t^{\prime})$. Now pair formation is governed by this logarithmic divergence and the system is close to be two-dimensional. The crossover between anisotropic 3D and pure 2D regimes occurs at $t^{\prime} \simeq 0.002 \, t$ [\onlinecite{Kornilovitch2024}], section 9.1. The second notable feature of Fig.~\ref{MLH:fig:Bone} is the threshold dependence on pair momentum ${\bf P}$. It is the common property of all lattice models that moving pairs are easier to form than stationary ones [\onlinecite{Kornilovitch2024},\onlinecite{Kornilovitch2004}], and the present model is no exception. This effect may be related to formation of pair density waves. It will not be discussed further in this work.      

When attraction exceeds the threshold, pair energy can be found from Eq.~(\ref{MLH:eq:btwelve}) by a root-searching algorithm. Again, if $P_x = P_y$, an analytical formula for $M_{11}$ exists even for arbitrary $E$, albeit it is very cumbersome [\onlinecite{Joyce2001a,Joyce2001b,Joyce2003}]. For arbitrary ${\bf P}$, the triple integral $M_{11}$ can be computed as is. Alternatively, either one or two integrations can be performed analytically, which significantly speeds up computation. 

Pair effective mass $m^{\ast}$ is now discussed. From now on, we are going to be interested only in the effective mass near the bottom of the band. The three masses are defined as follows:
\begin{equation}
E({\bf P} \ll 1 ) = E_{0} + \frac{\hbar^2 P^2_{x}}{2 m^{\ast}_{x}}
+ \frac{\hbar^2 P^2_{y}}{2 m^{\ast}_{y}} + \frac{\hbar^2 P^2_{z}}{2 m^{\ast}_{z}} \: ,    
\label{MLH:eq:bsixteen}
\end{equation}
where $E_{0} = E({\bf P} = 0)$. In all our models, $m^{\ast}_{x} = m^{\ast}_{y}$, so only two masses will be needed. There are two principle methods of deriving $m^{\ast}_{i}$. The first method is purely numerical. $E$ is computed for ${\bf P} = 0$ and for some ${\bf P} \ll 1$, and then the second derivative is obtained as a finite difference. However, in simple cases like the $n = 1$ Hubbard model, analytical expansion of $E$ at small ${\bf P}$ is feasible, which provides an alternative route. Expanding Eq.~(\ref{MLH:eq:btwelve}) for small $P_{i}$ one obtains
\begin{equation}
\frac{m^{\ast}_{x}}{m_0} = 2
\frac{ \int\limits^{\pi}_{-\pi} \!\!\! \int\limits^{\pi}_{-\pi} \!\!\! \int\limits^{\pi}_{-\pi} 
\frac{ {\rm d}q_x \, {\rm d}q_y \, {\rm d}q_z }{ (2\pi)^3 }
\frac{1}{ [ \vert E_{0} \vert - 4 t \, ( \cos{q_x} + \cos{q_y} ) - 4 t^{\prime} \cos{q_z} ) ]^2 } }
{ \int\limits^{\pi}_{-\pi} \!\!\! \int\limits^{\pi}_{-\pi} \!\!\! \int\limits^{\pi}_{-\pi} 
\frac{ {\rm d}q_x \, {\rm d}q_y \, {\rm d}q_z }{ (2\pi)^3 }
\frac{\cos{q_x}}{ [ \vert E_{0} \vert - 4 t \, ( \cos{q_x} + \cos{q_y} ) - 4 t^{\prime} \cos{q_z} ) ]^2 } } \: ,  
\label{MLH:eq:bseventeen}
\end{equation}
\begin{equation}
\frac{m^{\ast}_{z}}{m_0} = 2 \: \left( \frac{t}{t^{\prime}} \right)
\frac{ \int\limits^{\pi}_{-\pi} \!\!\! \int\limits^{\pi}_{-\pi} \!\!\! \int\limits^{\pi}_{-\pi} 
\frac{ {\rm d}q_x \, {\rm d}q_y \, {\rm d}q_z }{ (2\pi)^3 }
\frac{1}{ [ \vert E_{0} \vert - 4 t \, ( \cos{q_x} + \cos{q_y} ) - 4 t^{\prime} \cos{q_z} ) ]^2 } }
{ \int\limits^{\pi}_{-\pi} \!\!\! \int\limits^{\pi}_{-\pi} \!\!\! \int\limits^{\pi}_{-\pi} 
\frac{ {\rm d}q_x \, {\rm d}q_y \, {\rm d}q_z }{ (2\pi)^3 }
\frac{\cos{q_z}}{ [ \vert E_{0} \vert - 4 t \, ( \cos{q_x} + \cos{q_y} ) - 4 t^{\prime} \cos{q_z} ) ]^2 } } \: ,  
\label{MLH:eq:beighteen}
\end{equation}
where $m_{0} = \hbar^2/(2 t a^2)$ is the bare $(xy)$ mass. Once $E_{0}$ is known, each mass requires two triple integrations. Notice that both integrals can be expressed as derivatives in $\vert E_{0} \vert$ of simpler triple integrals for which some analytical integration can be performed. This way, several alternative representations of the masses can be derived, which can be used together to debug the method.

\begin{figure}[t]
\includegraphics[width=0.98\textwidth]{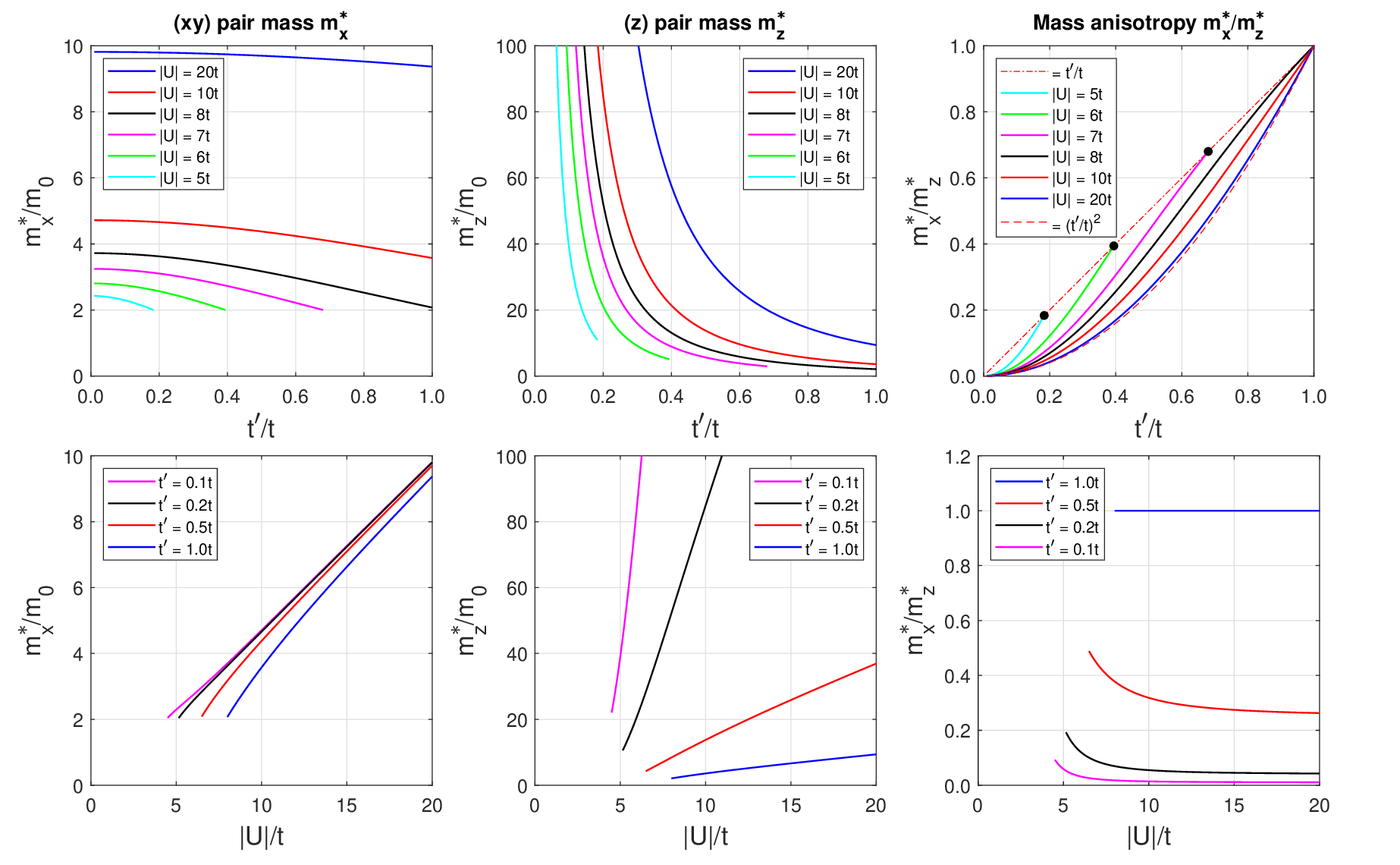}
\caption{Effective pair masses in the $n = 1$ attractive Hubbard model. The top row shows dependencies on the anisotropy parameter $t^{\prime}/t$ and the bottom row dependencies on attractive strength $\vert U \vert$.} 
\label{MLH:fig:Btwo}
\end{figure}

Pair masses computed from Eqs.~(\ref{MLH:eq:bseventeen}) and (\ref{MLH:eq:beighteen}) are shown in Fig.~\ref{MLH:fig:Btwo}. The graphs reveal several trends. (i) The masses always approach two free-particle masses near paring threshold and increase away from it. (ii) At large $\vert U \vert$, the masses increase as $\propto \vert U \vert / t$. This follows from the fact that pair motion is a second-order process in hopping and that {\em the intermediate state has a higher energy that the bound state}. The latter property is unique to the attractive Hubbard model. In more realistic models with longer-range attraction or longer-range hopping \cite{Alexandrov2002,Hague2007a,Hague2007b,Hague2008,Adebanjo2022,Adebanjo2024b}, a pair moves in the first-order in $t$ which leads to the mass saturating to a finite value even in the limit of infinitely strong attraction. (iii) Pair dispersion is more anisotropic than free-particle dispersion, see the top-right panel of Fig.~\ref{MLH:fig:Btwo}.      

Next, pair size is discussed. One begins with the pair wave function, Eq.~(\ref{MLH:eq:beight}). Since total momentum ${\bf P}$ is fixed, $\psi_{11}$ is a function of only one argument ${\bf q}$:
\begin{equation}
\psi_{11, {\bf P}}( {\bf q} ) = 
\frac{ 1 }{ E - \varepsilon_{{\bf q}} - \varepsilon_{ {\bf P} - {\bf q} } } \: . 
\label{MLH:eq:bnineteen}
\end{equation}
The normalization constant has been omitted. A real-space wave function follows from Eq.~(\ref{twopart:eq:five}) 
\begin{equation}
\Psi({\bf m}_1, {\bf m}_2) = \frac{1}{N_1} \sum_{\bf q}
\frac{ e^{ i {\bf q} {\bf m}_1 + i ( {\bf P} - {\bf q} ) {\bf m}_2 } }
     { E - \varepsilon_{\bf q} - \varepsilon_{{\bf P} - {\bf q}} } 
= e^{ i {\bf P} \frac{ ( {\bf m}_1 + {\bf m}_2 )}{2} } \frac{1}{N_1} \sum_{\bf q} 
\frac{ e^{ i {\bf q} ( {\bf m}_1 - {\bf m}_2 ) } }
     { E - \varepsilon_{ \frac{{\bf P}}{2} + {\bf q} } - \varepsilon_{ \frac{{\bf P}}{2} - {\bf q}} }  \: .      
\label{twopart:eq:twentyonethree}
\end{equation}
The first factor describes center-of-mass motion, while the integral over ${\bf q}$ describes internal structure of the pair. Effective radii $r^{\ast}_{j}$ are defined as follows
\begin{equation}
\left( r^{\ast}_{j} \right)^2 = \langle m^2_{j} \rangle = 
\frac{ \sum_{\bf m} m^2_{j} \: \Psi^{\ast}({\bf m},0) \Psi({\bf m},0) }
     { \sum_{\bf m}            \Psi^{\ast}({\bf m},0) \Psi({\bf m},0) } 
\equiv \frac{J_{j}}{J_{0}}     \: .   
\label{twopart:eq:twentytwothree}
\end{equation}
Substituting here Eq.~(\ref{twopart:eq:twentyonethree}) one obtains 
\begin{equation}
J_{0} = \frac{1}{N_1} \sum_{\bf q} 
\frac{1}{ [ E - \varepsilon_{ \frac{{\bf P}}{2} + {\bf q} } - \varepsilon_{ \frac{{\bf P}}{2} - {\bf q} } ]^2 } 
= \frac{1}{N_1} \sum_{\bf q} \frac{1}
{ [ \vert E \vert + \varepsilon_{ \frac{{\bf P}}{2} + {\bf q} } + \varepsilon_{ \frac{{\bf P}}{2} - {\bf q} } ]^2 } \: .   
\label{MLH:eq:btwentytwo}
\end{equation}
\begin{equation}
J_{j} = \sum_{\bf m} \left( \frac{1}{N_1} \sum_{{\bf q}_1} 
\frac{ \frac{\partial}{\partial q_{1j}} e^{ i {\bf q}_1 {\bf m} } }
{ E - \varepsilon_{ \frac{{\bf P}}{2} + {\bf q}_1 } - \varepsilon_{ \frac{{\bf P}}{2} - {\bf q}_1 } } \right) 
\left( \frac{1}{N_1} \sum_{{\bf q}_2} 
\frac{ \frac{\partial}{\partial q_{2j}} e^{ - i {\bf q}_2 {\bf m} } }
{ E - \varepsilon_{ \frac{{\bf P}}{2} + {\bf q}_2 } - \varepsilon_{ \frac{{\bf P}}{2} - {\bf q}_2 } } \right) .   
\label{MLH:eq:btwentythree}
\end{equation}
Both sums over ${\bf q}$ can be integrated by parts utilizing the Green's theorem for periodic functions, see for example [\onlinecite{Ashcroft1976}]. The result is
\begin{equation}
J_{j} = \frac{1}{N_1} \sum_{\bf q} \left[ \frac{\partial}{\partial q_{j}} 
\frac{ 1 }{ E - \varepsilon_{ \frac{{\bf P}}{2} + {\bf q} } - \varepsilon_{ \frac{{\bf P}}{2} - {\bf q} } }  
\right]^2 = 
\frac{1}{N_1} \sum_{\bf q} 
\frac{  \left[  \frac{\partial}{\partial q_{j}} 
( \varepsilon_{ \frac{{\bf P}}{2} + {\bf q} } + \varepsilon_{ \frac{{\bf P}}{2} - {\bf q} } ) \right]^2  }
{ \left[ \vert E \vert + \varepsilon_{ \frac{{\bf P}}{2} + {\bf q} } 
                       + \varepsilon_{ \frac{{\bf P}}{2} - {\bf q} } \right]^4 } \: .
\label{MLH:eq:btwentyfour}
\end{equation}
Equations (\ref{twopart:eq:twentytwothree}), (\ref{MLH:eq:btwentytwo}), and (\ref{MLH:eq:btwentyfour}) are sufficient to compute the pair size for {\em any} ${\bf P}$ once energy $E$ is known. Explicit expressions for the ground state, ${\bf P} = 0$, are derived below. Utilizing the one-particle dispersion, Eq.~(\ref{MLH:eq:bfive}), one obtains 
\begin{equation}
( r^{\ast}_{x})^2 = ( r^{\ast}_{y} )^2 = 16 t^2 \: 
\frac{ \int\limits^{\pi}_{-\pi} \!\!\! \int\limits^{\pi}_{-\pi} \!\!\! \int\limits^{\pi}_{-\pi} 
\frac{ \sin^2{q_{x}} \: {\rm d}q_x \, {\rm d}q_y \, {\rm d}q_z }
{ [ \vert E_0 \vert - 4 t \, ( \cos{q_x} + \cos{q_y} ) - 4 t^{\prime} \cos{q_z} ) ]^4 } }
{ \int\limits^{\pi}_{-\pi} \!\!\! \int\limits^{\pi}_{-\pi} \!\!\! \int\limits^{\pi}_{-\pi} 
\frac{ {\rm d}q_x \, {\rm d}q_y \, {\rm d}q_z}
{ [ \vert E_0 \vert - 4 t \, ( \cos{q_x} + \cos{q_y} ) - 4 t^{\prime} \cos{q_z} ) ]^2 } }  \: ,  
\label{twopart:eq:sixtyeightthree}
\end{equation}
\begin{equation}
( r^{\ast}_{z} )^2 = 16 t^{\prime 2} \: 
\frac{ \int\limits^{\pi}_{-\pi} \!\!\! \int\limits^{\pi}_{-\pi} \!\!\! \int\limits^{\pi}_{-\pi} 
\frac{ \sin^2{q_{z}} \: {\rm d}q_x \, {\rm d}q_y \, {\rm d}q_z }
{ [ \vert E_0 \vert - 4 t \, ( \cos{q_x} + \cos{q_y} ) - 4 t^{\prime} \cos{q_z} ) ]^4 } }
{ \int\limits^{\pi}_{-\pi} \!\!\! \int\limits^{\pi}_{-\pi} \!\!\! \int\limits^{\pi}_{-\pi} 
\frac{ {\rm d}q_x \, {\rm d}q_y \, {\rm d}q_z}
{ [ \vert E_0 \vert - 4 t \, ( \cos{q_x} + \cos{q_y} ) - 4 t^{\prime} \cos{q_z} ) ]^2 } }  \: .  
\label{MLH:eq:btwentysix}
\end{equation}
Pair sizes of the $n = 1$ model are shown in Fig.~\ref{MLH:fig:Bthree}. All components diverge near the threshold as expected on physical grounds. Another interesting feature apparent from the top-right panel of Fig.~\ref{MLH:fig:Bthree} that is $r^{\ast}_{z}/r^{\ast}_{x} \to \sqrt{t^{\prime}/t}$ near the threshold. In general, the pair aspect ratio is confined between $\sqrt{t^{\prime}/t}$ and $(t^{\prime}/t)$.

\begin{figure}[t]
\includegraphics[width=0.98\textwidth]{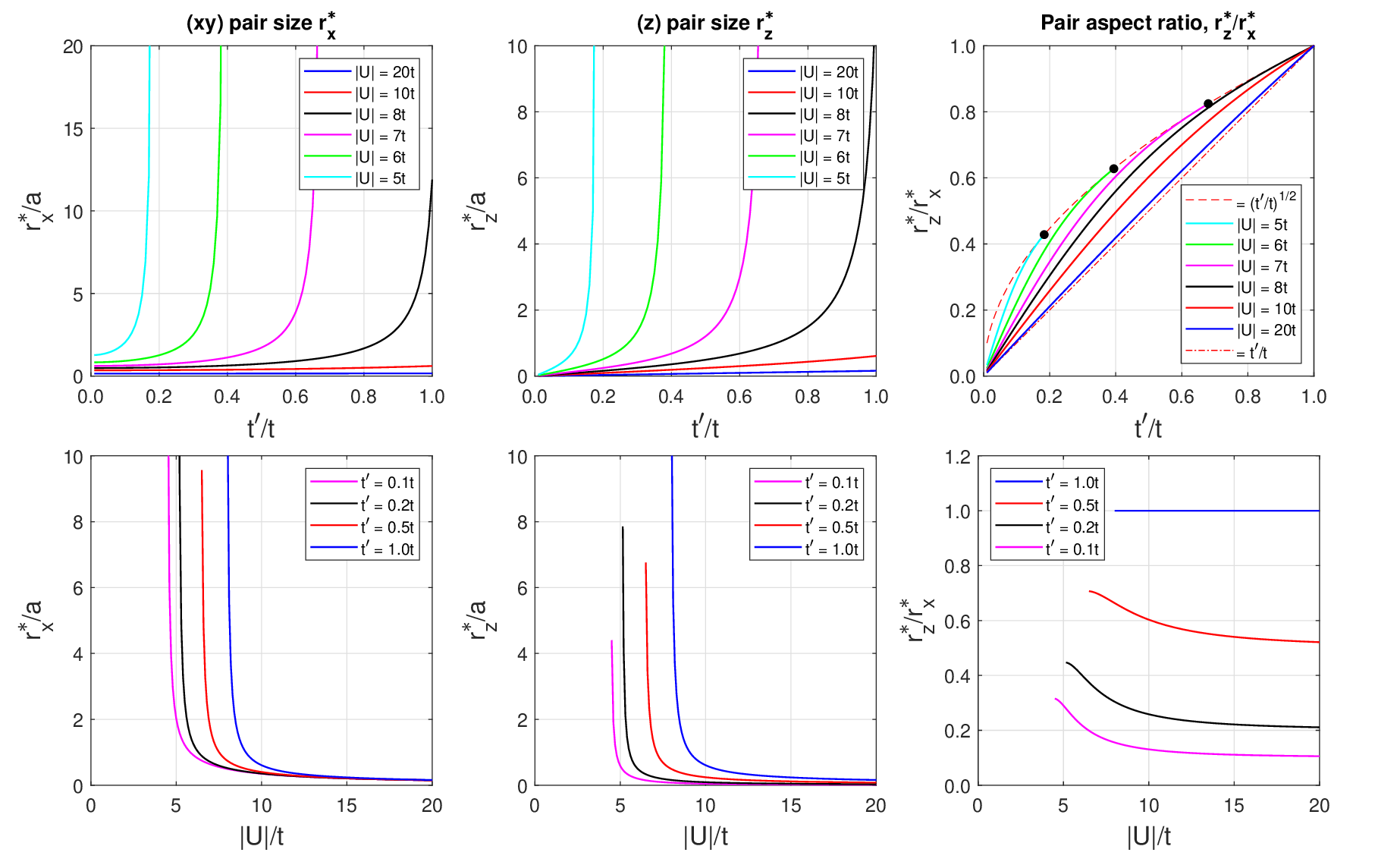}
\caption{Effective pair sizes in the $n = 1$ model. The top row shows dependencies on the anisotropy parameter $t^{\prime}/t$ and the bottom row dependencies on attractive strength $\vert U \vert$. Notice how $r^{\ast}_{j}$ diverge near binding thresholds.} 
\label{MLH:fig:Bthree}
\end{figure}

Finally, we discuss the close-packed critical temperature $T^{\ast}_{\rm cp}$. To prevent unphysical collapse of pair volume $\Omega_p$ to zero at strong coupling we use a regularized expression [\onlinecite{Kornilovitch2015}] 
\begin{equation}
\Omega_{p} = \sqrt{ ( 1 + r^{\ast 2}_{x} )( 1 + r^{\ast 2}_{y} )( 1 + r^{\ast 2}_{z} ) } \: . 
\label{MLH:eq:btwentyseven}
\end{equation}
The dimensionless close-packed critical temperature is given by 
\begin{equation}
{\cal T}^{\ast}_{\rm cp} = \frac{1}{t} \: k_{B} T^{\ast}_{\rm cp}( n_{\rm cp} ) = 
\frac{ 6.62 }{ ( m^{\ast}_{x} m^{\ast}_{y} m^{\ast}_{z} )^{1/3} \, \Omega^{2/3}_{p} } \: . 
\label{MLH:eq:btwentyeight}
\end{equation}
Here all masses are measured in units of $m_0 = \hbar^2/(2 t a^2)$ and pair volume is measured in $a^3$. The critical temperature is plotted in Fig.~\ref{MLH:fig:Bfour}. The most conspicuous feature of both dependencies, vs. $t^{\prime}$ and vs. $\vert U \vert$, are well-defined peaks whose origin is apparent from Eq.~(\ref{MLH:eq:btwentyeight}). Consider the dependence of $T^{\ast}_{\rm cp}$ on lattice anisotropy, Fig.~\ref{MLH:fig:Bfour}(a). At small $t^{\prime} \to 0$, the kinetic energy is small, particle movement is almost two-dimensional, and the pairs are well formed. At the same time, hopping between layers is impeded, $m^{\ast}_{z} \to \infty$, and $T^{\ast}_{\rm cp} \to 0$ as a result. Physically, there is no 3D phase coherence and no superconductivity. On the opposite end of larger $t^{\prime}$, the kinetic energy is large and the pairs are close to binding threshold. Pair volume $\Omega_p$ balloons to infinity and $T^{\ast}_{\rm cp} \to 0$ as well. Physically, the pairs are light (good for superconductivity) but because of the large pair volume maximal pair density is small (bad for superconductivity). Thus, {\em for any given attraction $\vert U \vert$ there is an optimal anisotropy where $T^{\ast}_{\rm cp}$ is maximal.} The same conclusion holds for more complex interaction profiles [\onlinecite{Kornilovitch2015}]. Next, let us look at the $T^{\ast}_{\rm cp}$ dependence on the attractive strength, see Fig.~\ref{MLH:fig:Bfour}(b). At small $\vert U \vert$, the pairs are close to pairing threshold, hence $\Omega_p \to \infty$ and $T^{\ast}_{\rm cp} \to 0$. Conversely, at strong coupling $\vert U \vert \to \infty$, the pairs are very compact, $\Omega \to 1$, but heavy, $m^{\ast} \to \infty$, and $T^{\ast}_{\rm cp}$ tends to zero as well. Thus, for {\em for any given anisotropy $t^{\prime}$ there is an optimal attraction where $T^{\ast}_{\rm cp}$ is maximal.}

\begin{figure}[t]
\includegraphics[width=0.90\textwidth]{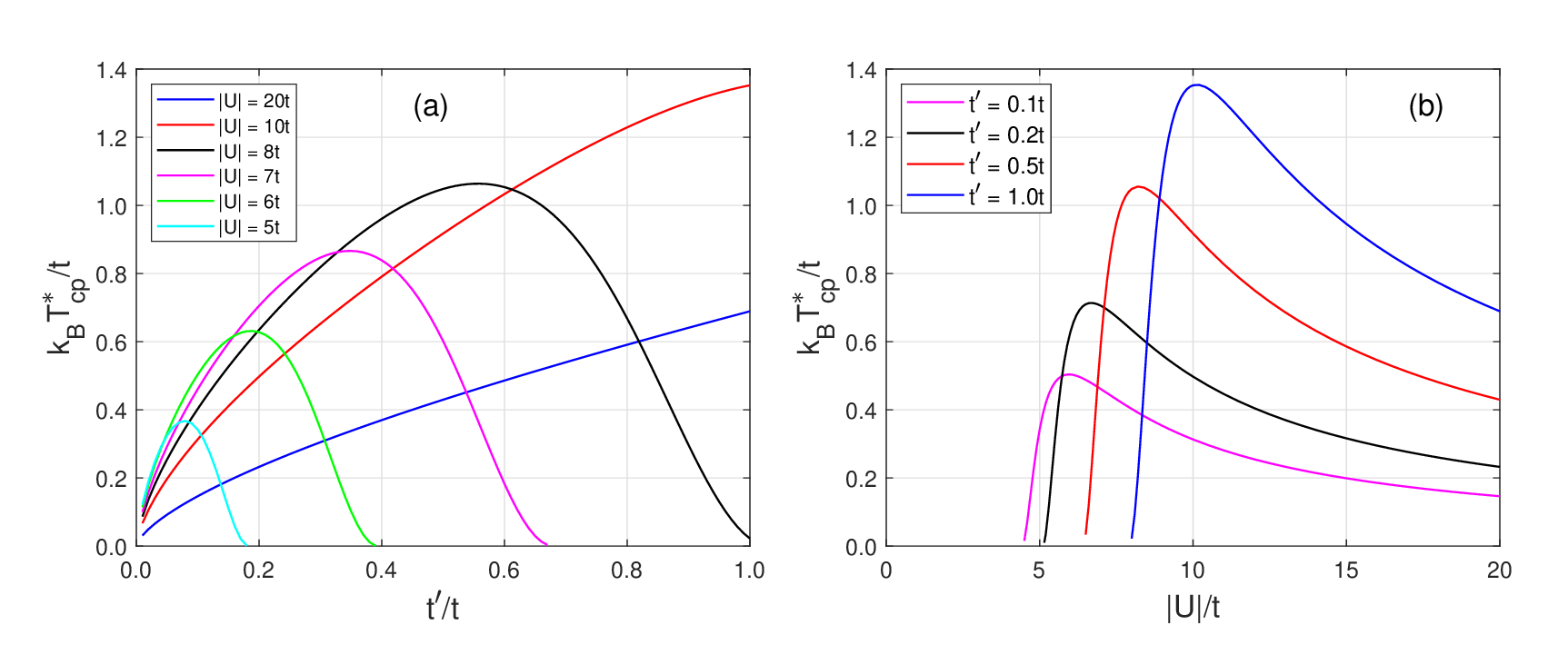}
\caption{Close-packed critical temperature, Eq.~(\ref{MLH:eq:btwentyeight}), in the $n = 1$ model. $T^{\ast}_{\rm cp}$ has well-defined maxima as a function of both $t^{\prime}$ and $\vert U \vert$. } 
\label{MLH:fig:Bfour}
\end{figure}

In real materials, such as the cuprates, the (effective) attraction is going to be small. Then it follows from Fig.~\ref{MLH:fig:Bfour}(a) that: (i) The system {\em must} be anisotropic, otherwise the pairs do not form and superconductivity does not exist at all; (ii) There is {\em always} an optimal anisotropy where $T^{\ast}_{\rm cp}$ is maximal. By fine-tuning $t^{\prime}$, this maximal $T^{\ast}_{\rm cp}$ can be realized; (iii) In general, the maximal $T^{\ast}_{\rm cp}$ increases with $\vert U \vert$. Thus, stronger attractions leads to larger $T^{\ast}_{\rm cp}$. However, this method of boosting $T^{\ast}_{\rm cp}$ is limited by formation of clusters and phase separation \cite{Kornilovitch2015,Kornilovitch2023}.

\section{\label{MLH:sec:app:c}
Two layers, $n = 2$
}

The $n = 2$ model is the first multi-band pairing model to be studied in this work. On one hand, the model is still simple enough to allow complete analytical solution. On the other hand, it possesses all features of the more complex $n \geq 3$ models. We will be using explicit $n = 2$ expressions to illustrate general properties of the more complex models which cannot be easily verified analytically. For that reason, the $n = 2$ model is presented below in detail.      

We begin with deriving the free one-particle dispersion. The one-particle wave function is a $(2 \times 1)$ array $\phi_{\alpha}({\bf m})$. The Schr\"odinger equation reads
\begin{eqnarray}
- t \left[ \phi_{1}( {\bf m} - {\bf x} ) + \phi_{1}( {\bf m} + {\bf x} ) 
         + \phi_{1}( {\bf m} - {\bf y} ) + \phi_{1}( {\bf m} + {\bf y} ) \right]
- t \, \phi_{2}( {\bf m} ) - t^{\prime} \phi_{2}( {\bf m} + 2 {\bf z} )          
& = & \varepsilon \phi_{1}( {\bf m} )                                                \: ,
\label{MLH:eq:cone} \\
- t \left[ \phi_{2}( {\bf m} - {\bf x} ) + \phi_{2}( {\bf m} + {\bf x} ) 
         + \phi_{2}( {\bf m} - {\bf y} ) + \phi_{2}( {\bf m} + {\bf y} ) \right]
- t \, \phi_{1}( {\bf m} ) - t^{\prime} \phi_{2}( {\bf m} - 2 {\bf z} )                   
& = & \varepsilon \phi_{2}( {\bf m} )                                                \: . 
\label{MLH:eq:ctwo}
\end{eqnarray}
The equations are diagonalized by Fourier transformation that we define as 
\begin{eqnarray}
\phi_{\alpha}( {\bf m} ) & = & \frac{1}{\sqrt{N_2}} \sum_{\bf k} \phi_{\alpha}( {\bf k} ) \, e^{   i {\bf k m } } \: ,
\label{MLH:eq:cthree} \\
\phi_{\alpha}( {\bf k} ) & = & \frac{1}{\sqrt{N_2}} \sum_{\bf m} \phi_{\alpha}( {\bf m} ) \, e^{ - i {\bf k m } } \: . 
\label{MLH:eq:cfour}
\end{eqnarray}
$N_2 = N_1/2$ used here is the number of unit cells of volume $\Omega_0 = 2a^3$. The result of the transformation is two one-particle bands
\begin{eqnarray}
\varepsilon_{1 , {\bf k}} & = & - 2 t ( \cos{k_x} + \cos{k_y} )
- \sqrt{ t^2 + 2 t t^{\prime} \cos{(2k_z)} + t^{\prime 2} }           \: ,
\label{MLH:eq:cfive} \\
\varepsilon_{2 , {\bf k}} & = & - 2 t ( \cos{k_x} + \cos{k_y} )
+ \sqrt{ t^2 + 2 t t^{\prime} \cos{(2k_z)} + t^{\prime 2} }           \: . 
\label{MLH:eq:csix}
\end{eqnarray}
With our neighbor-to-neighbor distance convention, the Brillouin zone dimensions are
\begin{equation}
- \pi           \leq k_x \leq + \pi            \: , 
\hspace{1.0cm}
- \pi           \leq k_y \leq + \pi            \: , 
\hspace{1.0cm}
- \frac{\pi}{2} \leq k_z \leq + \frac{\pi}{2}  \: .  
\label{MLH:eq:cseven}
\end{equation}

The {\em pair} wave function is a $( 2^2 \times 1 )$ array $\Psi_{\alpha\beta}( {\bf m}_1 , {\bf m}_2 )$, $\alpha , \beta = 1 , 2$. It is convenient to envision the ``first'' particle to be ``spin up'' and the ``second'' particle to be ``spin down''. Then, for example, $\Psi_{21}( {\bf m}_1 , {\bf m}_2 )$ is the probability amplitude to find a spin-up on site 2 of unit cell ${\bf m}_1$ and spin-down on site 1 of unit cell ${\bf m}_2$. The Schr\"odinger equation reads
\begin{eqnarray}
&& - t \left[ 
\Psi_{11}( {\bf m}_1 - {\bf x} , {\bf m}_2 ) + \Psi_{11}( {\bf m}_1 + {\bf x} , {\bf m}_2 ) + 
\Psi_{11}( {\bf m}_1 - {\bf y} , {\bf m}_2 ) + \Psi_{11}( {\bf m}_1 + {\bf y} , {\bf m}_2 ) \right]
\nonumber \\              
&& - t \left[ 
\Psi_{11}( {\bf m}_1 , {\bf m}_2 - {\bf x} ) + \Psi_{11}( {\bf m}_1 , {\bf m}_2 + {\bf x} ) + 
\Psi_{11}( {\bf m}_1 , {\bf m}_2 - {\bf y} ) + \Psi_{11}( {\bf m}_1 , {\bf m}_2 + {\bf y} ) \right]              
\nonumber \\
&& - t^{\prime} \left[ 
\Psi_{21}( {\bf m}_1 + 2 {\bf z} , {\bf m}_2 ) + \Psi_{12}( {\bf m}_1 , {\bf m}_2 + 2 {\bf z} ) \right] 
- t \left[ \Psi_{21}( {\bf m}_1 , {\bf m}_2 ) + \Psi_{12}( {\bf m}_1 , {\bf m}_2 ) \right]              
\nonumber \\
&& - \vert U \vert \: \delta_{ {\bf m}_1 , {\bf m}_2 } \Psi_{11}( {\bf m}_1 , {\bf m}_2 ) 
= E \, \Psi_{11}( {\bf m}_1 , {\bf m}_2 ) \: ,   
\label{MLH:eq:ceight}
\end{eqnarray}
\begin{eqnarray}
&& - t \left[ 
\Psi_{12}( {\bf m}_1 - {\bf x} , {\bf m}_2 ) + \Psi_{12}( {\bf m}_1 + {\bf x} , {\bf m}_2 ) + 
\Psi_{12}( {\bf m}_1 - {\bf y} , {\bf m}_2 ) + \Psi_{12}( {\bf m}_1 + {\bf y} , {\bf m}_2 ) \right]
\nonumber \\              
&& - t \left[ 
\Psi_{12}( {\bf m}_1 , {\bf m}_2 - {\bf x} ) + \Psi_{12}( {\bf m}_1 , {\bf m}_2 + {\bf x} ) + 
\Psi_{12}( {\bf m}_1 , {\bf m}_2 - {\bf y} ) + \Psi_{12}( {\bf m}_1 , {\bf m}_2 + {\bf y} ) \right]              
\nonumber \\
&& - t^{\prime} \left[ 
\Psi_{22}( {\bf m}_1 + 2 {\bf z} , {\bf m}_2 ) + \Psi_{11}( {\bf m}_1 , {\bf m}_2 - 2 {\bf z} ) \right] 
- t \left[ \Psi_{22}( {\bf m}_1 , {\bf m}_2 ) + \Psi_{11}( {\bf m}_1 , {\bf m}_2 ) \right]              
\nonumber \\
&& = E \, \Psi_{12}( {\bf m}_1 , {\bf m}_2 ) \: ,   
\label{MLH:eq:cnine}
\end{eqnarray}
\begin{eqnarray}
&& - t \left[ 
\Psi_{21}( {\bf m}_1 - {\bf x} , {\bf m}_2 ) + \Psi_{21}( {\bf m}_1 + {\bf x} , {\bf m}_2 ) + 
\Psi_{21}( {\bf m}_1 - {\bf y} , {\bf m}_2 ) + \Psi_{21}( {\bf m}_1 + {\bf y} , {\bf m}_2 ) \right]
\nonumber \\              
&& - t \left[ 
\Psi_{21}( {\bf m}_1 , {\bf m}_2 - {\bf x} ) + \Psi_{21}( {\bf m}_1 , {\bf m}_2 + {\bf x} ) + 
\Psi_{21}( {\bf m}_1 , {\bf m}_2 - {\bf y} ) + \Psi_{21}( {\bf m}_1 , {\bf m}_2 + {\bf y} ) \right]              
\nonumber \\
&& - t^{\prime} \left[ 
\Psi_{11}( {\bf m}_1 - 2 {\bf z} , {\bf m}_2 ) + \Psi_{22}( {\bf m}_1 , {\bf m}_2 + 2 {\bf z} ) \right] 
- t \left[ \Psi_{11}( {\bf m}_1 , {\bf m}_2 ) + \Psi_{22}( {\bf m}_1 , {\bf m}_2 ) \right]              
\nonumber \\
&& = E \, \Psi_{21}( {\bf m}_1 , {\bf m}_2 ) \: ,   
\label{MLH:eq:cten}
\end{eqnarray}
\begin{eqnarray}
&& - t \left[ 
\Psi_{22}( {\bf m}_1 - {\bf x} , {\bf m}_2 ) + \Psi_{22}( {\bf m}_1 + {\bf x} , {\bf m}_2 ) + 
\Psi_{22}( {\bf m}_1 - {\bf y} , {\bf m}_2 ) + \Psi_{22}( {\bf m}_1 + {\bf y} , {\bf m}_2 ) \right]
\nonumber \\              
&& - t \left[ 
\Psi_{22}( {\bf m}_1 , {\bf m}_2 - {\bf x} ) + \Psi_{22}( {\bf m}_1 , {\bf m}_2 + {\bf x} ) + 
\Psi_{22}( {\bf m}_1 , {\bf m}_2 - {\bf y} ) + \Psi_{22}( {\bf m}_1 , {\bf m}_2 + {\bf y} ) \right]              
\nonumber \\
&& - t^{\prime} \left[ 
\Psi_{12}( {\bf m}_1 - 2 {\bf z} , {\bf m}_2 ) + \Psi_{21}( {\bf m}_1 , {\bf m}_2 - 2 {\bf z} ) \right] 
- t \left[ \Psi_{12}( {\bf m}_1 , {\bf m}_2 ) + \Psi_{21}( {\bf m}_1 , {\bf m}_2 ) \right]              
\nonumber \\
&& - \vert U \vert \: \delta_{ {\bf m}_1 , {\bf m}_2 } \Psi_{22}( {\bf m}_1 , {\bf m}_2 ) 
= E \, \Psi_{22}( {\bf m}_1 , {\bf m}_2 ) \: .   
\label{MLH:eq:celeven}
\end{eqnarray}
Notice that the equations for $\psi_{12}$ and $\psi_{21}$ do not include interaction terms because in these cases the two particles do not occupy the same site. Inclusion of any attraction in $\psi_{12}$ or $\psi_{21}$ would automatically trigger addition of longer-range interactions to $\psi_{11}$ and $\psi_{22}$ which would significantly complicate the structure of the solution, which goes beyond the scope of this work. Next, we apply Fourier transformation
\begin{eqnarray}
\Psi_{\alpha\beta}({\bf m}_1 , {\bf m}_2) & = & \frac{1}{N_2} \sum_{{\bf k}_1 {\bf k}_2} 
\psi_{\alpha\beta}({\bf k}_1 , {\bf k}_2) \, e^{ i {\bf k}_1 {\bf m}_1 + i {\bf k}_2 {\bf m}_2 }   \, ,  
\label{twopart:eq:ctwelve} \\
\psi_{\alpha\beta}({\bf k}_1 , {\bf k}_2) & = & \frac{1}{N_2} \! \sum_{{\bf m}_1 {\bf m}_2} \!\! 
\Psi_{\alpha\beta}({\bf m}_1 , {\bf m}_2) \, e^{ - i {\bf k}_1 {\bf m}_1 - i {\bf k}_2 {\bf m}_2 } \, . 
\label{twopart:eq:cthirteen}
\end{eqnarray}
A transformed Schr\"odinger equation is 
\begin{equation}
\left( \begin{array}{cccc}
E - \xi_{ {\bf k}_1 {\bf k}_2 }        & g_2 & g_1  & 0    \\
g^{\ast}_2 & E - \xi_{ {\bf k}_1 {\bf k}_2 } & 0    & g_1  \\ 
g^{\ast}_1 & 0 & E - \xi_{ {\bf k}_1 {\bf k}_2 }    & g_2  \\
0 & g^{\ast}_1 & g^{\ast}_2 & E - \xi_{ {\bf k}_1 {\bf k}_2 } 
\end{array} \right) 
\left( \begin{array}{c}
\psi_{11}( {\bf k}_1 , {\bf k}_2 ) \\ \psi_{12}( {\bf k}_1 , {\bf k}_2 ) \\ 
\psi_{21}( {\bf k}_1 , {\bf k}_2 ) \\ \psi_{22}( {\bf k}_1 , {\bf k}_2 )  
\end{array} \right) =
\left( \begin{array}{c}
- \vert U \vert \Phi_{11}({\bf P}) \\ 0 \\ 0 \\ - \vert U \vert \Phi_{22}({\bf P}) 
\end{array} \right) ,
\label{MLH:eq:cseventeenone}
\end{equation}
where 
\begin{equation}
\xi_{ {\bf k}_1 {\bf k}_2 } \equiv - 2 t ( \cos{k_{1x}} + \cos{k_{1y}} ) - 2 t ( \cos{k_{2x}} + \cos{k_{2y}} ) \: ,  
\label{MLH:eq:ceighteen}
\end{equation}
\begin{equation}
g_{\alpha} \equiv t + t^{\prime} e^{ 2 i k_{\alpha z} } \: ,  
\hspace{1.0cm}
\vert g_{\alpha} \vert^2 = t^2 + 2 t t^{\prime} \cos{(2k_{\alpha z})} + t^{\prime 2} \: ,
\label{MLH:eq:cnineteen}
\end{equation}
\begin{equation}
\Phi_{\alpha\alpha}( {\bf P} )
\equiv \frac{1}{N_2} \sum_{\bf q} \psi_{\alpha\alpha}( {\bf q}, {\bf P} - {\bf q} ) 
=      \frac{1}{N_2} \sum_{\bf q} 
\psi_{\alpha\alpha} \left( \frac{\bf P}{2} + {\bf q}, \frac{\bf P}{2} - {\bf q} \right) .  
\label{MLH:eq:ctwenty}
\end{equation}
Equation~(\ref{MLH:eq:cseventeenone}) is an inhomogeneous system of linear equations for $\psi_{\alpha\beta}$. The latter can be expressed via $\Phi_{\alpha\alpha}$. For the energy and mass, we will only need $\psi_{11}$ and $\psi_{22}$; this is the consequence of the chosen form of interaction. Cramer's rule yields
\begin{eqnarray}
\psi_{11}( {\bf k}_1 , {\bf k}_2 ) & = & 
- \vert U \vert \Phi_{11}({\bf P}) \, \frac{\Delta_{11}({\bf k}_1,{\bf k}_2)}{\Delta_{0}({\bf k}_1,{\bf k}_2)} 
- \vert U \vert \Phi_{22}({\bf P}) \, \frac{\Delta_{12}({\bf k}_1,{\bf k}_2)}{\Delta_{0}({\bf k}_1,{\bf k}_2)}  \: ,  
\label{twopart:eq:ctwentyone} \\
\psi_{22}( {\bf k}_1 , {\bf k}_2 ) & = &   
- \vert U \vert \Phi_{11}({\bf P}) \, \frac{\Delta_{21}({\bf k}_1,{\bf k}_2)}{\Delta_{0}({\bf k}_1,{\bf k}_2)} 
- \vert U \vert \Phi_{22}({\bf P}) \, \frac{\Delta_{22}({\bf k}_1,{\bf k}_2)}{\Delta_{0}({\bf k}_1,{\bf k}_2)}  \: ,  
\label{twopart:eq:ctwentytwo}
\end{eqnarray}
where
\begin{eqnarray}
\Delta_{0}( {\bf k}_1 , {\bf k}_2 ) & = & 
\left( E - \xi_{ {\bf k}_1 {\bf k}_2 } \right)^4 
- 2 \left( \vert g_1 \vert^2 + \vert g_2 \vert^2 \right) \left( E - \xi_{ {\bf k}_1 {\bf k}_2 } \right)^2 
+ 2 \left( \vert g_1 \vert^2 - \vert g_2 \vert^2 \right)^2 ,  
\label{twopart:eq:ctwentythree} \\
\Delta_{11}( {\bf k}_1 , {\bf k}_2 ) = \Delta_{22}( {\bf k}_1 , {\bf k}_2 ) & = &   
\left( E - \xi_{ {\bf k}_1 {\bf k}_2 } \right)^3 - 
\left( \vert g_1 \vert^2 + \vert g_2 \vert^2 \right) \left( E - \xi_{ {\bf k}_1 {\bf k}_2 } \right) ,  
\label{twopart:eq:ctwentyfour}  \\
\Delta_{12}( {\bf k}_1 , {\bf k}_2 ) = \Delta^{\ast}_{21}( {\bf k}_1 , {\bf k}_2 ) & = &   
- 2 g_1 g_2 \left( E - \xi_{ {\bf k}_1 {\bf k}_2 } \right) . 
\label{twopart:eq:ctwentyfive}
\end{eqnarray}
Wave function components are defined by ratios of $n^2 - 1 = 3$ degree polynomials in $E$ to $n^2 = 4$ degree polynomials in $E$. This is a general rule that will be valid for all $n$. The quantity $\Delta_{0}$ corresponds to free motion of two particles when the right-hand side of Eq.~(\ref{MLH:eq:cseventeenone}) is zero. Indeed, either by solving the biquadratic equation, Eq.~(\ref{twopart:eq:ctwentythree}), or by direct substitution one can verify that
\begin{equation}
\Delta_{0}( {\bf k}_1 , {\bf k}_2 ) = 
\left( E - \varepsilon_{1 {\bf k}_1} - \varepsilon_{1 {\bf k}_2} \right) 
\left( E - \varepsilon_{1 {\bf k}_1} - \varepsilon_{2 {\bf k}_2} \right) 
\left( E - \varepsilon_{2 {\bf k}_1} - \varepsilon_{1 {\bf k}_2} \right) 
\left( E - \varepsilon_{2 {\bf k}_1} - \varepsilon_{2 {\bf k}_2} \right) ,  
\label{MLH:eq:ctwentysix}
\end{equation}
where $\varepsilon_{1{\bf k}}$ and $\varepsilon_{2{\bf k}}$ are two branches of the one-particle dispersion given by Eqs.~(\ref{MLH:eq:cfive}) and (\ref{MLH:eq:csix}). Next, we substitute wave functions $\psi_{\alpha\alpha}$ back into definitions of $\Phi_{\alpha\alpha}$, Eq.~(\ref{MLH:eq:ctwenty}), which yields $n = 2$ homogeneous linear equations
\begin{eqnarray}
\Phi_{11}({\bf P}) & = & 
- \vert U \vert \left[
\frac{1}{N_2} \sum_{\bf q} \frac{\Delta_{11}({\bf q},{\bf P}-{\bf q})}{\Delta_{0}({\bf q},{\bf P}-{\bf q})} 
\right] \Phi_{11}({\bf P})   
- \vert U \vert \left[
\frac{1}{N_2} \sum_{\bf q} \frac{\Delta_{12}({\bf q},{\bf P}-{\bf q})}{\Delta_{0}({\bf q},{\bf P}-{\bf q})} 
\right] \Phi_{22}({\bf P})   \: , 
\label{twopart:eq:ctwentyseven} \\
\Phi_{22}({\bf P}) & = & 
- \vert U \vert \left[
\frac{1}{N_2} \sum_{\bf q} \frac{\Delta_{21}({\bf q},{\bf P}-{\bf q})}{\Delta_{0}({\bf q},{\bf P}-{\bf q})} 
\right] \Phi_{11}({\bf P})   
- \vert U \vert \left[
\frac{1}{N_2} \sum_{\bf q} \frac{\Delta_{22}({\bf q},{\bf P}-{\bf q})}{\Delta_{0}({\bf q},{\bf P}-{\bf q})} 
\right] \Phi_{22}({\bf P})   \: . 
\label{twopart:eq:ctwentyeight}
\end{eqnarray}
As in the $n = 1$ case, it is the dependence of $\Phi$ only on the sum of two momenta that ensures conversion to linear equations. After shifting the internal variable by ${\bf P}/2$, the system is rendered in its final form
\begin{equation}
\left( \begin{array}{cc}
\vert U \vert M_{11}({\bf P})  &  \vert U \vert M_{12}({\bf P})   \\
\vert U \vert M_{21}({\bf P})  &  \vert U \vert M_{22}({\bf P})
\end{array} \right) 
\left( \begin{array}{c}
\Phi_{11}({\bf P})  \\  \Phi_{22}({\bf P})  
\end{array} \right) =
\left( \begin{array}{c}
\Phi_{11}({\bf P})  \\  \Phi_{22}({\bf P}) 
\end{array} \right) ,
\label{MLH:eq:ctwentynine}
\end{equation}
\begin{equation}
M_{\alpha\beta}({\bf P}) \equiv - \frac{1}{N_2} \sum_{\bf q}
\frac{\Delta_{\alpha\beta} \left( \frac{\bf P}{2} + {\bf q}, \frac{\bf P}{2} - {\bf q} \right)}
     {\Delta_{0} \left( \frac{\bf P}{2} + {\bf q}, \frac{\bf P}{2} - {\bf q} \right) } = 
- \int\limits^{\pi}_{-\pi} \!\! \int\limits^{\pi}_{-\pi} \!\! \int\limits^{\frac{\pi}{2}}_{-\frac{\pi}{2}}
2 \, \frac{ {\rm d} q_x {\rm d} q_y {\rm d} q_z}{ ( 2 \pi )^3 }  
\frac{\Delta_{\alpha\beta} \left( \frac{\bf P}{2} + {\bf q}, \frac{\bf P}{2} - {\bf q} \right)}
     {\Delta_{0} \left( \frac{\bf P}{2} + {\bf q}, \frac{\bf P}{2} - {\bf q} \right) } \: .  
\label{MLH:eq:cthirty}
\end{equation}

The consistency condition of Eq.~(\ref{MLH:eq:ctwentynine}) determines pair dispersion $E({\bf P})$ as well as pair mass in the $n = 2$ model. Let us note a general property of pair band structure. Because there are two non-equivalent sites 1 and 2, one should expect two pair states for each pair momentum ${\bf P}$. This doubling is compensated by half the number of ${\bf P}$ values because the Brillouin zone is half the size of the $n = 1$ Brillouin zone. In other words, one should expect two pair branches for ${\bf P}$ limited by the same conditions as in Eq.~(\ref{MLH:eq:cseven}). Similarly, one should expect two pair sizes, two binding conditions, and so on, for each ${\bf P}$. 

Let us derive binding thresholds. The lowest energy of two free particles with total momentum ${\bf P}$ is $E_{11}({\bf P}) = 2\varepsilon_{1,\frac{\bf P}{2}}$. Substituting $E = E_{11}({\bf P})$ in Eq.~(\ref{MLH:eq:cthirty}) produces certain values of integrals $M$ which will be denoted as $M^{0}_{\alpha\beta}({\bf P})$. Additionally, we have $M^{0}_{11}({\bf P}) = M^{0}_{22}({\bf P})$ and $M^{0}_{12}({\bf P}) = M^{0\ast}_{21}({\bf P})$ by virtue of Eqs.~(\ref{twopart:eq:ctwentyfour}) and (\ref{twopart:eq:ctwentyfive}). So, there are only two independent values: $M^{0}_{11}({\bf P})$ and $M^{0}_{12}({\bf P})$. Substituting them in Eq.~(\ref{MLH:eq:ctwentynine}) leads to a quadratic equation for $\vert U \vert$ that has two simple solutions:
\begin{equation}
\vert U_{\rm cr} \vert_{1} = \frac{1}{ M^{0}_{11}({\bf P}) + \vert M^{0}_{12}({\bf P}) \vert } \: ,  
\hspace{1.0cm}
\vert U_{\rm cr} \vert_{2} = \frac{1}{ M^{0}_{11}({\bf P}) - \vert M^{0}_{12}({\bf P}) \vert } \: .  
\label{MLH:eq:cthirtyone}
\end{equation}
Note that both $\vert U_{\rm cr} \vert$ are ${\bf P}$-dependent. At threshold, the integrands in Eq.~(\ref{MLH:eq:cthirty}) are singular. The singularity is integrable, which leads to finite $\vert U_{\rm cr} \vert$. Nonetheless, {\em numerical} evaluation of threshold values presents a technical difficulty. In the $n = 1$ model, the difficulty was avoided by utilizing a known analytical expression for the Watson integral, Eq.~(\ref{MLH:eq:bthirteen}). In $n \geq 2$ models, the integrals are not known analytically and the singularity needs to be isolated by purely numerical means. This method is described in subsequent Appendices. However, $n = 2$ is a special case where the one-particle dispersion is explicitly known. As a consequence, the singularity can be isolated analytically, which is now discussed. By virtue of the factorized form of $\Delta_0$, Eq.~(\ref{MLH:eq:ctwentysix}), the ratios $\Delta_{11}/\Delta_{0}$ and $\Delta_{12}/\Delta_{0}$ can be expanded into a sum of elementary fractions:
\begin{eqnarray}
\frac{\Delta_{11}( {\bf k}_1 , {\bf k}_2 )}{\Delta_{0}( {\bf k}_1 , {\bf k}_2 )} & = &  
\frac{1}{4} \left( 
\frac{1}{ E - \varepsilon_{1 {\bf k}_1} - \varepsilon_{1 {\bf k}_2} } +  
\frac{1}{ E - \varepsilon_{1 {\bf k}_1} - \varepsilon_{2 {\bf k}_2} } + 
\frac{1}{ E - \varepsilon_{2 {\bf k}_1} - \varepsilon_{1 {\bf k}_2} } + 
\frac{1}{ E - \varepsilon_{2 {\bf k}_1} - \varepsilon_{2 {\bf k}_2} }  \right) ,  
\label{MLH:eq:cthirtytwo} \\
\frac{\Delta_{12}( {\bf k}_1 , {\bf k}_2 )}{\Delta_{0}( {\bf k}_1 , {\bf k}_2 )} & = &  
\frac{1}{4} \frac{g_1 g_2}{ \vert g_1 \vert \vert g_2 \vert } \left(  -
\frac{1}{ E - \varepsilon_{1 {\bf k}_1} - \varepsilon_{1 {\bf k}_2} } +  
\frac{1}{ E - \varepsilon_{1 {\bf k}_1} - \varepsilon_{2 {\bf k}_2} } + 
\frac{1}{ E - \varepsilon_{2 {\bf k}_1} - \varepsilon_{1 {\bf k}_2} } - 
\frac{1}{ E - \varepsilon_{2 {\bf k}_1} - \varepsilon_{2 {\bf k}_2} }  \right) ,  
\label{MLH:eq:cthirtythree}
\end{eqnarray}
where $\varepsilon_{1{\bf k}}$ and $\varepsilon_{2{\bf k}}$ are given by Eqs.~(\ref{MLH:eq:cfive}) and (\ref{MLH:eq:csix}). The threshold singularity is contained only in the first fraction that involves $\varepsilon_{1{\bf k}_{1}}$ and $\varepsilon_{1{\bf k}_{2}}$. Note that the ratios $g_{\alpha}/|g_{\alpha}|$ are nowhere singular and can be cast into a regular expression by standard trigonometry. Also, it is important that $g_{\alpha}/|g_{\alpha}|$ is a function of $k_{z}$ only. Upon substitution of Eqs.~(\ref{MLH:eq:cthirtytwo}) and (\ref{MLH:eq:cthirtythree}) in Eg.~(\ref{MLH:eq:cthirty}), $M_{\alpha\beta}$ become linear combinations of triple integrals  
\begin{equation}
{\cal M} = \int\limits^{\pi}_{-\pi} \!\! \int\limits^{\pi}_{-\pi} \!\! \int\limits^{\frac{\pi}{2}}_{-\frac{\pi}{2}}
2 \, \frac{ {\rm d} q_x {\rm d} q_y {\rm d} q_z}{ ( 2 \pi )^3 }  
\frac{f(P_z,q_z)}{ \vert E \vert - c \cos{q_x} - b \cos{q_y} - w(P_z,q_z) } \: ,  
\label{MLH:eq:cthirtyfour}
\end{equation}
where $c = 4t \cos{(P_x/2)}$, $b = 4t \cos{(P_y/2)}$, and $f , w$ are even functions of $q_z$. (More accurately, only the sum of two ${\cal M}$'s that involve both $\varepsilon_1$ and $\varepsilon_2$ is even under $q_z \to - q_z$, which still enables replacing an integral over $[-\frac{\pi}{2} , \frac{\pi}{2}]$ with twice an integral over $[ 0 , \frac{\pi}{2}]$ for {\em all} ${\bf P}$.) Integration over $q_x, q_y$ is now explicitly isolated, and the result is known analytically (see for example [\onlinecite{Kornilovitch2024}], section B.5)
\begin{equation}
{\cal M} = \int\limits^{\frac{\pi}{2}}_{0}
2 \, \frac{ {\rm d} q_z}{ \pi }  \frac{2}{\pi} 
\frac{f(P_z,q_z) \, {\bf K}(\kappa)}{ \sqrt{ \left[ \vert E \vert - w(P_z,q_z) \right]^2 - ( c - b )^2 } } \: ,  
\hspace{1.0cm}
\kappa \equiv \sqrt{ \frac{ 4 c b }{ \left[ \vert E \vert - w(P_z,q_z) \right]^2 - ( c - b )^2 } } \: .
\label{MLH:eq:cthirtyfive}
\end{equation}
One of the ${\cal M}$'s is still singular at $q_{z} = 0$ but because the remaining integration is one-dimensional, the result is finite, and the singularity occurs at one of the integration limits, the integration is handled perfectly by standard numerical routines.

\begin{figure}[t]
\includegraphics[width=0.48\textwidth]{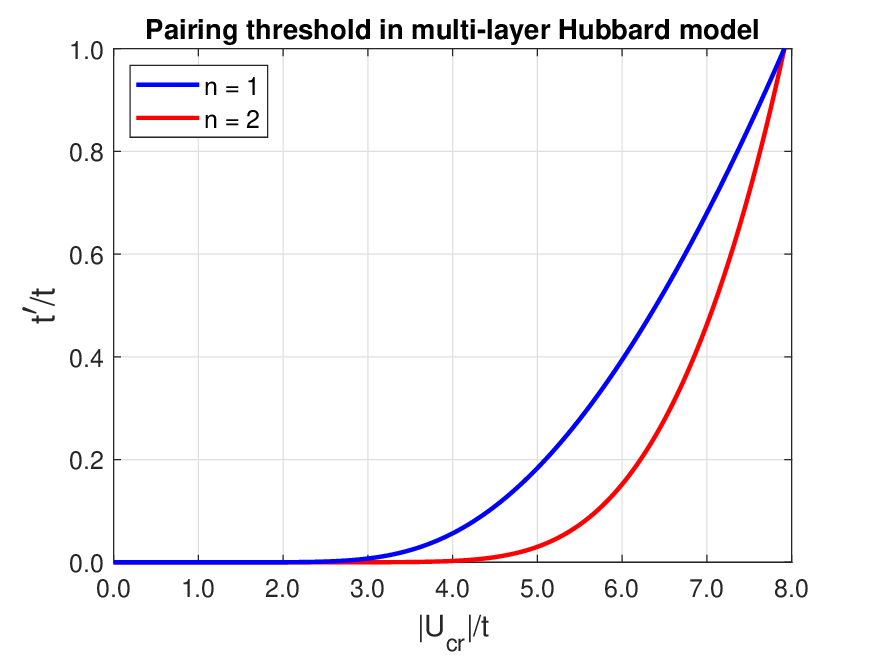}
\caption{Binding threshold in the $n = 1$ and $n = 2$ multilayer attractive Hubbard models for ${\bf P} = 0$. The $n = 1$ curve is the same as the ${\bf P} = 0$ curve in Fig.~\ref{MLH:fig:Bone}. For a fixed $t^{\prime}$, the $n = 2$ threshold is {\em always larger} than the $n = 1$ threshold because the $n = 2$ lattice is less anisotropic and has larger kinetic energy. Notice that at $t^{\prime} = t$, the $n = 2$ and $n = 1$ thresholds coincide because both models are equivalent to the same isotropic cubic model.} 
\label{MLH:fig:Cone}
\end{figure}

The $n = 2$ pairing threshold computed from Eq.~(\ref{MLH:eq:cthirtyone}) is shown in Fig.~\ref{MLH:fig:Cone} where it is also compared with the $n = 1$ threshold. For any fixed $t^{\prime}$, $\vert U_{\rm cr}(2) \vert > \vert U_{\rm cr}(1) \vert$. Physically, the $n = 2$ has larger kinetic energy and is {\em less} anisotropic than the $n = 1$ model. As a result, binding requires a stronger attraction. This is our first example where the number of conductive CuO$_2$ layers adjusts the level of lattice anisotropy.    

For $|U|$ exceeding the pairing threshold, the consistency condition of Eq.~(\ref{MLH:eq:ctwentynine}) provides pair dispersion $E({\bf P})$. Energy is determined numerically by a root-searching algorithm with ${\bf P}$ fixed. There are two values of $E$ for each ${\bf P}$. A typical pair dispersion is shown in Fig.~\ref{MLH:fig:Ctwo}.

\begin{figure}[t]
\includegraphics[width=0.48\textwidth]{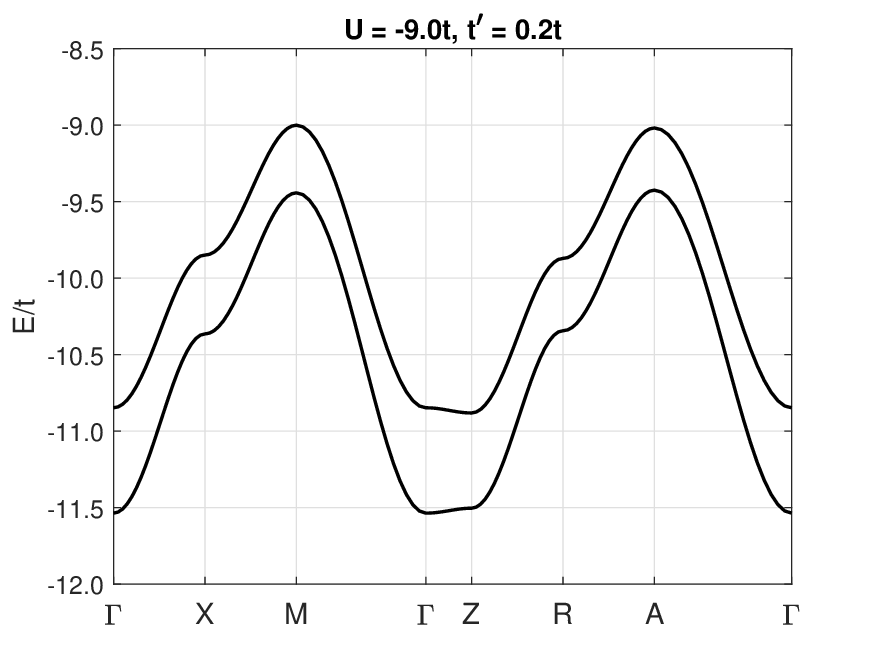}
\caption{Pair dispersion in the $n = 2$ attractive Hubbard model for $U = -9.0 \, t$ and $t^{\prime} = 0.2\,t$. Notice the difference in curvature along the $\Gamma-X$ and $\Gamma-Z$ lines, which reflects a difference between in-plane and out-of-plane pair masses.} 
\label{MLH:fig:Ctwo}
\end{figure}

The same equation (\ref{MLH:eq:ctwentynine}) allows calculation of pair mass. Unlike the $n = 1$ model, no further analytical progress is feasible, and one is forced to proceed numerically. From now on, we will be interested only in the effective masses of the lowest energy branch near the ground state, ${\bf P} = 0$. Restoring temporarily the nearest-neighbor distance $a$, keeping in mind that ${\bf P}$ is dimensionless, and expanding for small ${\bf P}$ one obtains
\begin{equation}
E( P_x , P_y , P_z \ll 1 ) = E_{0} + \frac{\hbar^2 P^2_x}{2 a^2 m^{\ast}_{x}} 
                                   + \frac{\hbar^2 P^2_y}{2 a^2 m^{\ast}_{y}} 
                                   + \frac{\hbar^2 P^2_z}{2 a^2 m^{\ast}_{z}} \: .
\label{MLH:eq:cthirtysix}
\end{equation}
Let $\triangle P \ll 1$. Approximating the second derivative as a finite difference yields
\begin{equation}
\frac{m^{\ast}_{x}}{m_0} = \frac{m^{\ast}_{y}}{m_0} = \frac{t (\triangle P)^2}{ E( \triangle P , 0 , 0 ) - E_0 } \: ,
\hspace{1.0cm}
\frac{m^{\ast}_{z}}{m_0} = \frac{t (\triangle P)^2}{ E( 0 , 0 , \triangle P ) - E_0 } \: ,
\label{MLH:eq:cthirtyseven}
\end{equation}
where $m_0 = \hbar^2/( 2 t a^2 )$. 

Pair effective masses in the $n = 2$ model are shown in Fig.~\ref{MLH:fig:Cthree}. The top row shows the dependence vs. $t^{\prime}$ whereas the bottom row vs. $\vert U \vert$. In panels (a), (b), (d), and (e), the $n = 2$ mass is compared with the corresponding $n = 1$ mass. It can be seen that the $n = 2$ mass is consistently smaller than the $n = 1$ mass, which is expected on physical grounds. For each $\vert U \vert$ and $t^{\prime}$, the $n = 2$ system is less anisotropic than the $n = 1$ system, the kinetic energy is larger, the attraction is relatively weaker, and the pairs are lighter as a result. Here we can see for the first time that the number of layers $n$ serves as a tuning parameter for pair mass. Notice that $m^{\ast}_{z}$ is affected by $n$ to a larger extent than $m^{\ast}_{x}$.

\begin{figure}[t]
\includegraphics[width=0.8\textwidth]{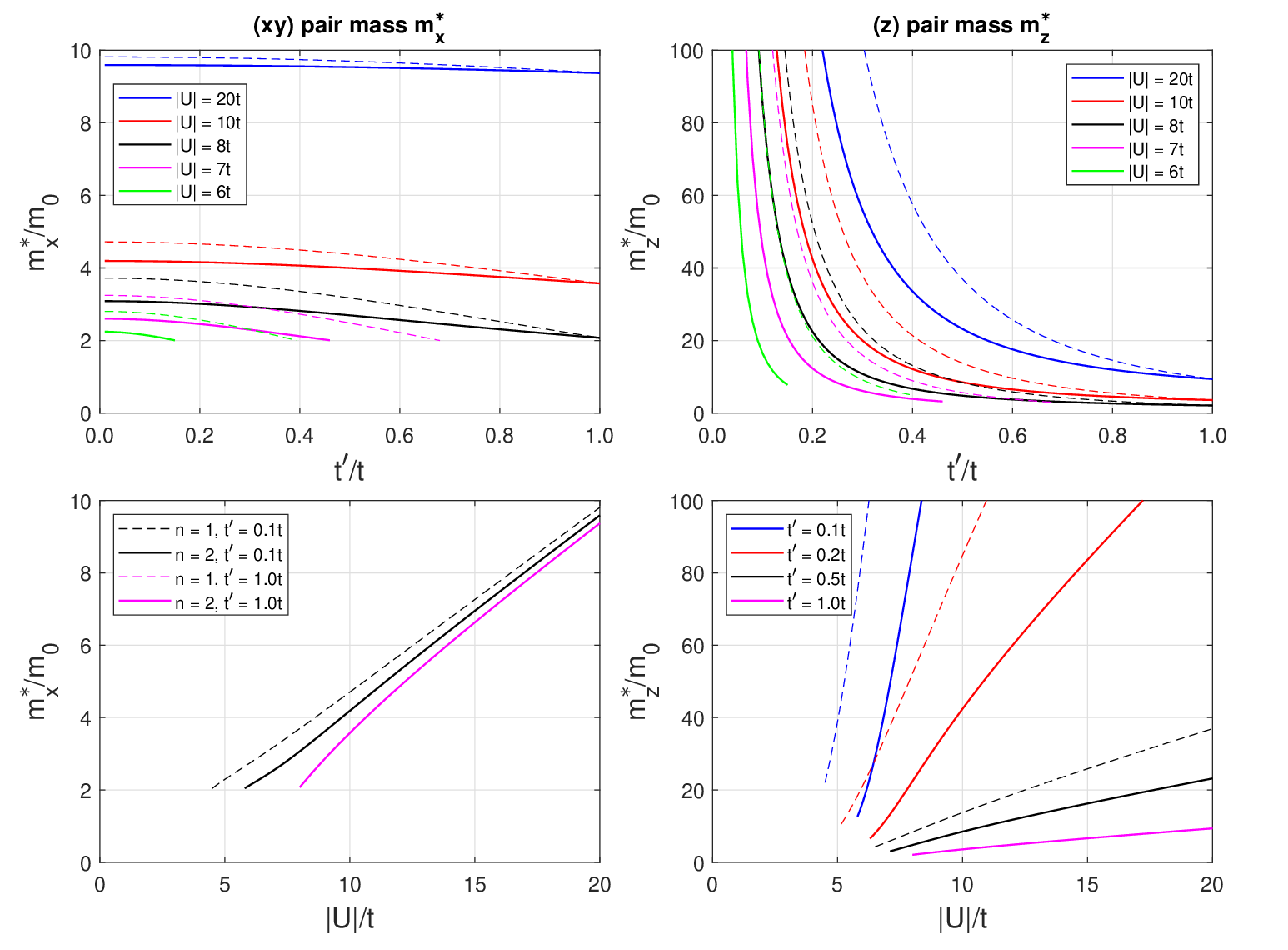}
\caption{Solid lines: effective pair masses in the $n = 2$ attractive Hubbard model. The top row shows dependencies on the anisotropy parameter $t^{\prime}/t$ and the bottom row dependencies on attractive strength $\vert U \vert$. The dashed lines of the same color show the corresponding $n = 1$ masses. Notice that $m^{\ast}_{z}$ is significantly reduced in going from $n = 1$ to $n = 2$, indicating the lesser anisotropy of the $n = 2$ model.} 
\label{MLH:fig:Cthree}
\end{figure}
\begin{figure}[t]
\includegraphics[width=0.98\textwidth]{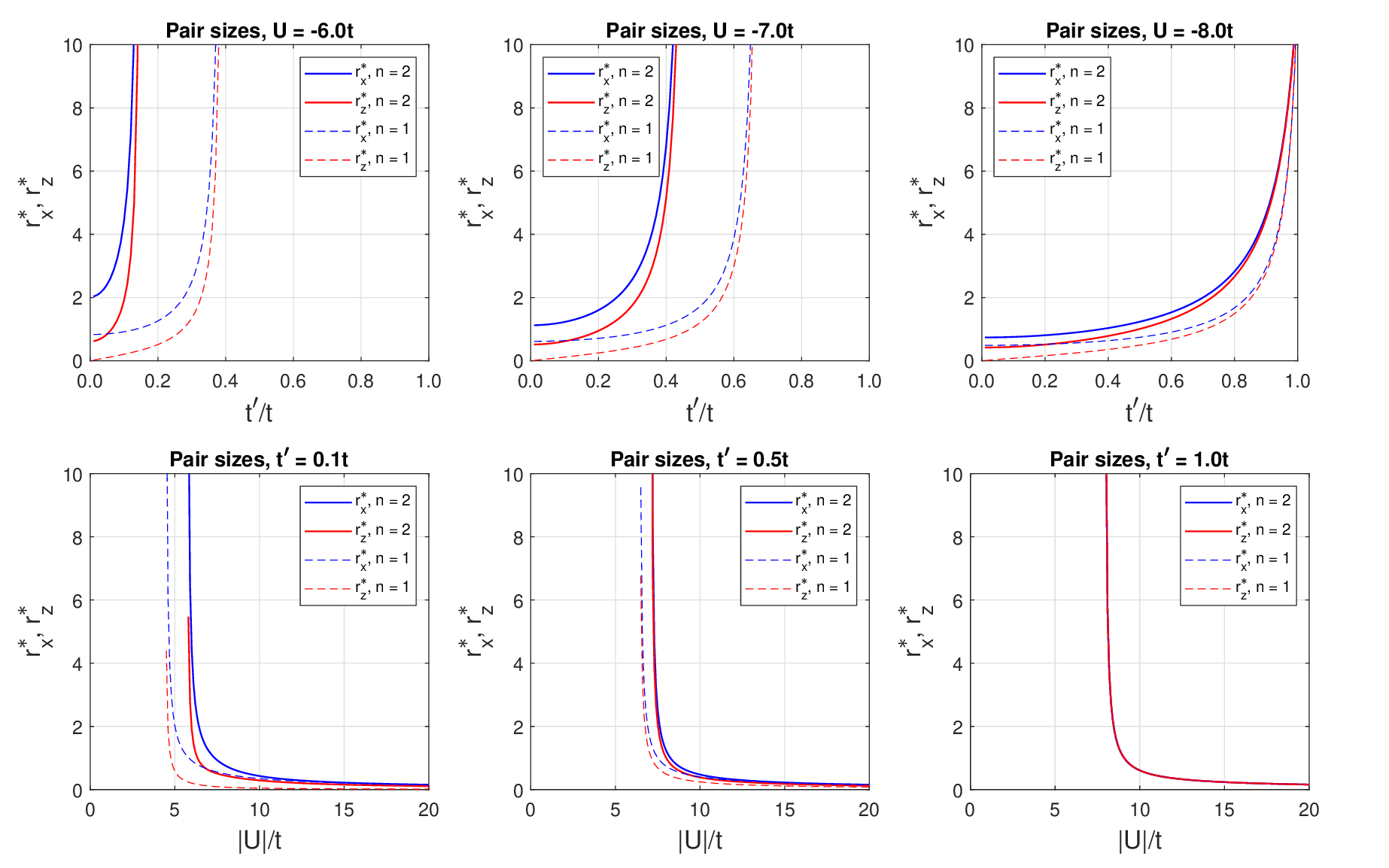}
\caption{Pair sizes in the $n = 2$ attractive Hubbard model. The top row shows dependencies on the anisotropy parameter $t^{\prime}/t$ and the bottom row dependencies on attractive strength $\vert U \vert$. The radii diverge near the pairing threshold. Solid lines show $n = 2$ sizes and dashed lines show $n = 1$ sizes.} 
\label{MLH:fig:Cfour}
\end{figure}

Next, we proceed to calculating the pair size. First, the in-plane effective radius, $r^{\ast}_{x}$, is discussed. For all wave function components, $\Psi_{\alpha\beta}( {\bf m}_1 , {\bf m}_2 )$, the $x$-axis physical distance between the two particles is given by $\Delta m_x = m_{2x} - m_{1x}$, in units of $a$. From the definition of wave function
\begin{equation}
\left( r^{\ast}_{x} \right)^2 = \langle ( \Delta m_{x} )^2 \rangle = 
\frac{ \sum_{\alpha\beta} \sum_{\bf m} m^2_{x} \: \Psi^{\ast}_{\alpha\beta}(0,{\bf m}) \Psi_{\alpha\beta}(0,{\bf m}) }
     { \sum_{\alpha\beta} \sum_{\bf m}            \Psi^{\ast}_{\alpha\beta}(0,{\bf m}) \Psi_{\alpha\beta}(0,{\bf m}) } 
\equiv \frac{J_{x}}{J_{0}}     \: .   
\label{MLH:eq:cthirtyeight}
\end{equation}
Substituting here Eq.~(\ref{twopart:eq:ctwelve}) under constraint ${\bf k}_1 + {\bf k}_2 = {\bf P}$, integrating by parts, and shifting the integration variable by $\frac{1}{2} {\bf P}$, one obtains
\begin{equation}
J_{0} = \sum_{\alpha\beta} \frac{1}{N_2} \sum_{\bf q} 
        \psi_{\alpha\beta} \left( \frac{\bf P}{2} + {\bf q} , \frac{\bf P}{2} - {\bf q} \right) 
        \psi^{\ast}_{\alpha\beta} \left( \frac{\bf P}{2} + {\bf q} , \frac{\bf P}{2} - {\bf q} \right) ,   
\label{MLH:eq:cthirtynine}
\end{equation}
\begin{equation}
J_{x} = \sum_{\alpha\beta} \frac{1}{N_2} \sum_{\bf q} 
       \frac{\partial}{\partial q_x}
       \left[ \psi_{\alpha\beta} \left( \frac{\bf P}{2} + {\bf q} , \frac{\bf P}{2} - {\bf q} \right) \right] 
       \frac{\partial}{\partial q_x}
       \left[ \psi^{\ast}_{\alpha\beta} \left( \frac{\bf P}{2} + {\bf q} , \frac{\bf P}{2} - {\bf q} \right) \right] .   
\label{MLH:eq:cforty}
\end{equation}
One makes two important observations about these equations. (i) The fact that we are dealing with the $n = 2$ model has been used nowhere. Therefore, Eqs.~(\ref{MLH:eq:cthirtynine}) and (\ref{MLH:eq:cforty}) is valid for {\em all} $n$. (ii) All four wave function components $\psi_{\alpha\beta}$ are needed to compute $r^{\ast}_{x}$. This is different from the pair energy and mass which required only the two diagonal components $\psi_{\alpha\alpha}$.     

We now switch to the pair's $z$-size. For off-diagonal wave function components, the physical distance between two particles along the $z$ axis differs from $( m_{2z} - m_{1z} )$. It is equal to $( m_{2z} - m_{1z} - 1)$ for $\Psi_{12}( {\bf m}_1 , {\bf m}_2 )$ and $( m_{2z} - m_{1z} + 1)$ for $\Psi_{21}( {\bf m}_1 , {\bf m}_2 )$. Taking this into account, the $z$ effective radius becomes   
\begin{equation}
\left( r^{\ast}_{z} \right)^2 
= \frac{\sum_{\bf m} \left\{ ( m_{z}     )^2  \vert \Psi_{11}(0,{\bf m}) \vert^2 + 
                             ( m_{z} - 1 )^2  \vert \Psi_{12}(0,{\bf m}) \vert^2 +  
                             ( m_{z} + 1 )^2  \vert \Psi_{21}(0,{\bf m}) \vert^2 +  
                             ( m_{z}     )^2  \vert \Psi_{22}(0,{\bf m}) \vert^2    \right\}}
       {\sum_{\alpha\beta} \sum_{\bf m} \vert \Psi_{\alpha\beta}(0,{\bf m}) \vert^2 }
\equiv \frac{J_{z}}{J_{0}}     \: .   
\label{MLH:eq:cfortyone}
\end{equation}
Substituting here Eq.~(\ref{twopart:eq:ctwelve}), $J_z$ is transformed into the following:
\begin{eqnarray}
J_{z} & = & \frac{1}{N_2} \sum_{\bf q} 
  \frac{\partial}{\partial q_z}
  \left[ \psi_{11} \left( \frac{\bf P}{2} + {\bf q} , \frac{\bf P}{2} - {\bf q} \right) \right] 
  \frac{\partial}{\partial q_z}
  \left[ \psi^{\ast}_{11} \left( \frac{\bf P}{2} + {\bf q} , \frac{\bf P}{2} - {\bf q} \right) \right] + 
\nonumber \\      
      &   & \frac{1}{N_2} \sum_{\bf q} 
  \frac{\partial}{\partial q_z}
  \left[ e^{- i q_z } \psi_{12} \left( \frac{\bf P}{2} + {\bf q} , \frac{\bf P}{2} - {\bf q} \right) \right] 
  \frac{\partial}{\partial q_z}
  \left[ e^{  i q_z } \psi^{\ast}_{12} \left( \frac{\bf P}{2} + {\bf q} , \frac{\bf P}{2} - {\bf q} \right) \right] + 
\nonumber \\      
      &   & \frac{1}{N_2} \sum_{\bf q} 
  \frac{\partial}{\partial q_z}
  \left[ e^{  i q_z } \psi_{21} \left( \frac{\bf P}{2} + {\bf q} , \frac{\bf P}{2} - {\bf q} \right) \right] 
  \frac{\partial}{\partial q_z}
  \left[ e^{- i q_z } \psi^{\ast}_{21} \left( \frac{\bf P}{2} + {\bf q} , \frac{\bf P}{2} - {\bf q} \right) \right] + 
\nonumber \\      
      &   & \frac{1}{N_2} \sum_{\bf q} 
  \frac{\partial}{\partial q_z}
  \left[ \psi_{22} \left( \frac{\bf P}{2} + {\bf q} , \frac{\bf P}{2} - {\bf q} \right) \right] 
  \frac{\partial}{\partial q_z}
  \left[ \psi^{\ast}_{22} \left( \frac{\bf P}{2} + {\bf q} , \frac{\bf P}{2} - {\bf q} \right) \right] .
\label{MLH:eq:cfortytwo}
\end{eqnarray}
$J_z$ has a more complicated structure than $J_x$. The real-space shifts in off-diagonal wave function components resulted in extra factors $e^{\pm i k_z}$ to be differentiated in Eq.~(\ref{MLH:eq:cfortytwo}). Next, one needs off-diagonal wave function components. Returning to Eq.~(\ref{MLH:eq:cseventeenone}), they are
\begin{eqnarray}
\psi_{12}( {\bf k}_1 , {\bf k}_2 ) & = & 
- \vert U \vert \Phi_{11}({\bf P}) \, \frac{\Delta_{12,11}({\bf k}_1,{\bf k}_2)}{\Delta_{0}({\bf k}_1,{\bf k}_2)} 
- \vert U \vert \Phi_{22}({\bf P}) \, \frac{\Delta_{12,22}({\bf k}_1,{\bf k}_2)}{\Delta_{0}({\bf k}_1,{\bf k}_2)}  \: ,  
\label{twopart:eq:cfortythree} \\
\psi_{21}( {\bf k}_1 , {\bf k}_2 ) & = &   
- \vert U \vert \Phi_{11}({\bf P}) \, \frac{\Delta_{21,11}({\bf k}_1,{\bf k}_2)}{\Delta_{0}({\bf k}_1,{\bf k}_2)} 
- \vert U \vert \Phi_{22}({\bf P}) \, \frac{\Delta_{21,22}({\bf k}_1,{\bf k}_2)}{\Delta_{0}({\bf k}_1,{\bf k}_2)}  \: ,  
\label{twopart:eq:cfortyfour}
\end{eqnarray}
\begin{eqnarray}
\Delta_{12,11}( {\bf k}_1 , {\bf k}_2 ) & = & \Delta^{\ast}_{21,22}( {\bf k}_1 , {\bf k}_2 ) = 
- g^{\ast}_{2} \left[ \left( E - \xi_{ {\bf k}_1 {\bf k}_2 } \right)^2 + \vert g_1 \vert^2 - \vert g_2 \vert^2 \right] ,  
\label{twopart:eq:cfortyfive} \\
\Delta_{12,22}( {\bf k}_1 , {\bf k}_2 ) & = & \Delta^{\ast}_{21,11}( {\bf k}_1 , {\bf k}_2 ) =  
- g_{1} \left[ \left( E - \xi_{ {\bf k}_1 {\bf k}_2 } \right)^2 - \vert g_1 \vert^2 + \vert g_2 \vert^2 \right] .      
\label{twopart:eq:cfortysix}
\end{eqnarray}
These ratios $\Delta/\Delta_0$ can also be reduced to elementary fractions but for our purposes here it is not necessary. Equations (\ref{MLH:eq:cthirtyeight})-(\ref{twopart:eq:cfortysix}) and (\ref{MLH:eq:cthirtytwo})-(\ref{MLH:eq:cthirtythree}) enable calculation of the size of {\em any} pair, that is, of any momentum ${\bf P}$ or dispersion branch. However, the corresponding expressions can be quite complicated. From now on, we will be interested only in the size of the ground state pair, that is ${\bf P} = 0$ in the lowest branch. That brings about two significant simplifications. First, only wave function components with opposite momenta are needed. Second, in the ground state, $\Phi_{11}(0) = \Phi_{22}(0)$. This is a consequence of the lattice to be symmetric with respect to reflections in $z$-plane. The ground state pair wave function corresponds to the symmetric representation of this symmetry group. We will be using notation $\psi^{0}_{\alpha\beta}( {\bf q} , - {\bf q} )$ to distinguish those. The general expressions reduce to
\begin{eqnarray}
\psi^{0}_{11}( {\bf q} , - {\bf q} ) & = & 
- \vert U \vert \Phi_{11}(0) \, \frac{ E - \xi_{\bf q} }{ ( E - \xi_{\bf q} )^2 - 4 \vert g_{\bf q} \vert^2 } \: ,  
\label{twopart:eq:cfortyseven} \\
\psi^{0}_{12}( {\bf q} , - {\bf q} ) & = &   
- \vert U \vert \Phi_{11}(0) \, \frac{( - 2 g_{\bf q} )}{ ( E - \xi_{\bf q} )^2 - 4 \vert g_{\bf q} \vert^2 } \: ,  
\label{twopart:eq:cfortyeight} \\
\psi^{0}_{21}( {\bf q} , - {\bf q} ) & = &   
- \vert U \vert \Phi_{11}(0) \, \frac{( - 2 g^{\ast}_{\bf q} )}{ ( E - \xi_{\bf q} )^2 - 4 \vert g_{\bf q} \vert^2 } \: ,     
\label{twopart:eq:cfortynine}  \\
\psi^{0}_{22}( {\bf q} , - {\bf q} ) & = &   
- \vert U \vert \Phi_{11}(0) \, \frac{ E - \xi_{\bf q} }{ ( E - \xi_{\bf q} )^2 - 4 \vert g_{\bf q} \vert^2 } \: ,     
\label{twopart:eq:cfifty}
\end{eqnarray}
where
\begin{eqnarray}
    \xi_{\bf q}          & = & - 4 t ( \cos{q_x a} + \cos{q_y a} )                    \: ,  
\label{twopart:eq:cfiftyone}  \\
      g_{\bf q}          & = &  t + t^{\prime} e^{2 i q_z a}                          \: ,     
\label{twopart:eq:cfiftytwo}  \\
\vert g_{\bf q} \vert^2  & = &  t^2 + 2 t t^{\prime} \cos{( 2 q_z a )} + t^{\prime 2} \: .     
\label{twopart:eq:cfiftythree}
\end{eqnarray}
Note that $q_x$ enters only via $\xi_{\bf q}$ whereas $q_z$ only via $g_{\bf q}$, which simplifies differentiation. The common constant factor $- \vert U \vert \Phi_{11}(0)$ cancels out in the ratios $J_{x}/J_{0}$ and $J_{z}/J_{0}$ and can be omitted. With that in mind, the final expressions for $J$'s read 
\begin{equation}
J_{0} = \frac{1}{N_2} \sum_{\bf q} 2 
       \frac{ ( E - \xi_{\bf q} )^2 + 4 \vert g_{\bf q} \vert^2 }
     { \left[ ( E - \xi_{\bf q} )^2 - 4 \vert g_{\bf q} \vert^2 \right]^2 } \: ,   
\label{MLH:eq:cfiftyfour}
\end{equation}
\begin{equation}
J_{x} = ( 4 t a )^2 \frac{1}{N_2} \sum_{\bf q} 2 \sin^2{( q_x a )}
 \frac{ \left[ ( E - \xi_{\bf q} )^2 + 4 \vert g_{\bf q} \vert^2 \right]^2 + 
                                 16 \vert g_{\bf q} \vert^2 ( E - \xi_{\bf q} )^2 }
      { \left[ ( E - \xi_{\bf q} )^2 - 4 \vert g_{\bf q} \vert^2 \right]^4 } \: ,   
\label{MLH:eq:cfiftyfive}
\end{equation}
\begin{equation}
J_{z} = ( 8 a^2 ) \frac{1}{N_2} \sum_{\bf q} 
 \frac{ \left[ ( E - \xi_{\bf q} )^2 - 4 \vert g_{\bf q} \vert^2 \right]^2 \cdot
        \left[ t^2 - 2 t t^{\prime} \cos{( 2 q_z a )} + t^{\prime 2} \right] + 
        128 \, t^2 t^{\prime 2} ( E - \xi_{\bf q} )^2 \sin^2{( 2 q_z a )} }
      { \left[ ( E - \xi_{\bf q} )^2 - 4 \vert g_{\bf q} \vert^2 \right]^4 } \: .   
\label{MLH:eq:cfiftysix}
\end{equation}
Pair sizes of the $n = 2$ model computed from Eqs.~(\ref{MLH:eq:cthirtyeight}), (\ref{MLH:eq:cfortyone}), (\ref{twopart:eq:cfiftyone})-(\ref{MLH:eq:cfiftysix}), are shown in Fig.~\ref{MLH:fig:Cfour} and compared with the $n = 1$ sizes for the same $|U|$ and $t^{\prime}$. 

The close-packing temperature $\cal{T}^{\ast}_{\rm cp}$ defined in Eq.~(\ref{MLH:eq:btwentyeight}) for the $n = 2$ and $n = 1$ models is shown in Fig.~\ref{MLH:fig:Cfive}.

\begin{figure}[t]
\includegraphics[width=0.90\textwidth]{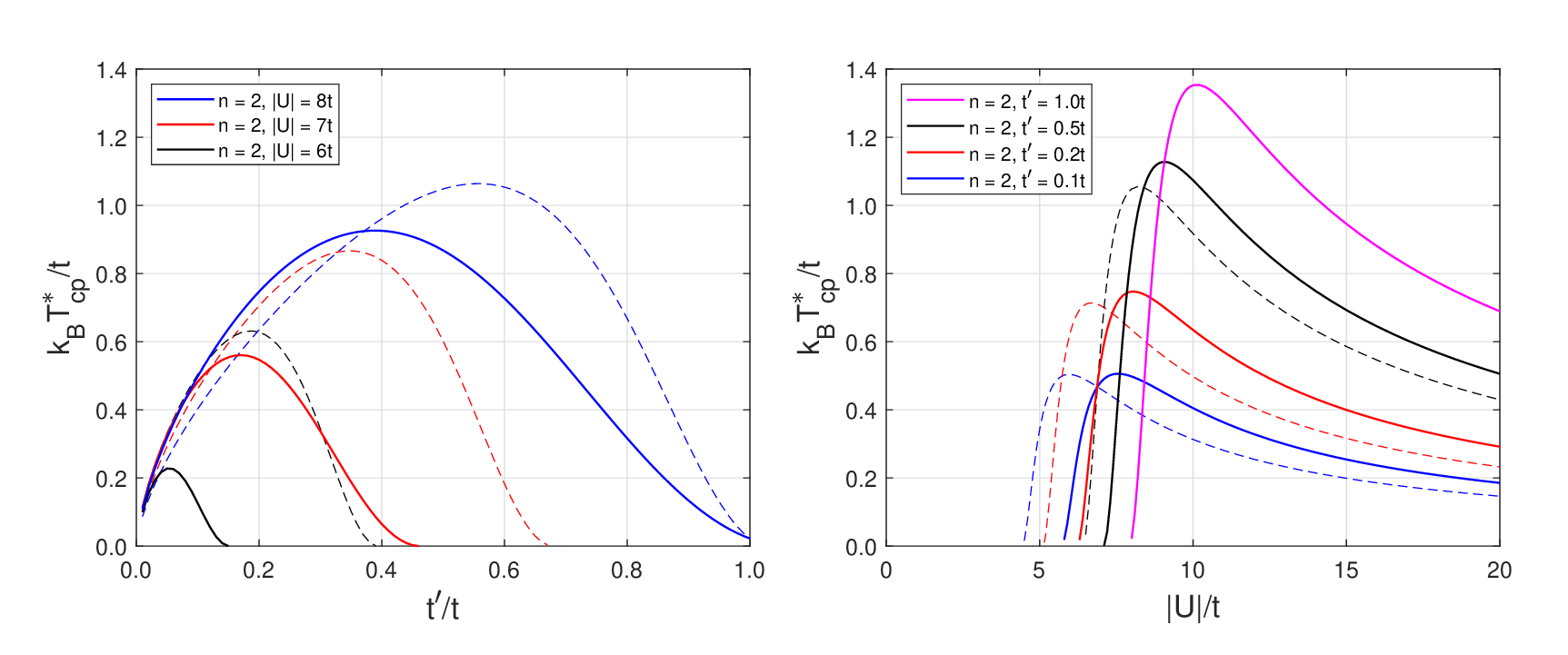}
\caption{Close-packed critical temperature, Eq.~(\ref{MLH:eq:btwentyeight}). Solid lines are the $n = 2$ model. Dashed lines are the $n = 1$ model.} 
\label{MLH:fig:Cfive}
\end{figure}

\section{\label{MLH:sec:app:e}
Arbitrary number of layers, $n \geq 3$
}

In this Appendix, we formulate the process of solving the two-body problem for arbitrary $n$. Almost nothing can be derived analytically, so the solution is going to be purely numerical. One begins with the one-particle dispersion $\varepsilon_{\bf k}$. One must allow for each layer to have its own on-site energy $\epsilon_{\alpha}$. The one-particle wave function is an $( n \times 1 )$ array $\phi_{\alpha}({\bf m})$. After Fourier transformation, the Schr\"odinger equation becomes a matrix equation:    
\begin{equation}
\left( \begin{array}{cccccccc}
 \tilde{\varepsilon}_{\bf k} - \epsilon_{1} &  t  &  0 &  0   &  \cdots  &   0   &   0   &  t^{\prime} e^{i n k_z}  \\
 t       &  \tilde{\varepsilon}_{\bf k} - \epsilon_{2} &  t   &  0    &  \cdots  &   0   &  0   &   0               \\
 0       &  t  &  \tilde{\varepsilon}_{\bf k} - \epsilon_{3}  &  t    &  \cdots  &   0   &  0   &   0               \\
 0       &  0  &  t  &  \tilde{\varepsilon}_{\bf k} - \epsilon_{4}    &  \cdots  &   0   &  0   &   0               \\
 \vdots  &  \vdots   &  \vdots   &  \vdots  &  \ddots  &   \vdots  &  \vdots   &  \vdots                            \\
 0       &  0  &  0  &  0  &  \dots    &   \tilde{\varepsilon}_{\bf k} - \epsilon_{n-2}  &  t   &  0                \\
 0       &  0  &  0  &  0  &  \dots    &   t  &  \tilde{\varepsilon}_{\bf k} - \epsilon_{n-1}   &  t                \\
 t^{\prime} e^{- i n k_z}  &  0  &  0  &  0  &  \dots  &   0   &  t   &  \tilde{\varepsilon}_{\bf k} - \epsilon_{n} 
\end{array} \right) 
\left( \begin{array}{c}
\phi_{1}( {\bf k} )      \\
\phi_{2}( {\bf k} )      \\
\phi_{3}( {\bf k} )      \\
\phi_{4}( {\bf k} )      \\
    \vdots               \\
\phi_{n-2}( {\bf k} )    \\
\phi_{n-1}( {\bf k} )    \\
\phi_{n}  ( {\bf k} ) 
\end{array} \right) = 
\left( \begin{array}{c}
    0                                \\
    0                                \\
    0                                \\
    0                                \\
 \vdots                              \\
    0                                \\
    0                                \\
    0  
\end{array} \right) ,  
\label{MLH:eq:eone}
\end{equation}
where 
\begin{equation}
\tilde{\varepsilon}_{\bf k} \equiv {\varepsilon}_{\bf k} + 2 t \cos{k_x} + 2 t \cos{k_y} \: .  
\label{MLH:eq:efour}
\end{equation}
In the $n = 1$ and $n = 2$ models, all the CuO$_2$ planes were equivalent so $\epsilon_{\alpha}$ could be set to zero. Beginning with $n = 3$, there are non-equivalent planes, and the differences between $\epsilon_{\alpha}$ have physical significance. The consistency condition of Eq.~(\ref{MLH:eq:eone}) provides $n$ one-particle energy bands $\varepsilon_{\alpha}({\bf k})$. 

The two-particle wave function is an $( n^2 \times 1 )$ array $\Psi_{\alpha\beta}( {\bf m}_1 , {\bf m}_2 )$. We arrange wave function components by the second index changing fastest: $\hat{\boldsymbol\psi} = ( \psi_{11} , \psi_{12} , \ldots , \psi_{1n} ; \psi_{21} , \psi_{22} , \ldots , \psi_{2n} ; \ldots ; \psi_{n1} , \psi_{n2} , \ldots , \psi_{nn} )^{T} $.  The Fourier-transformed Schr\"odinger equation is an $( n^2 \times n^2 )$ inhomogeneous matrix equation that can be written in block form:
\begin{equation}
\left( \! \begin{array}{cccccccc}
{\bf g_2} - {\boldsymbol\epsilon}_{1} &  {\bf t} & {\bf 0} &  {\bf 0} &  \cdots  & {\bf 0} & {\bf 0} & {\bf g_1} \\
{\bf t} & {\bf g_2} - {\boldsymbol\epsilon}_{2}  & {\bf t} &  {\bf 0} &  \cdots  & {\bf 0} & {\bf 0} &  {\bf 0}  \\
{\bf 0} & {\bf t}  & {\bf g_2} - {\boldsymbol\epsilon}_{3} &  {\bf t} &  \cdots  & {\bf 0} & {\bf 0} &  {\bf 0}  \\
{\bf 0} & {\bf 0}  & {\bf t} & {\bf g_2} - {\boldsymbol\epsilon}_{4}  &  \cdots  & {\bf 0} & {\bf 0} &  {\bf 0}  \\
\vdots     &  \vdots     &  \vdots    &  \vdots  & \ddots  &   \vdots    &  \vdots    &  \vdots      \\
{\bf 0} & {\bf 0} & {\bf 0} & {\bf 0} &  \dots   & {\bf g_2} - {\boldsymbol\epsilon}_{n-2} & {\bf t} & {\bf 0}   \\
{\bf 0} & {\bf 0} & {\bf 0} & {\bf 0} &  \dots   & {\bf t} &  {\bf g_2} - {\boldsymbol\epsilon}_{n-1} & {\bf t}  \\
{\bf g^{\ast}_1} & {\bf 0} & {\bf 0} & {\bf 0} & \dots & {\bf 0} & {\bf t}  &  {\bf g_2} - {\boldsymbol\epsilon}_{n} 
\end{array} \! \right) 
\!\! \left( \!\! \begin{array}{c}
\hat{\boldsymbol\psi}_{1i}( {\bf k}_1 , {\bf k}_2 )      \\
\hat{\boldsymbol\psi}_{2i}( {\bf k}_1 , {\bf k}_2 )      \\
\hat{\boldsymbol\psi}_{3i}( {\bf k}_1 , {\bf k}_2 )      \\
\hat{\boldsymbol\psi}_{4i}( {\bf k}_1 , {\bf k}_2 )      \\
    \vdots                                               \\
\hat{\boldsymbol\psi}_{n-2,i}( {\bf k}_1 , {\bf k}_2 )   \\
\hat{\boldsymbol\psi}_{n-1,i}( {\bf k}_1 , {\bf k}_2 )   \\
\hat{\boldsymbol\psi}_{ni}  ( {\bf k}_1 , {\bf k}_2 ) 
\end{array} \!\! \right) =  
\left( \!\! \begin{array}{c}
 - \vert U \vert \Phi_{11}( {\bf P} )                    \\
  {\bf 0}_{n}                                            \\
 - \vert U \vert \Phi_{22}( {\bf P} )                    \\
  {\bf 0}_{n}                                            \\
    \vdots                                               \\
 - \vert U \vert \Phi_{n-1 , n-1}( {\bf P} )             \\
  {\bf 0}_{n}                                            \\
 - \vert U \vert \Phi_{nn}( {\bf P} ) 
\end{array} \!\! \right)  
\label{MLH:eq:efive}
\end{equation}
where ${\bf 0}_{n} = ( 0 , 0 , \ldots , 0 )^{T}$ is an $( n \times 1 )$ array of $n$ zeros, $\hat{\boldsymbol\psi}_{1i} = ( \psi_{11} , \psi_{12} , \ldots , \psi_{1n} )^{T}$, $\hat{\boldsymbol\psi}_{2i} = ( \psi_{21} , \psi_{22} , \ldots , \psi_{2n} )^{T}$, and so on. Each element in the main matrix of Eq.~(\ref{MLH:eq:efive}) is an $( n \times n )$ matrix by itself. The block ${\bf g_2}$ describes hopping of particle 2 when particle 1 is stationary. Therefore, ${\bf g_2}$ has the same structure as the matrix in Eq.~(\ref{MLH:eq:eone}) but with $k_z$ replaced with $k_{2z}$ and one-particle energy $\tilde{\varepsilon}$ replaced with $\tilde{E}$: 
\begin{equation}
\tilde{E} \equiv E + 2 t ( \cos{k_{1x}} + \cos{k_{1y}} ) + 2 t ( \cos{k_{2x}} + \cos{k_{2y}} ) \: ,  
\label{MLH:eq:esix}
\end{equation}
where $E$ is the total two-particle energy. For example, for $n = 5$, the block ${\bf g_2}$ is
\begin{equation}
{\bf g_2}( n = 5 ) = 
\left( \begin{array}{ccccc}
\tilde{E}                      &     t       &     0       &     0       &  t^{\prime} e^{i n k_{2z}}  \\
   t                           &  \tilde{E}  &     t       &     0       &     0                       \\
   0                           &     t       &  \tilde{E}  &     t       &     0                       \\
   0                           &     0       &     t       &  \tilde{E}  &     t                       \\
t^{\prime} e^{ - i n k_{2z} }  &     0       &     0       &     t       &  \tilde{E}
\end{array} \right)  ,  
\label{MLH:eq:eseven}
\end{equation}
where $n = 5$ should be used. Additionally, the diagonal $( n \times n )$ matrices ${\boldsymbol\epsilon}_{\alpha}$ account for the differences in on-site energies: 
\begin{equation}
{\boldsymbol\epsilon}_{1} = 
\left( \begin{array}{ccccc}
\epsilon_{1} + \epsilon_{1}    &     0       &     \cdots  &     0       &     0       \\
   0       &  \epsilon_{1} + \epsilon_{2}    &     \cdots  &     0       &     0       \\
   \vdots  &    \vdots         &  \ddots     &     \vdots  &     \cdots                \\
   0       &     0             &  \cdots     &  \epsilon_{1} + \epsilon_{n-1}  &    0  \\
   0       &     0             &  \cdots     &     0       &  \epsilon_{1} + \epsilon_{n}
\end{array} \right)  ,  
\hspace{0.8cm}
{\boldsymbol\epsilon}_{2} = 
\left( \begin{array}{ccccc}
\epsilon_{2} + \epsilon_{1}    &     0       &     \cdots  &     0       &     0       \\
   0       &  \epsilon_{2} + \epsilon_{2}    &     \cdots  &     0       &     0       \\
   \vdots  &    \vdots         &  \ddots     &     \vdots  &     \cdots                \\
   0       &     0             &  \cdots     &  \epsilon_{2} + \epsilon_{n-1}  &    0  \\
   0       &     0             &  \cdots     &     0       &  \epsilon_{2} + \epsilon_{n}
\end{array} \right)  ,  
\label{MLH:eq:esevenone}
\end{equation}
and so on. Off-diagonal blocks of the matrix in Eq.~(\ref{MLH:eq:efive}) describe hopping of particle 1 when particle 2 is stationary. Each block is a diagonal $( n \times n )$ matrix:
\begin{equation}
{\bf t} = 
\left( \begin{array}{cccc}
    t      &  0      & \ldots   &     0      \\
    0      &  t      & \ldots   &     0      \\  
 \vdots    &  \vdots & \ddots   &  \vdots    \\
    0      &  \ldots &    0     &     t                                
\end{array} \right)  ;  
\hspace{1.0cm}
{\bf 0} = 
\left( \begin{array}{cccc}
    0      &  0      & \ldots   &     0      \\
    0      &  0      & \ldots   &     0      \\  
 \vdots    &  \vdots & \ddots   &  \vdots    \\
    0      &  \ldots &    0     &     0                                
\end{array} \right)  ;  
\hspace{1.0cm}
{\bf g_1} = 
\left( \begin{array}{cccc}
    t^{\prime} e^{i n k_{1z}}  &  0                          & \ldots   &     0      \\
    0                          &  t^{\prime} e^{i n k_{1z}}  & \ldots   &     0      \\  
 \vdots                        &  \vdots                     & \ddots   &  \vdots    \\
    0                          &  \ldots                     &    0     &     t^{\prime} e^{i n k_{1z}}                                
\end{array} \right)  .  
\label{MLH:eq:eeight}
\end{equation}
Finally, functions $\Phi_{\alpha\alpha}({\bf P})$ in the right-hand-side of Eq.~(\ref{MLH:eq:efive}) are  
\begin{equation}
\Phi_{\alpha\alpha}( {\bf P} )
\equiv \frac{1}{N_n} \sum_{\bf q} \psi_{\alpha\alpha}( {\bf q}, {\bf P} - {\bf q} ) 
=      \frac{1}{N_n} \sum_{\bf q} 
\psi_{\alpha\alpha} \left( \frac{\bf P}{2} + {\bf q}, \frac{\bf P}{2} - {\bf q} \right) ,  
\label{MLH:eq:enine}
\end{equation}
where $N_n = N/n$ is the number of unit cells of volume $\Omega_0 = na^3$. 

The solution process begins with solving Eq.~(\ref{MLH:eq:efive}) as a inhomogeneous system of linear equations and expressing $\psi_{\alpha\beta}$ as a linear combination of $\Phi_{\gamma\gamma}$
\begin{equation}
\psi_{\alpha\beta}( {\bf k}_1 , {\bf k}_2 ) = - \vert U \vert \sum^{n}_{\gamma = 1}
\Phi_{\gamma\gamma}({\bf P}) \, 
\frac{\Delta_{\alpha\beta,\gamma\gamma}({\bf k}_1,{\bf k}_2)}{\Delta_{0}({\bf k}_1,{\bf k}_2)} \: .  
\label{MLH:eq:eten}
\end{equation}
Here, $\Delta_{0}$ is an $n^2$-degree polynomial in $\tilde{E}$ and $\Delta_{\alpha\beta,\gamma\gamma}$ in an $( n^2 - 1 )$-degree polynomial in $\tilde{E}$. For pair energy and mass, only the diagonal wave function components $\psi_{\alpha\alpha}$ are needed. Substitution of Eq.~(\ref{MLH:eq:eten}) in Eq.~(\ref{MLH:eq:enine}) leads to a homogeneous system of equations for $\Phi_{\alpha\alpha}$
\begin{equation}
\left( \begin{array}{cccc}
- \vert U \vert M_{11}({\bf P})  &  - \vert U \vert M_{12}({\bf P})  & \ldots & - \vert U \vert M_{1n}({\bf P}) \\
- \vert U \vert M_{21}({\bf P})  &  - \vert U \vert M_{22}({\bf P})  & \ldots & - \vert U \vert M_{2n}({\bf P}) \\
  \vdots                         &    \vdots                         & \ddots &   \vdots                        \\ 
- \vert U \vert M_{n1}({\bf P})  &  - \vert U \vert M_{n2}({\bf P})  & \ldots & - \vert U \vert M_{nn}({\bf P}) 
\end{array} \right) 
\left( \begin{array}{c}
\Phi_{11}({\bf P})  \\  \Phi_{22}({\bf P})  \\  \vdots \\ \Phi_{nn}({\bf P})     
\end{array} \right) =
\left( \begin{array}{c}
\Phi_{11}({\bf P})  \\  \Phi_{22}({\bf P})  \\  \vdots \\ \Phi_{nn}({\bf P}) 
\end{array} \right) ,
\label{MLH:eq:eeleven}
\end{equation}
\begin{equation}
M_{\alpha\gamma}({\bf P}) \equiv \frac{1}{N_n} \sum_{\bf q}
\frac{\Delta_{\alpha\alpha,\gamma\gamma} \left( \frac{\bf P}{2} + {\bf q}, \frac{\bf P}{2} - {\bf q} \right)}
     {\Delta_{0} \left( \frac{\bf P}{2} + {\bf q}, \frac{\bf P}{2} - {\bf q} \right) } = 
  \int\limits^{\pi}_{-\pi} \!\! \int\limits^{\pi}_{-\pi} \!\! \int\limits^{\frac{\pi}{n}}_{-\frac{\pi}{n}}
n \, \frac{ {\rm d} q_x {\rm d} q_y {\rm d} q_z}{ ( 2 \pi )^3 }  
\frac{\Delta_{\alpha\alpha,\gamma\gamma} \left( \frac{\bf P}{2} + {\bf q}, \frac{\bf P}{2} - {\bf q} \right)}
     {\Delta_{0} \left( \frac{\bf P}{2} + {\bf q}, \frac{\bf P}{2} - {\bf q} \right) } \: .  
\label{MLH:eq:etwelve}
\end{equation}
The consistency condition of Eq.~(\ref{MLH:eq:eeleven}) defines pair energy $E$ as a function of pair momentum ${\bf P}$. There are $n$ dispersion bands. Pair momentum is restricted to  
\begin{equation}
- \pi           \leq P_x \leq + \pi            \: , 
\hspace{1.0cm}
- \pi           \leq P_y \leq + \pi            \: , 
\hspace{1.0cm}
- \frac{\pi}{n} \leq P_z \leq + \frac{\pi}{n}  \: .  
\label{MLH:eq:ethirteen}
\end{equation}
Since $q_x$ and $q_y$ enter the equations only via $\tilde{E}$, the integrands in Eq.~(\ref{MLH:eq:etwelve}) are explicitly even functions of $q_x$ and $q_y$ for any ${\bf P}$. The integration domain can therefore be restricted to $q_x, q_y > 0$:
\begin{equation}
M_{\alpha\gamma}({\bf P}) = 
  \int\limits^{\pi}_{0} \!\! \int\limits^{\pi}_{0} \!\! \int\limits^{\frac{\pi}{n}}_{-\frac{\pi}{n}}
(4n) \, \frac{ {\rm d} q_x {\rm d} q_y {\rm d} q_z}{ ( 2 \pi )^3 }  
\frac{\Delta_{\alpha\alpha,\gamma\gamma} \left( \frac{\bf P}{2} + {\bf q}, \frac{\bf P}{2} - {\bf q} \right)}
     {\Delta_{0} \left( \frac{\bf P}{2} + {\bf q}, \frac{\bf P}{2} - {\bf q} \right) } \: .  
\label{MLH:eq:efourteen}
\end{equation}
Calculation of $M_{\alpha\gamma}$ is the main technical challenge of the method. Since everything is done numerically, it is convenient to use the sum-over-${\bf q}$ representation in Eq.~(\ref{MLH:eq:etwelve}). In practice, infinite sums are replaces with finite sums. For each ${\bf q}$, Eq.~(\ref{MLH:eq:efive}) is solved numerically with $-|U| \Phi_{\alpha\alpha}$ in the right-hand-side replaced by 1's, to determine the corresponding $\Delta/\Delta_0$'s and the contribution of this ${\bf q}$ to all $M_{\alpha\gamma}$.    

To obtain pairing thresholds, pair energy is set to be equal to the minimum energy of two free particles with the same total momentum, $E = E_{11} = 2 \varepsilon_{1 , \frac{\bf P}{2}}$ where $\varepsilon_{1 {\bf k}}$ is the lowest one-particle energy band defined by Eq.~(\ref{MLH:eq:eone}). With that, Eq.~(\ref{MLH:eq:eeleven}) defines threshold values $|U_{\rm cr}|$ as functions of pair momentum ${\bf P}$. At threshold, all the integrands in Eq.~(\ref{MLH:eq:etwelve}) are singular. The threshold values $M^{0}_{\alpha\gamma}$ can be obtained by computing $M_{\alpha\gamma}$ for a series of energies $E = E_{11} - \delta_{i}$, $\delta_{i} > 0$, and then numerically extrapolating to $\delta = 0$.       

Knowing pair energy vs. momentum, pair masses are computed numerically using Eq.~(\ref{MLH:eq:cthirtyseven}). 

For calculating pair size, all wave function components $\psi_{\alpha\beta}$ are needed. The $(xy)$ pair size is computed from Eqs.~(\ref{MLH:eq:cthirtyeight})-(\ref{MLH:eq:cforty}), where $N_n$ should be used in place of $N_2$. Note that $q_x$ only enters via $\tilde{E}$, which simplifies calculation of $q_x$ derivatives. 

To calculate $r^{\ast}_z$, we generalize the derivation of Eqs.~(\ref{MLH:eq:cfortyone})-(\ref{MLH:eq:cfortytwo}) to arbitrary $n$. Let us assume the first particle resides in an $\alpha$th layer of unit cell ${\bf m}$. Then its $z$ coordinate in units of $a$ is $z_1 = m_{1z} - \alpha + 1$. Similarly for the the second particle, $z_2 = m_{2z} - \beta + 1$. The $z$ difference is therefore $z_2 - z_1 = m_{2z} - m_{1z} + \alpha - \beta$. Because of translational invariance, one can set ${\bf m}_1 = 0$ and only vary ${\bf m}_2$. By definition of the wave function, 
\begin{equation}
J_z = \sum_{\bf m} \sum_{\alpha\beta} ( m_{z} + \alpha - \beta )^2 \vert \Psi_{\alpha\beta} ( 0 , {\bf m} ) \vert^2 .   
\label{MLH:eq:efifteen}
\end{equation}
We apply the Fourier transformation, Eq.~(\ref{twopart:eq:ctwelve}), and express $( m_{z} + \alpha - \beta )$ as a $k_z$-derivative of $\exp{ \{ \pm {\bf k} ( {\bf m} + \alpha {\bf z} - \beta {\bf z}) \} }$ supplemented by an appropriate shift of the exponent. After that, integration by parts and summation over ${\bf m}$ yields:
\begin{equation}
J_z = \sum_{\alpha\beta} \frac{1}{N} \sum_{\bf q} 
\frac{\partial}{\partial q_z} \left[  e^{  i q_z ( \alpha - \beta ) } 
\psi_{\alpha\beta}\left( \frac{\bf P}{2} + {\bf q} , \frac{\bf P}{2} - {\bf q} \right) \right]
\frac{\partial}{\partial q_z} \left[  e^{- i q_z ( \alpha - \beta ) } 
\psi^{\ast}_{\alpha\beta}\left( \frac{\bf P}{2} + {\bf q} , \frac{\bf P}{2} - {\bf q} \right) \right]  .   
\label{MLH:eq:esixteen}
\end{equation}
This final expression can be used for numerical evaluation of $r^{\ast}_z$.    

Finally, the close-packed critical temperature is computed from Eqs.~(\ref{MLH:eq:btwentyseven})-(\ref{MLH:eq:btwentyeight}).

\begin{figure}[t]
\includegraphics[width=0.48\textwidth]{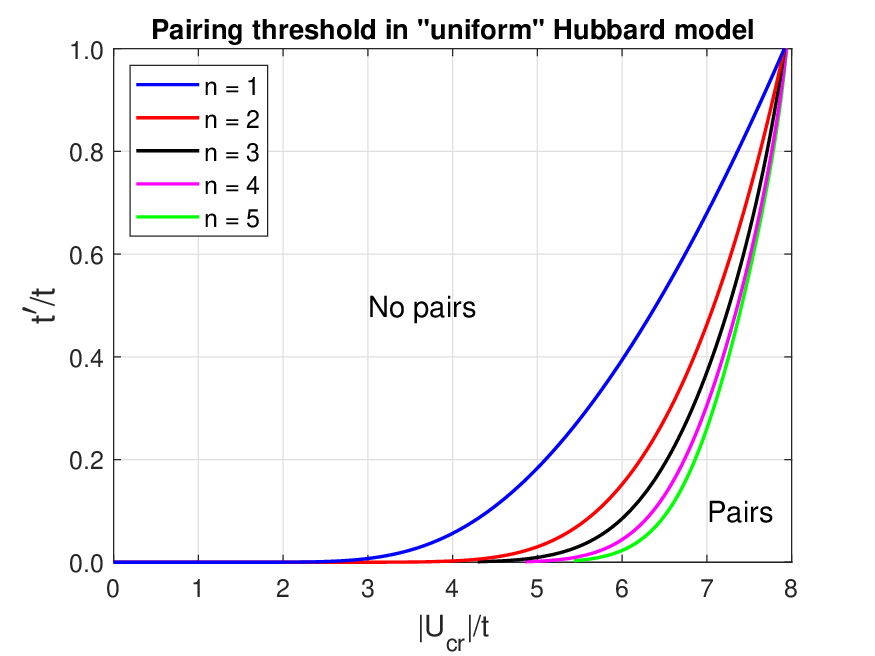}
\caption{Binding threshold in the $n = 1, 2, 3, 4$ and $5$ ``uniform'' attractive Hubbard models for ${\bf P} = 0$. The $n = 1$ and $2$ curves are the same as in Fig.~\ref{MLH:fig:Cone}. Notice that at $t^{\prime}/t = 1.0$, all models are equivalent to the isotropic cubic attractive Hubbard model ($n = 1$), with its binding threshold $7.913552 \, t$ \cite{Kornilovitch2024}.} 
\label{MLH:fig:Eone}
\end{figure}

\section{\label{MLH:sec:app:g}
Uniform model, $\epsilon_{\alpha} = 0$
}

In this Appendix, we discuss the properties of the ``uniform'' model in which all the one-particle energies are equal and therefore can be set to zero, $\epsilon_{\alpha} = 0$. This is the simplest multi-band pairing model and is the logical first step in trying to explain $T_{c,{\rm max}}(n)$. There are only two parameters, $U$ and $t^{\prime}$, which form a relatively small parameter space. The model therefore admits complete characterization.    

The pairing threshold is shown in Fig.~\ref{MLH:fig:Eone}. For a fixed interstack hopping $t^{\prime}$, there is systematic increase in the threshold value, which reflects the main physical trend: addition of each new layer adds more kinetic energy along the $z$ direction and makes the overall system more isotropic. The growing kinetic energy requires progressively larger attraction to form a pair. At the same time, the relative changes decrease with $n$, as expected on physical grounds. Indeed, the relative increase in kinetic energy between, for example, $n = 4$ and $n = 5$ is smaller than between $n = 1$ and $n = 2$. Correspondingly, the relative change in threshold is smaller, too. As can be deduced from Fig.~\ref{MLH:fig:Eone}, further increase of $n$ will result in physically insignificant changes of the threshold. The most significant variations of physical properties are expected for $n \leq 5$, which matches the cuprate phenomenology.

\begin{figure}[t]
\includegraphics[width=0.95\textwidth]{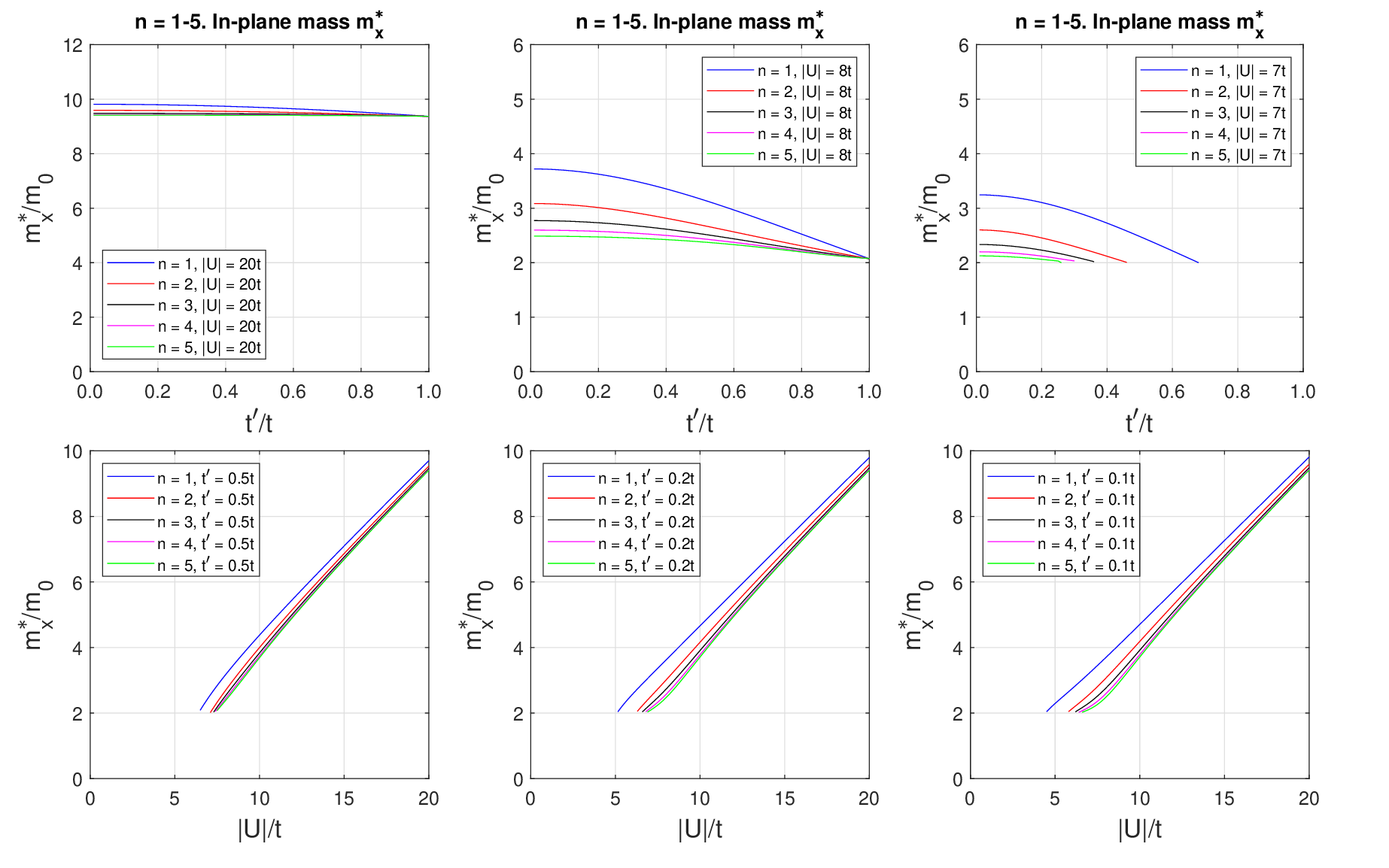}
\caption{In-plane pair mass $m^{\ast}_x$ in the $n = 1, 2, 3, 4, 5$ uniform attractive Hubbard models. The top row shows dependencies on the anisotropy parameter $t^{\prime}/t$ and the bottom row dependencies on attractive strength $\vert U \vert$.} 
\label{MLH:fig:Etwo}
\end{figure}

Figure~\ref{MLH:fig:Etwo} shows the in-plane pair mass $m^{\ast}_{x}$. The most notable feature is the weak dependence of $m^{\ast}_{x}$ on $n$. Additionally, there is a relatively weak dependence on inter-stack hopping $t^{\prime}$. It is understandable physically since hopping along the $z$ axis does not affect in-plane motion much. In contrast, the $U$ dependence of $m^{\ast}_{x}$ is approximately linear. This feature is specific to Hubbard interaction. With on-site attraction, any pair motion, including in-plane, proceeds with pair breaking via higher-energy intermediate states. As a result, pair movement is second-order in $t$ and $m^{\ast}_{x} \propto |U|/t^2$. In more realistic $UV$ models, pair movement is often first-order in $t$ and $m^{\ast}_{x}$ is approximately independent of attraction. Another notable feature of the in-plane mass is its absolute value. It will be clear from the subsequent discussion that the most relevant interval of attraction is $|U| = (7 - 10)t$. As can be seen in Fig.~\ref{MLH:fig:Etwo}, that places the pair mass in the $(3-5)m_0$ interval. This is consistent with some cuprate experiments. For example, observation of Shubnikov -- de Haas oscillation in underdoped YBa$_{2}$Cu$_{4}$O$_{8}$ produced a {\em single}-carrier mass of $2.7 \pm 0.3 \, m_e$ \cite{Bangura2008}.

\begin{figure}[t]
\includegraphics[width=0.95\textwidth]{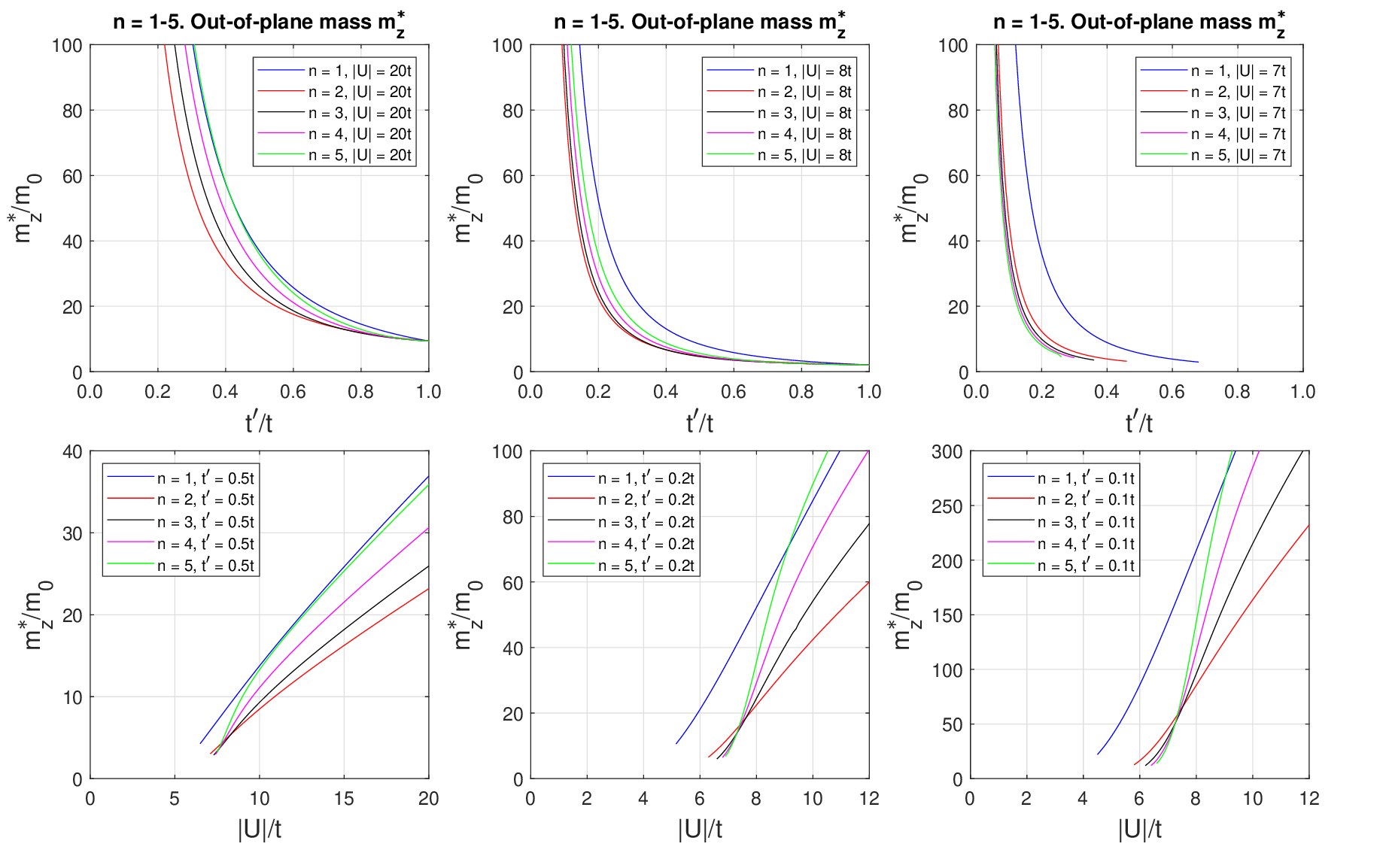}
\caption{Out-of-plane pair mass $m^{\ast}_z$ in the $n = 1, 2, 3, 4, 5$ uniform attractive Hubbard models. The top row shows dependencies on the anisotropy parameter $t^{\prime}/t$ and the bottom row dependencies on attractive strength $\vert U \vert$. In the bottom row, notice how $m^{\ast}_z$ sharply increases with $n$ for $|U| > 8t$. This is caused by increased localization of pairs within inner planes. See the text for more explanation.} 
\label{MLH:fig:Ethree}
\end{figure}

Out-of-plane pair mass $m^{\ast}_{z}$ is shown in Fig.~\ref{MLH:fig:Ethree}. It is much more sensitive to the model parameters, especially to $t^{\prime}$. $m^{\ast}_{z}$ diverges as $t^{\prime} \to 0$ for all models. We recall that it is precisely the divergence of out-of-plane mass that limits $T^{\ast}_{\rm cp}$ at high anisotropy. The $U$ dependence of $m^{\ast}_{z}$ is shown in the bottom row of Fig.~\ref{MLH:fig:Ethree}. The mass increases with $|U|$, following the same physical mechanism of pair movement via high-energy intermediate states. At the same time, $m^{\ast}_{z}(|U|)$ reveals an interesting, nonmonotonic dependence on the number of CuO$_{2}$ layers, $n$. At weak attraction close to the threshold, $m^{\ast}_{z}$ decreases with $n$. This behavior is a consequence of growing kinetic energy. Insertion of each new layer increases the pairing threshold and brings the pair closer to unbinding. As a result, $m^{\ast}_{z}$ systematically decreases and approaches the two free particle masses in the $z$ direction. However, this behavior only holds close to threshold, when the pair wave function is mostly composed of propagating plane waves. At stronger coupling, $|U| > 8t$, another effect comes into play: increasing localization of the pair wave function in the inner planes. Indeed, weak inter-stack hopping $t^{\prime}$ constitutes a potential barrier, thereby confining the pair within the stack of planes. The pair can freely propagate within the stack along $z$ direction but the tunneling barriers at the stack's edges decrease the wave function at the outer planes. The result is progressive localization of the pair in the inner planes. The effect becomes more pronounced at stronger coupling which deepens the potential well and effectively elevates the barriers. At larger $n$, there are more inner planes where the pair tends to reside. The wave function at the outer planes continues to decrease, reducing the overall probability to tunnel across the barrier. That is why $m^{\ast}_{z}$ sharply increases with $n$ for $n > 2$ and $|U| > 8t$.

\begin{figure}[t]
\includegraphics[width=0.95\textwidth]{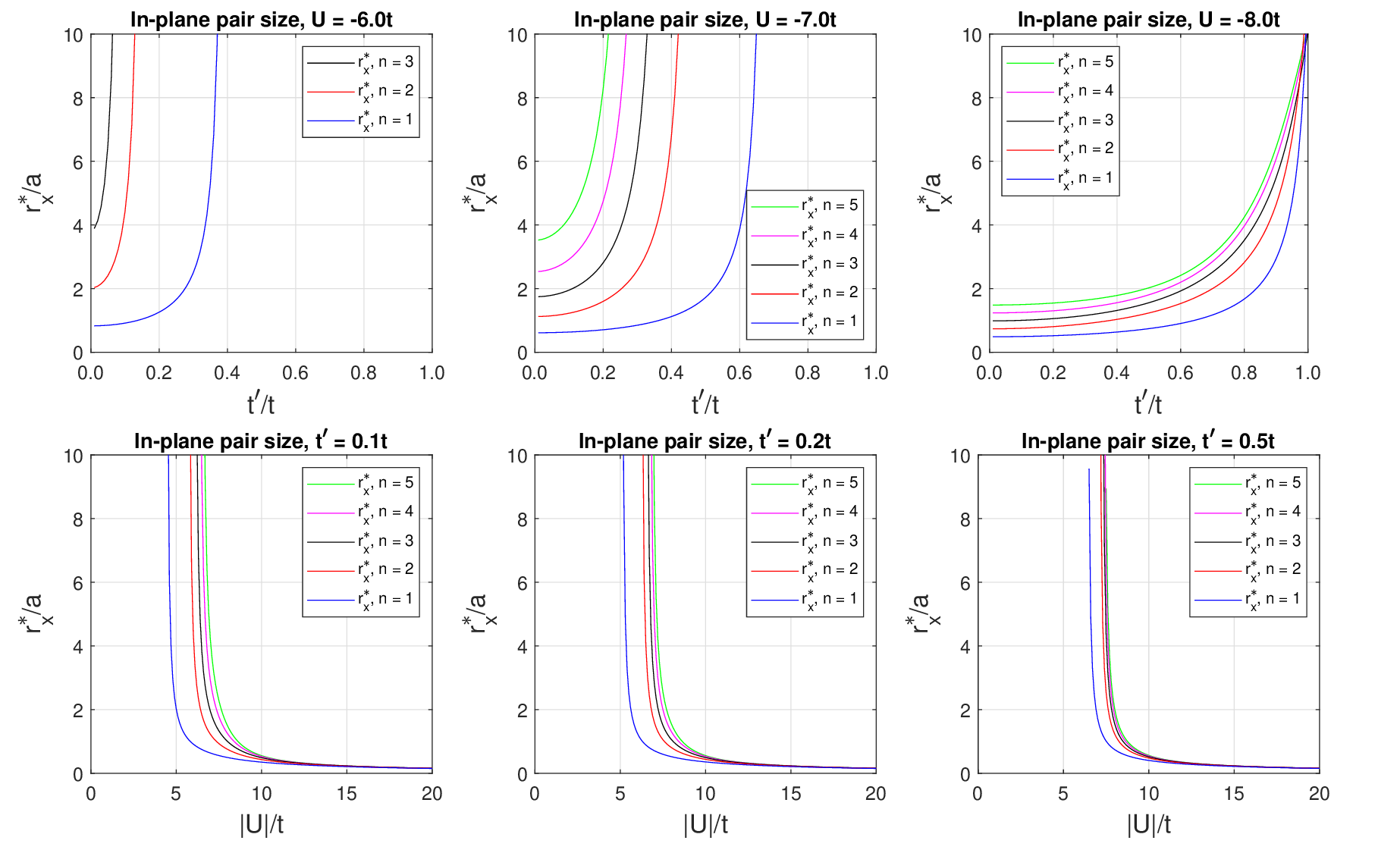}
\caption{In-plane pair size $r^{\ast}_x$ in the $n = 1, 2, 3, 4, 5$ uniform attractive Hubbard models. The top row shows dependencies on the anisotropy parameter $t^{\prime}/t$ and the bottom row dependencies on attractive strength $\vert U \vert$.} 
\label{MLH:fig:Efour}
\end{figure}
\begin{figure}[t]
\includegraphics[width=0.95\textwidth]{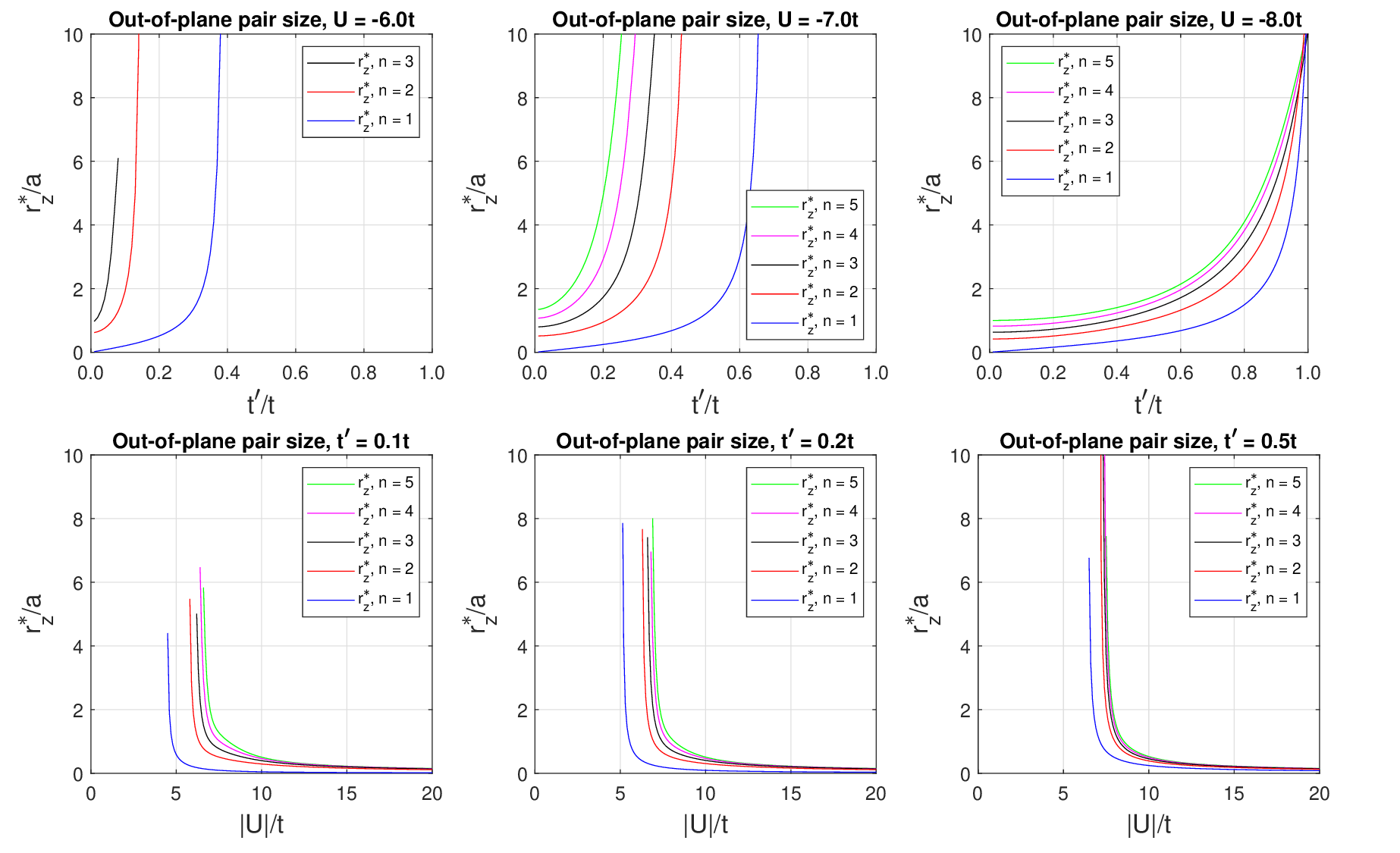}
\caption{Out-of-plane pair size $r^{\ast}_z$ in the $n = 1, 2, 3, 4, 5$ uniform attractive Hubbard models. The top row shows dependencies on the anisotropy parameter $t^{\prime}/t$ and the bottom row dependencies on attractive strength $\vert U \vert$.} 
\label{MLH:fig:Efive}
\end{figure}

Pair sizes are now discussed. The in-plane effective radius, $r^{\ast}_x$, and out-of-plane effective radius, $r^{\ast}_z$, are shown in Figs.~\ref{MLH:fig:Efour} and \ref{MLH:fig:Efive}, respectively. The most conspicuous feature of both properties is the divergence near the threshold. The divergence is rather sharp, as is apparent in the plots. We recall that the divergence of $r^{\ast}_{x,z}$ and hence of pair volume $\Omega_{p}$ is an essential ingredient of the physical picture developed in this paper. Near the threshold, the pair is loosely bound, $\Omega_{p}$ is large, and the corresponding packing density is small. As a result, $T^{\ast}_{\rm cp}$ is suppressed. This limits superconductivity on the side of high kinetic energy (that is, low anisotropy). Insertion of more planes shifts the threshold and increases $\Omega_{p}$ in a step-wise fashion. That is why, $T^{\ast}_{\rm cp}$ eventually goes down as $n$ increases.

\begin{figure}[t]
\includegraphics[width=0.98\textwidth]{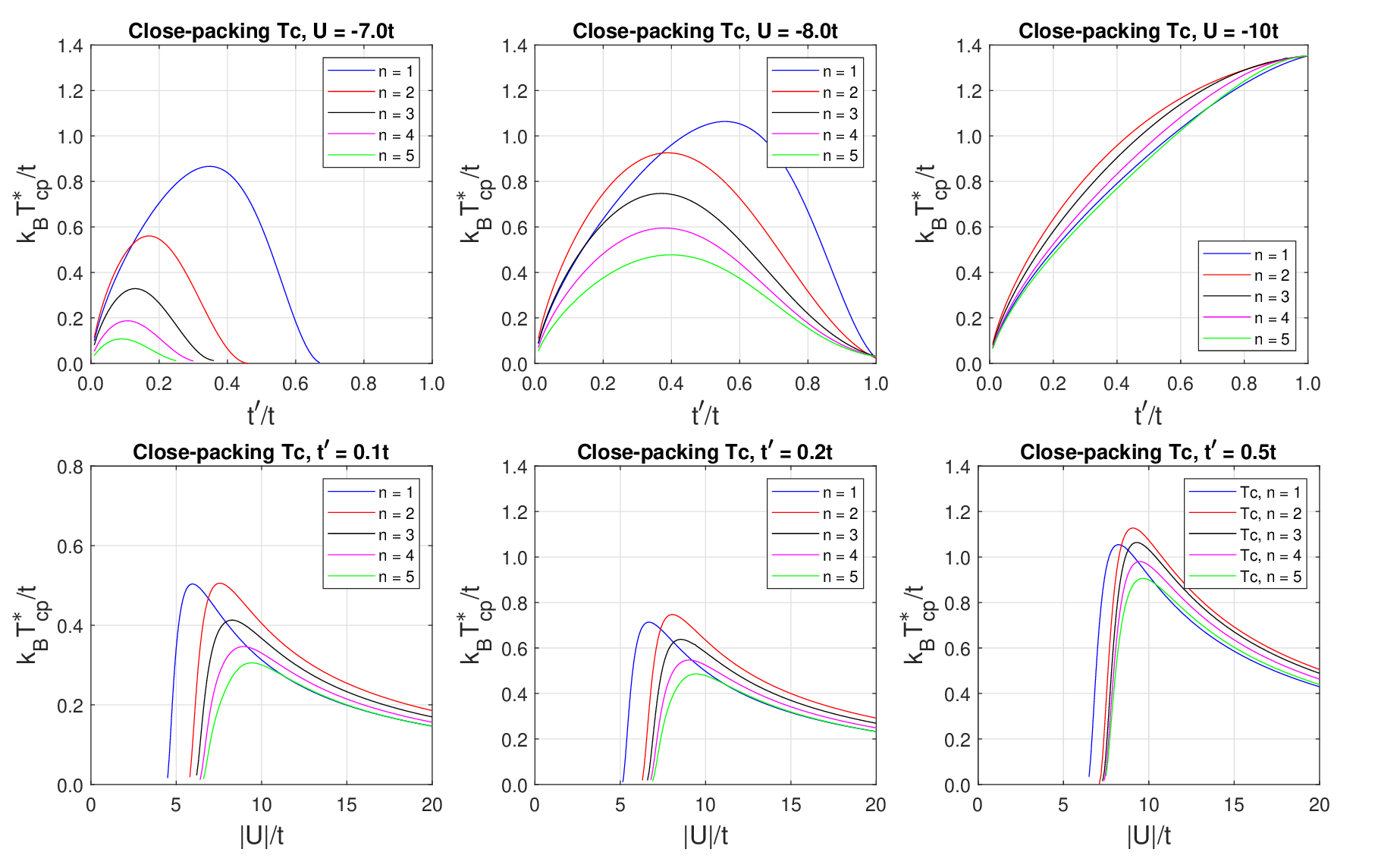}
\caption{Close-packing temperature $T^{\ast}_{\rm cp}$ in the $n = 1, 2, 3, 4, 5$ uniform attractive Hubbard models. The top row shows dependencies on the anisotropy parameter $t^{\prime}/t$ and the bottom row dependencies on attractive strength $\vert U \vert$. Observe how, for example, for $t^{\prime} = 0.1t$ and $U = -8.0t$, $T^{\ast}_{\rm cp}(n = 2) > T^{\ast}_{\rm cp}(n = 1)$ in agreement with experiment, but $T^{\ast}_{\rm cp}(n = 3) < T^{\ast}_{\rm cp}(n = 2)$ in disagreement with experiment.} 
\label{MLH:fig:Esix}
\end{figure}

Finally, we show the close-packing critical temperature in Fig.~\ref{MLH:fig:Esix}. The curves have characteristic peak shapes indicating the existence of optimal parameters. As discussed above, on the low-$t^{\prime}$, high-$|U|$ side, $T^{\ast}_{\rm cp}$ is limited by the large out-of-plane mass $m^{\ast}_{z}$, whereas on the high-$t^{\prime}$, low-$|U|$ side (close to the threshold), $T^{\ast}_{\rm cp}$ is limited by a high pair volume and correspondingly low packing density. For fixed $U$ and $t^{\prime}$, increasing the number of layers $n$ changes the pairing threshold in a discrete fashion, producing a discrete sequence $T^{\ast}_{\rm cp}(n)$ that spans both sides of the peak. Thus, this model is capable of producing a non-monotonic $T^{\ast}_{\rm cp}(n)$ in agreement with cuprate phenomenology. However, examination of Fig.~\ref{MLH:fig:Esix} reveals (see for example the low-left panel) that quite generally $T^{\ast}_{\rm cp}(n)$ peaks at $n = 2$ rather than at $n = 3$ as observed experimentally. The decrease of $T^{\ast}_{\rm cp}(n \geq 3)$ relative to $T^{\ast}_{\rm cp}(n = 2)$ can be traced back to the sharp increase of the pair out-of-plane mass due to pair localization on inner planes discussed above. Thus, while the uniform model, $\epsilon_{\alpha} = 0$, explains non-monotonic $T^{\ast}_{\rm cp}(n)$, it cannot produce a $T^{\ast}_{\rm cp}(n)$ that peaks at $n = 3$. We now examine a more general model that is fully consistent with cuprate physics.

\section{\label{MLH:sec:app:h}
``Low-outer-planes'' model, $\epsilon_{1} = \epsilon_{n} = \Delta$
}

Out of all possible models describing stacks of non-equivalent planes, the one with low one-particle energy of outer planes is of special interest. This model is supported by the recent observation of a strong polaron effect in three-layer cuprate HgBa$_2$Ba$_2$Cu$_3$O$_{8+\delta}$ \cite{Hong2025}. Apical oxygens are displaced in response to attraction from the positively charged holes residing in outer planes. In turn, the displaced oxygens lower the holes' energy forming polarons. The energy lowering, also known as {\em polaron shift}, is larger for outer planes than for inner planes since the former are closer to the oxygens. This difference in polaron shifts can be modeled by assigning an extra negative energy to one-particle levels of the outer planes. The model is parameterized by $\epsilon_{1} = \epsilon_{n} = \Delta$ with $\Delta < 0$. Hereafter, it will be called the {\em low outer planes} model. This section is dedicated to analysis of its properties.

\begin{figure*}[t]
\includegraphics[width=0.95\textwidth]{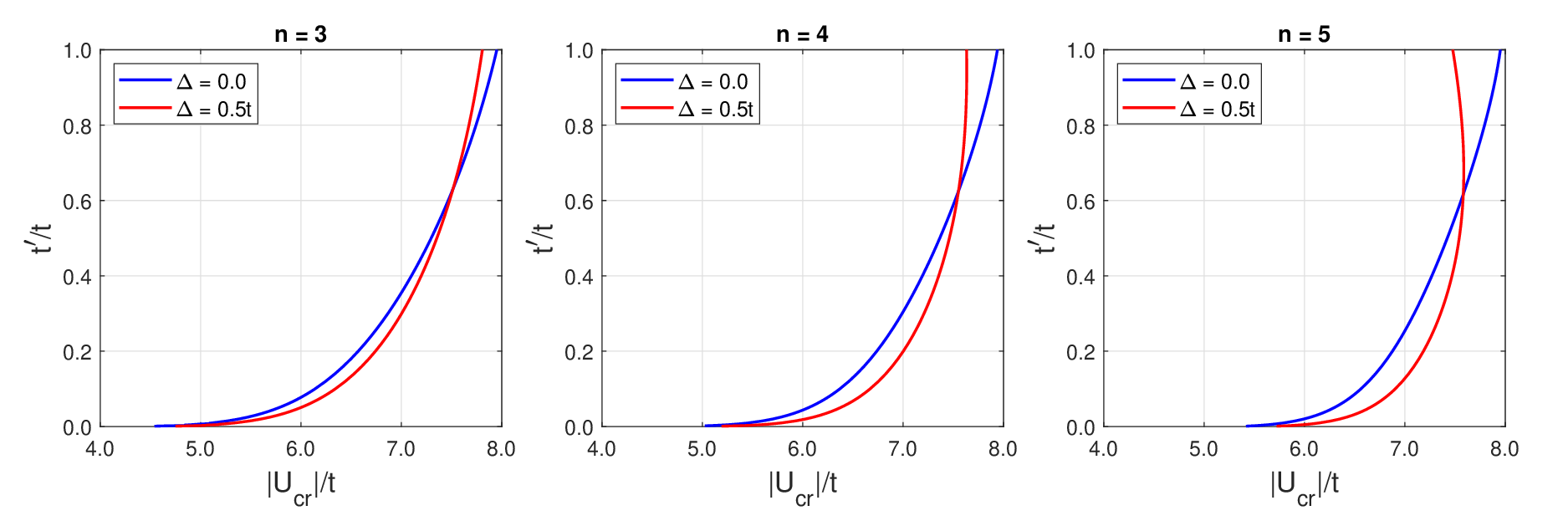}
\caption{Pairing threshold in the low-outer-planes model: one-particle energy of outer planes is pulled down by $\Delta$. The $\Delta = 0.0$ curves are the same as in Fig.~\ref{MLH:fig:Eone}. Notice different behaviors in $t^{\prime} \ll t$ and $t^{\prime} \approx t$ regimes.} 
\label{MLH:fig:Hone}
\end{figure*}

First, we discuss the pairing threshold shown in Fig.~\ref{MLH:fig:Hone}. The behaviors at weak and strong anisotropies are different. At weak anisotropy, $t^{\prime} \approx t$ and $\Delta = 0$, one-particle states are uniformly delocalized across all planes while a negative $\Delta$ localizes them on outer planes. That effectively reduces their kinetic energy and as a consequence the amount of attraction needed to overcome it to form a pair. As a result, the pairing threshold goes down. At strong anisotropy, $t^{\prime} \ll t$, one-particle states are localized on {\em inner} planes already at $\Delta = 0$. A negative $\Delta$ redistributes the wave function from inner to outer planes, effectively delocalizing the states. That increases the states' kinetic energy and the amount of attraction needed to form a pair. Thus, the pairing threshold increases from its $\Delta = 0$ value.

\begin{figure*}[t]
\includegraphics[width=0.98\textwidth]{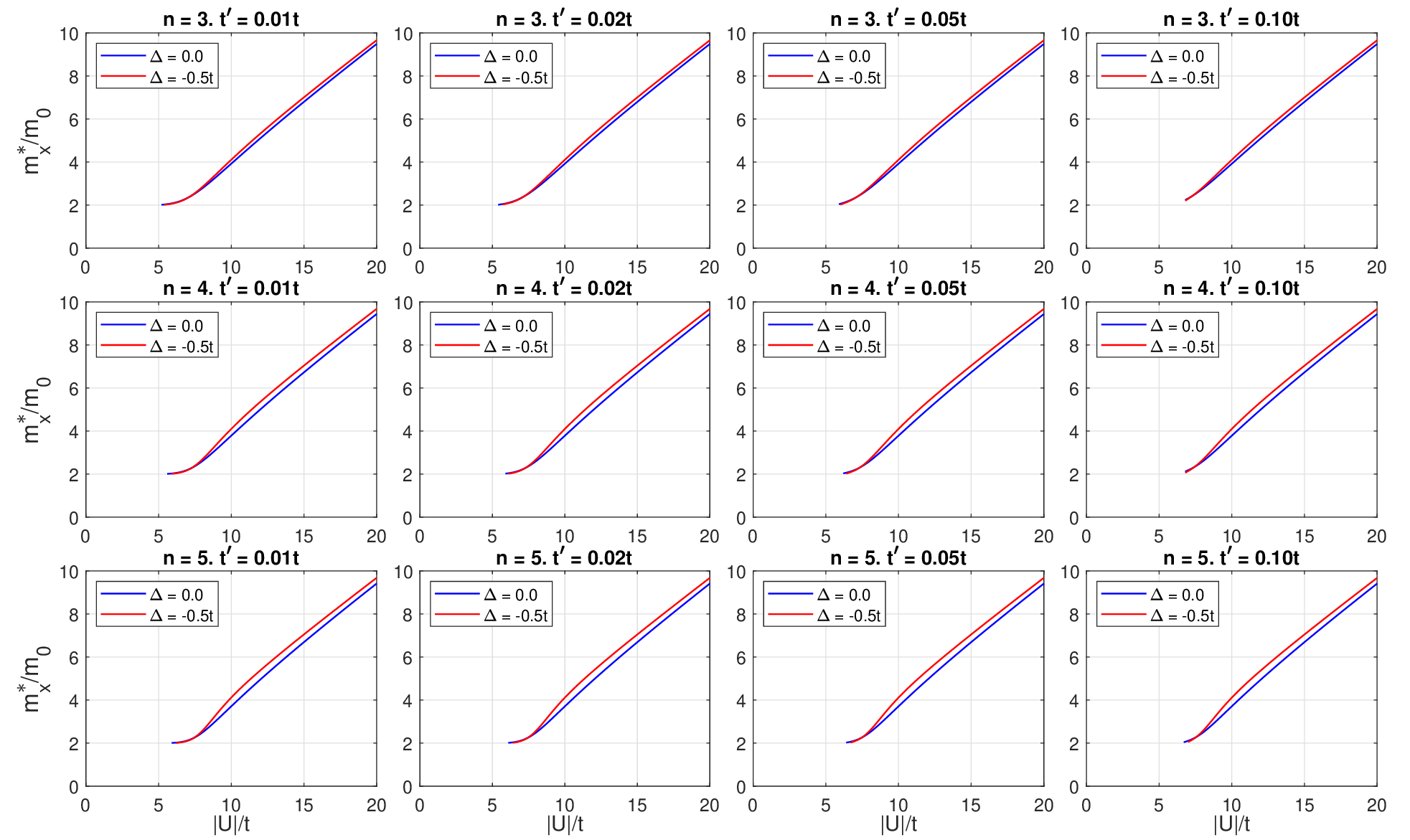}
\caption{In-plane pair mass $m^{\ast}_{x}$ in the low-outer-planes models, $n = 3 , 4 , 5$. Notice how weakly $m^{\ast}_{x}$ depends on the energy shift $\Delta$.} 
\label{MLH:fig:Htwo}
\end{figure*}
\begin{figure*}[t]
\includegraphics[width=0.98\textwidth]{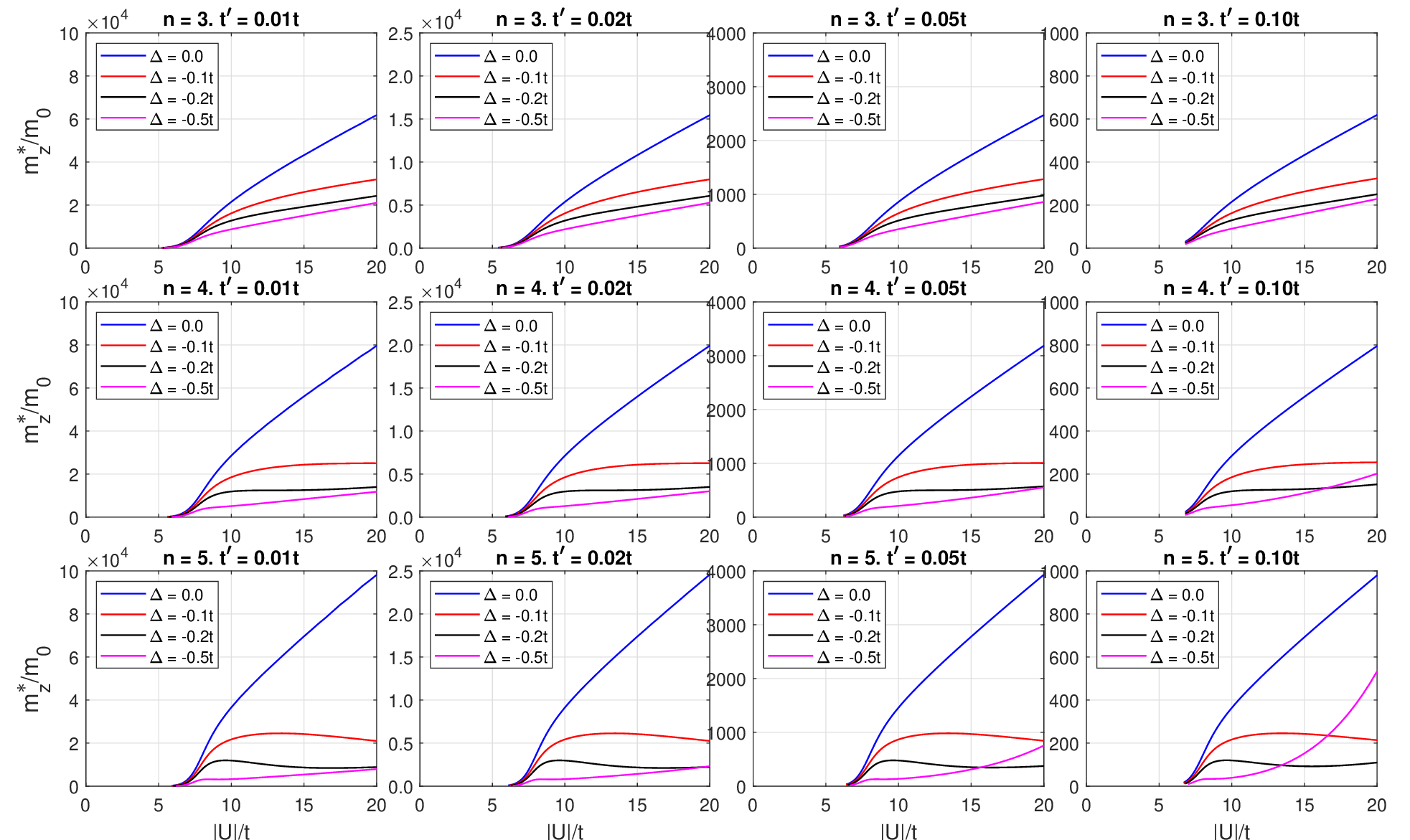}
\caption{Out-of-plane pair mass $m^{\ast}_{z}$ in the low-outer-planes models, $n = 3 , 4 , 5$. There is a significant decrease of mass at negative $\Delta$ for all $n$. Note different scales of $y$ axes.} 
\label{MLH:fig:Hthree}
\end{figure*}

Since $\Delta$ adds a third model parameter, complete characterization of the entire parameter space is challenging. In the following, we focus on the most physically interesting region of strong anisotropy, $t^{\prime} \leq 0.1 t$. The in-plane and out-of-plane pair masses are shown in Figs.~\ref{MLH:fig:Htwo} and \ref{MLH:fig:Hthree}, respectively. The two masses display two different $\Delta$ behaviors. The in-plane mass $m^{\ast}_{x}$ is almost insensitive to $\Delta$. Physically, this is understandable: the energy landscape remains uniform in the $(xy)$ plane so that in-plane movement of a pair at $\Delta \neq 0$ is governed by the same energetics as at $\Delta = 0$. In contrast, in the $z$ direction, the pair wave function is already non-uniform because of the weak coupling between the multi-plane stacks. At $\Delta = 0$, the wave function at the outer planes is small relative to the inner planes, reducing the probability to tunnel between the stacks and producing a relatively high $m^{\ast}_{z}$. A negative $\Delta$ pulls the pair more on the outer planes. Since the residence time on the outer planes increases, the probability to tunnel between plane stacks goes up and $m^{\ast}_{z}$ drops. The mass reduction seen in Fig.~\ref{MLH:fig:Hthree} is significant enough to elevate $T^{\ast}_{\rm cp}(n = 3)$ above $T^{\ast}_{\rm cp}(n = 2)$ and produce $T^{\ast}_{\rm cp}(n)$ consistent with the empirical cuprates data.

\begin{figure*}[t]
\includegraphics[width=0.98\textwidth]{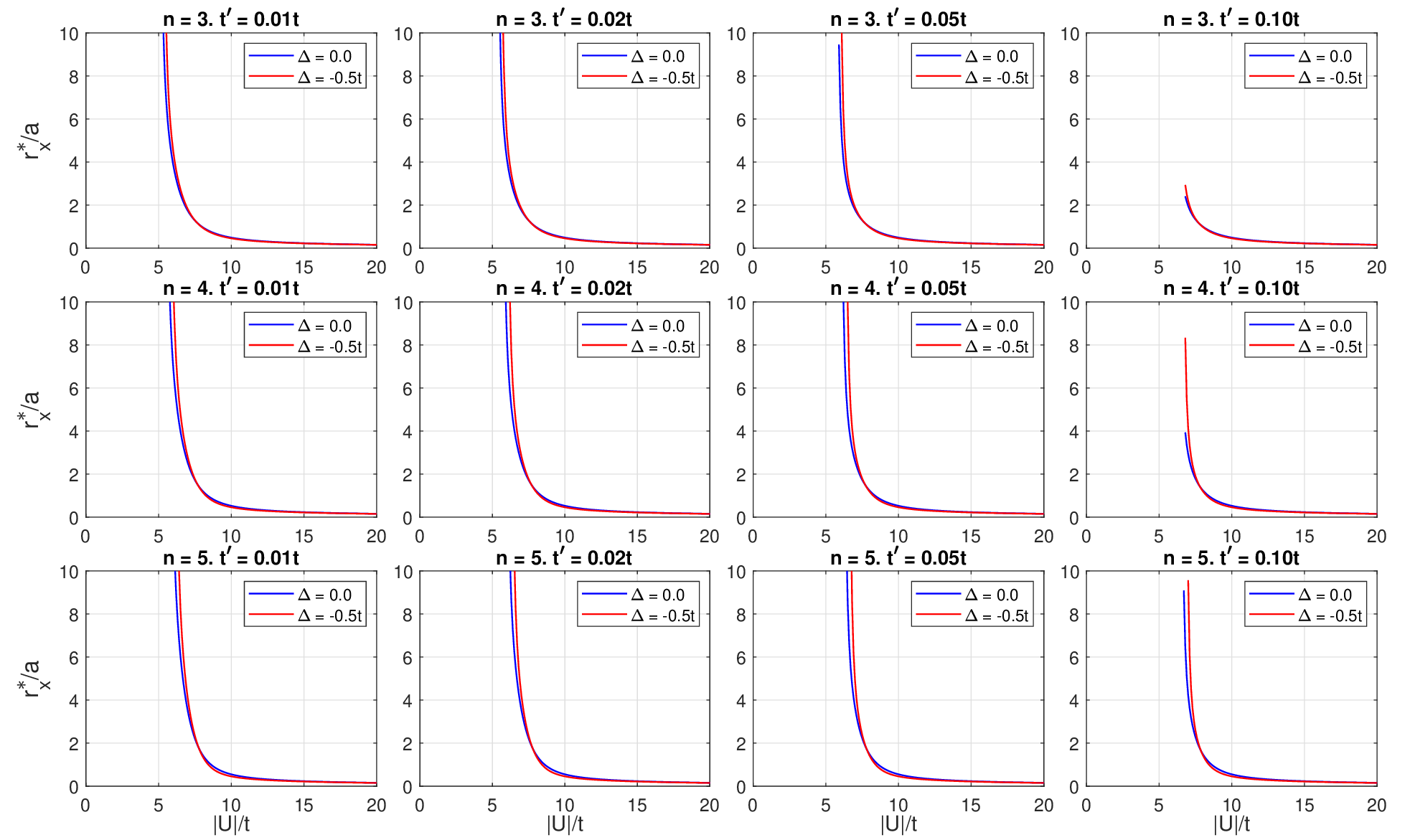}
\caption{In-plane pair size $r^{\ast}_{x}$ in the low-outer-planes models, $n = 3 , 4 , 5$. } 
\label{MLH:fig:Hfour}
\end{figure*}
\begin{figure*}[t]
\includegraphics[width=0.98\textwidth]{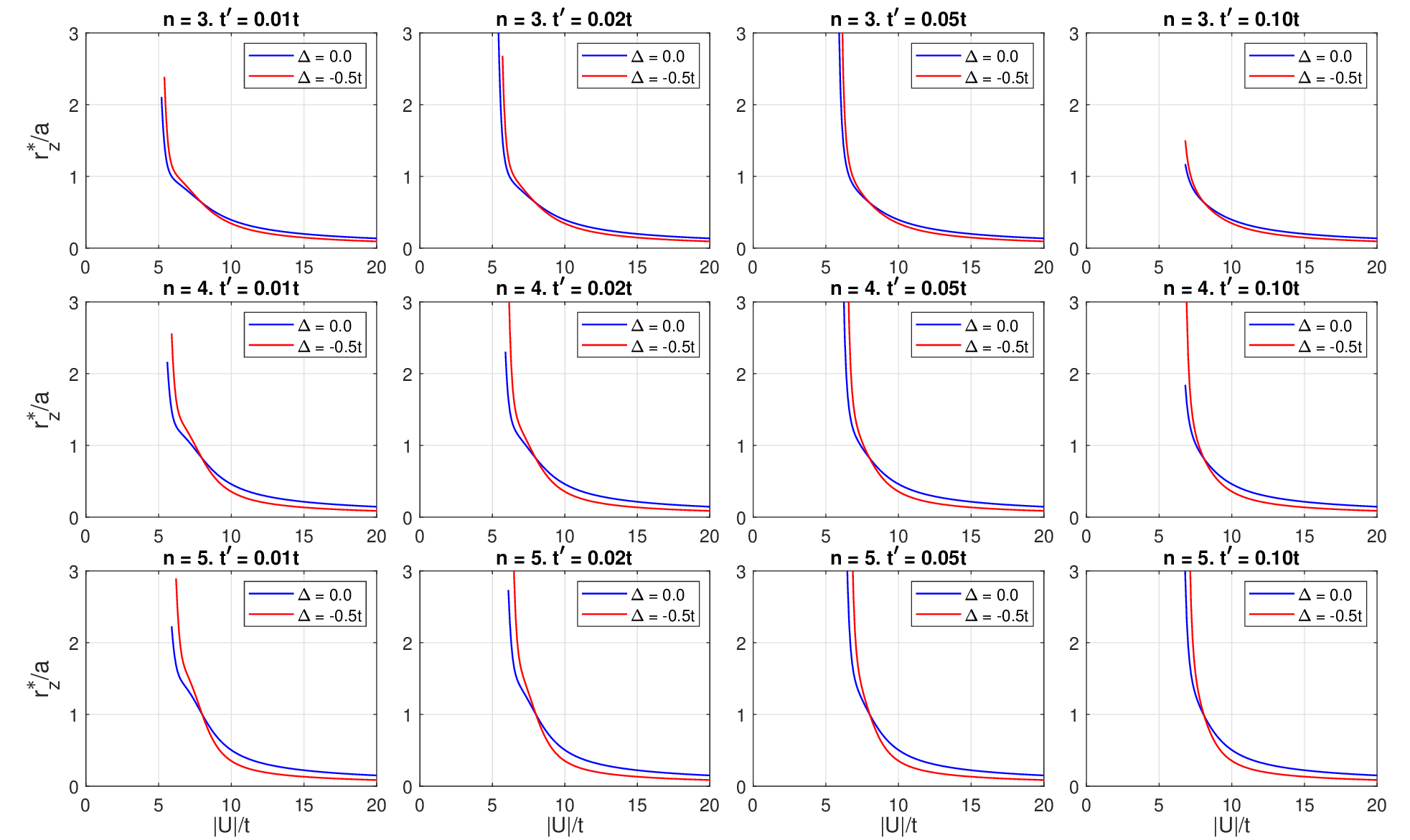}
\caption{Out-of-plane pair size $r^{\ast}_{z}$ in the low-outer-planes models, $n = 3 , 4 , 5$. } 
\label{MLH:fig:Hfive}
\end{figure*}

Next, pair sizes are discussed. The in-plane and out-of-plane pair effective radii are shown in Figs.~\ref{MLH:fig:Hfour} and \ref{MLH:fig:Hfive}, respectively. We note a weak dependence of both sizes on $\Delta$ except small shifts of the pairing thresholds discussed previously. Since $r^{\ast}_{x,z}$ are singular at threshold, the size change with $\Delta$ may be significant close to the threshold. This effect contributes to the $\Delta$-dependence of $T^{\ast}_{\rm cp}$.

\begin{figure}[t]
\includegraphics[width=0.98\textwidth]{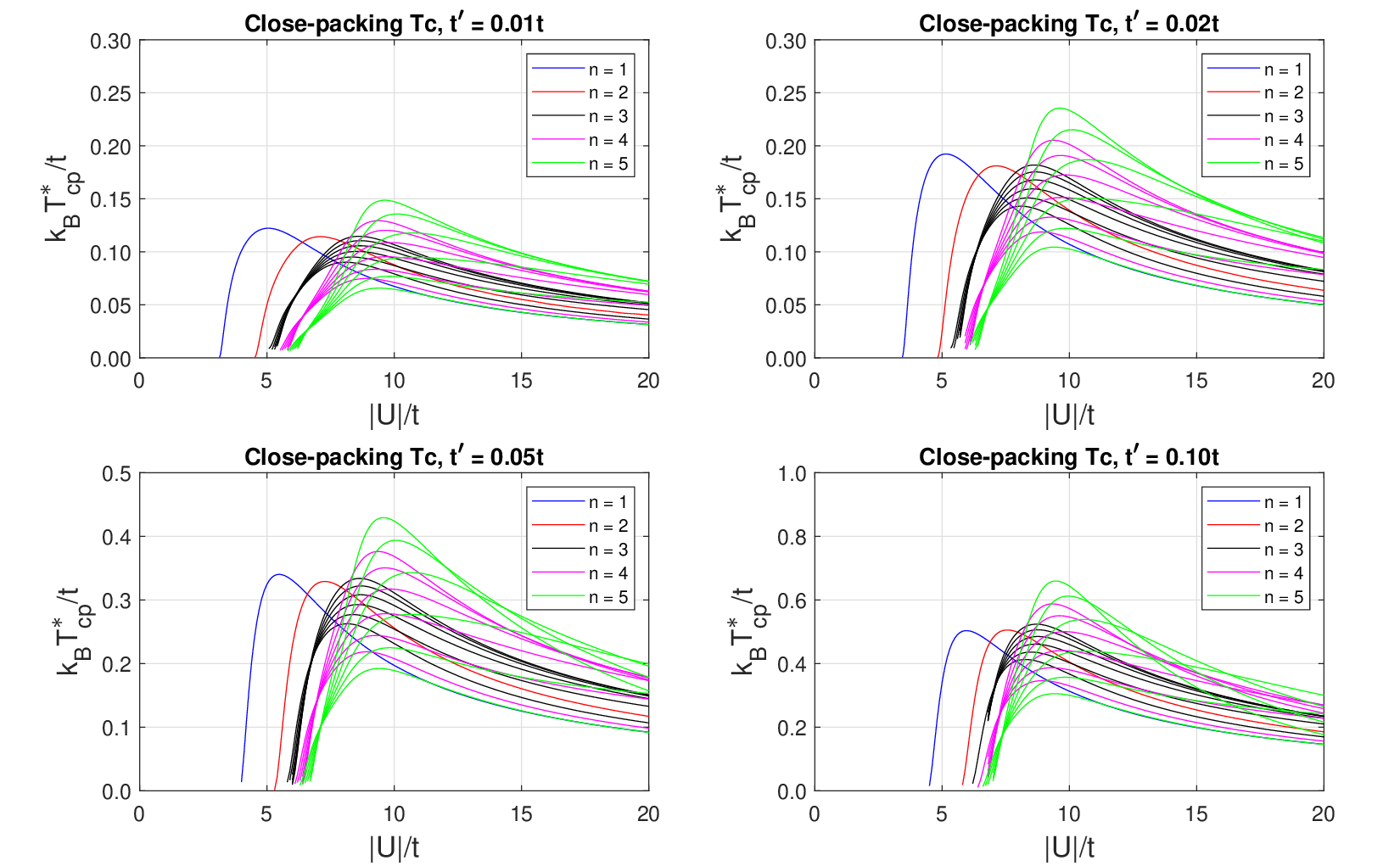}
\caption{Close-packing temperature $T^{\ast}_{\rm cp}$ in the $n = 1, 2, 3, 4, 5$ attractive Hubbard models. The $n = 3, 4$ and $5$ families show six traces each. The bottom one is $\Delta = 0.0$ and the rest is, from bottom to top, are $\Delta = -0.1t$, $-0.2t$, $-0.3t$, $-0.4t$, and $-0.5t$. } 
\label{MLH:fig:Hsix}
\end{figure}

Finally, close-packing critical temperature is shown in Fig.~\ref{MLH:fig:Hsix}.

\end{widetext}

\end{appendix}

\end{document}